\documentclass[runningheads]{llncs}
\pdfoutput=1

\usepackage{graphicx}
\usepackage{subfigure}
\usepackage{comment} 
\usepackage{microtype}
\usepackage{amsopn}
\usepackage{todo}
\usepackage{color, colortbl}
\usepackage{textcomp}                  
\usepackage{mathptmx}                  
\usepackage{times}                     
\usepackage{cite}                      
\usepackage{tabu}                      
\usepackage{booktabs}                  
\usepackage{algpseudocode}
\usepackage{algorithm}
\usepackage{amssymb}
\usepackage{amsmath}
\usepackage{xcolor}
\usepackage{graphicx}
\usepackage{microtype} 
\usepackage{subfigure} 
\usepackage{amsfonts}
\usepackage{hyperref}
\usepackage{comment}
\usepackage{wrapfig}
\usepackage{setspace}
\usepackage{amsmath}
\usepackage{multirow}

\DeclareGraphicsExtensions{.pdf,.png,.jpg}

\long\def\remove#1{}

\newcommand {\mm}[1] {\ifmmode{#1}\else{\mbox{\(#1\)}}\fi}

\newcommand{\reals}	{{\rm I\!\hspace{-0.025em} R}}

\newcommand{\HH}		{{{H}}}

\newcommand{\XX}		{{{X}}}

\DeclareMathOperator\Dg{Dg} 
\DeclareMathOperator\eDg{ExDg} 

\definecolor{red}{rgb}{0.6, 0, 0}
\definecolor{blue}{rgb}{0, 0, 0.6}
\definecolor{green}{rgb}{0.16, 0.435, 0.16}

\renewcommand{\paragraph}[1]	{{\vspace*{0.1in}\noindent {\bf #1.~}}}

\newcommand{\ignore}[1]{}

\newcommand{\figref}[1]{Fig.~\ref{#1}}

\begin{document}

\title{Propagate and Pair: A Single-Pass Approach to Critical Point Pairing in Reeb Graphs}

\titlerunning{Propagate and Pair}

\author{Junyi Tu\inst{1}\orcidID{0000-0001-7026-7454} \and
Mustafa Hajij\inst{2} \and
Paul Rosen\inst{1}\orcidID{0000-0002-0873-9518}}

\authorrunning{J. Tu et al.}

\institute{University of South Florida, Tampa FL 33620, USA\\
\email{\{junyi,prosen\}@mail.usf.edu} \and
The Ohio State University, Columbus 43210, USA}

\maketitle

\begin{abstract}
With the popularization of Topological Data Analysis, the Reeb graph has found new applications as a summarization technique in the analysis and visualization of large and complex data, whose usefulness extends beyond just the graph itself. Pairing critical points enables forming topological fingerprints, known as persistence diagrams, that provides insights into the structure and noise in data. Although the body of work addressing the efficient calculation of Reeb graphs is large, the literature on pairing is limited. In this paper, we discuss two algorithmic approaches for pairing critical points in Reeb graphs, first a multipass approach, followed by a new single-pass algorithm, called Propagate and Pair. 


\end{abstract}

\section{Introduction}
\label{sec:introduction}

The last two decades have witnessed great advances in methods using topology to analyze data, in a process called Topological Data Analysis (TDA). Their popularity is due in large part to their robustness and applicability to a variety of domains~\cite{munch2017user}. The \textit{Reeb graph}~\cite{reeb1946points}, which encodes the evolution of the connectivity of the level sets induced by a scalar function defined on a data domain, was originally proposed as a data structure to encode the geometric skeleton of 3D objects, but recently it has been re-purposed as a tool in TDA. Beside their usefulness in handling large data~\cite{edelsbrunner2004time}, Reeb graphs and their non-looping relative, contour trees~\cite{boyell1963hybrid}, have been successfully used in feature detection~\cite{takahashi2004topological}, data reduction and simplification~\cite{CarSnoPan2004,rosen2017using}, image processing~\cite{kweon1994extracting}, shape understanding~\cite{attene2003shape}, visualization of isosurfaces~\cite{bajaj1997contour} and many other applications.

\begin{figure}[!t]
	\centering
    {\begin{minipage}{0.440\linewidth}
	\subfigure[Data\label{fig.pipeline.mesh}]{\hspace{0pt}\includegraphics[trim=0 0 0 0, clip, height=4.25cm]{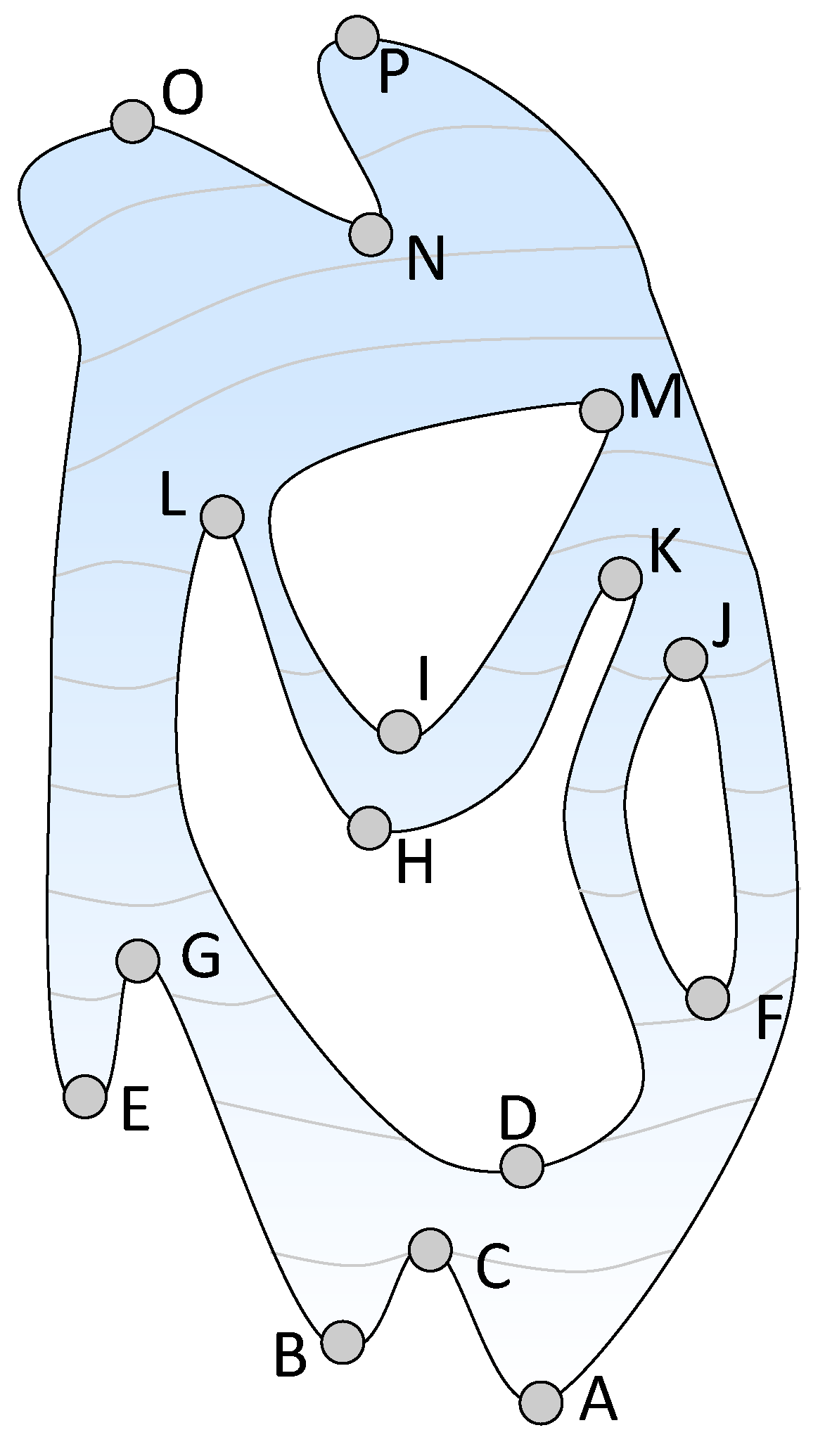}\hspace{0pt}}
	{\begin{minipage}{0.1\linewidth}
	    \vspace{-100pt}
	    \hspace{-5pt}
	    \includegraphics[trim=180pt 0 170pt 0, clip, height=2.25cm]{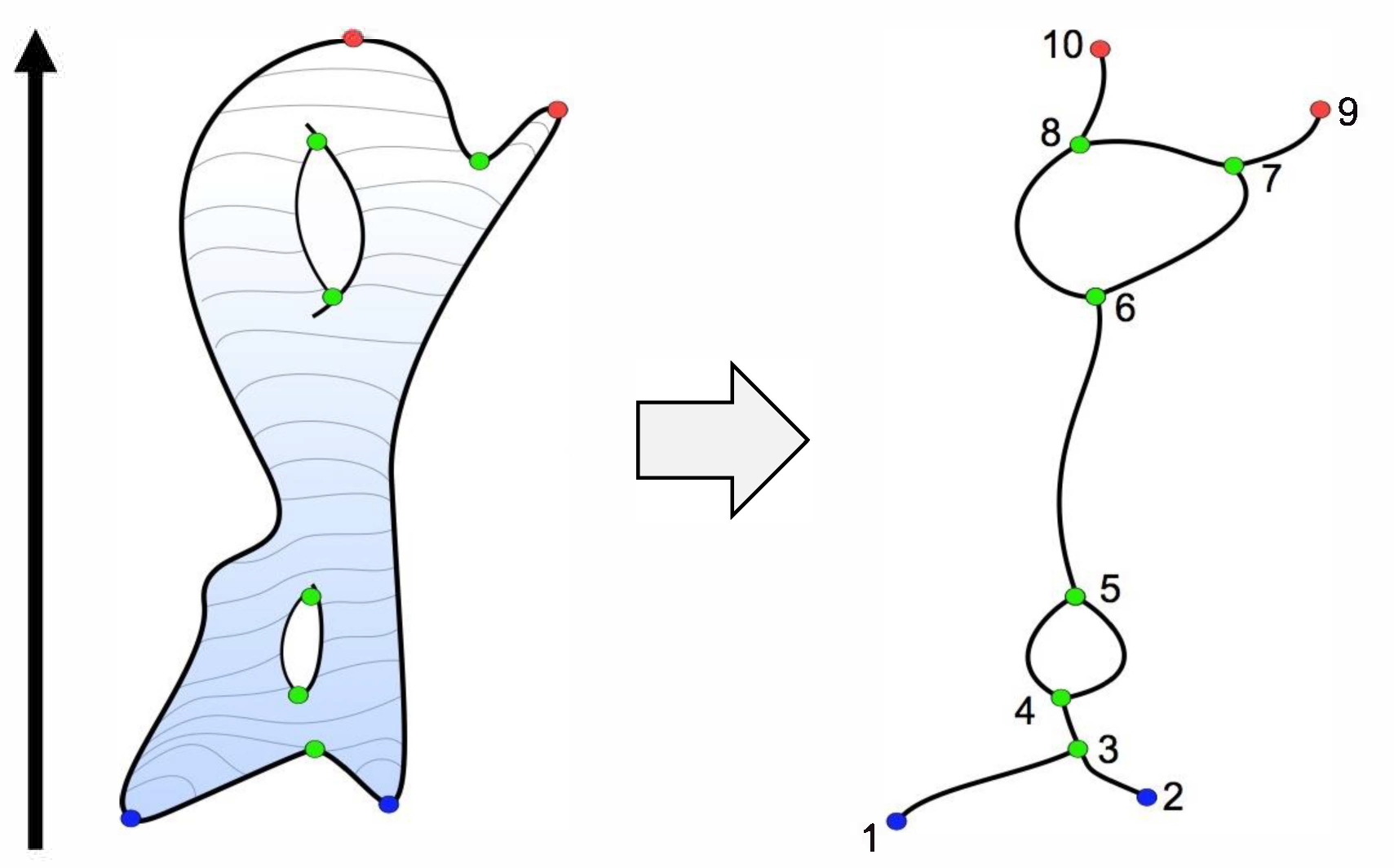}
	\end{minipage}}
    \hspace{-6pt}
    \subfigure[Reeb graph\label{fig.pipeline.reeb}]{\hspace{0pt}\includegraphics[height=4.25cm]{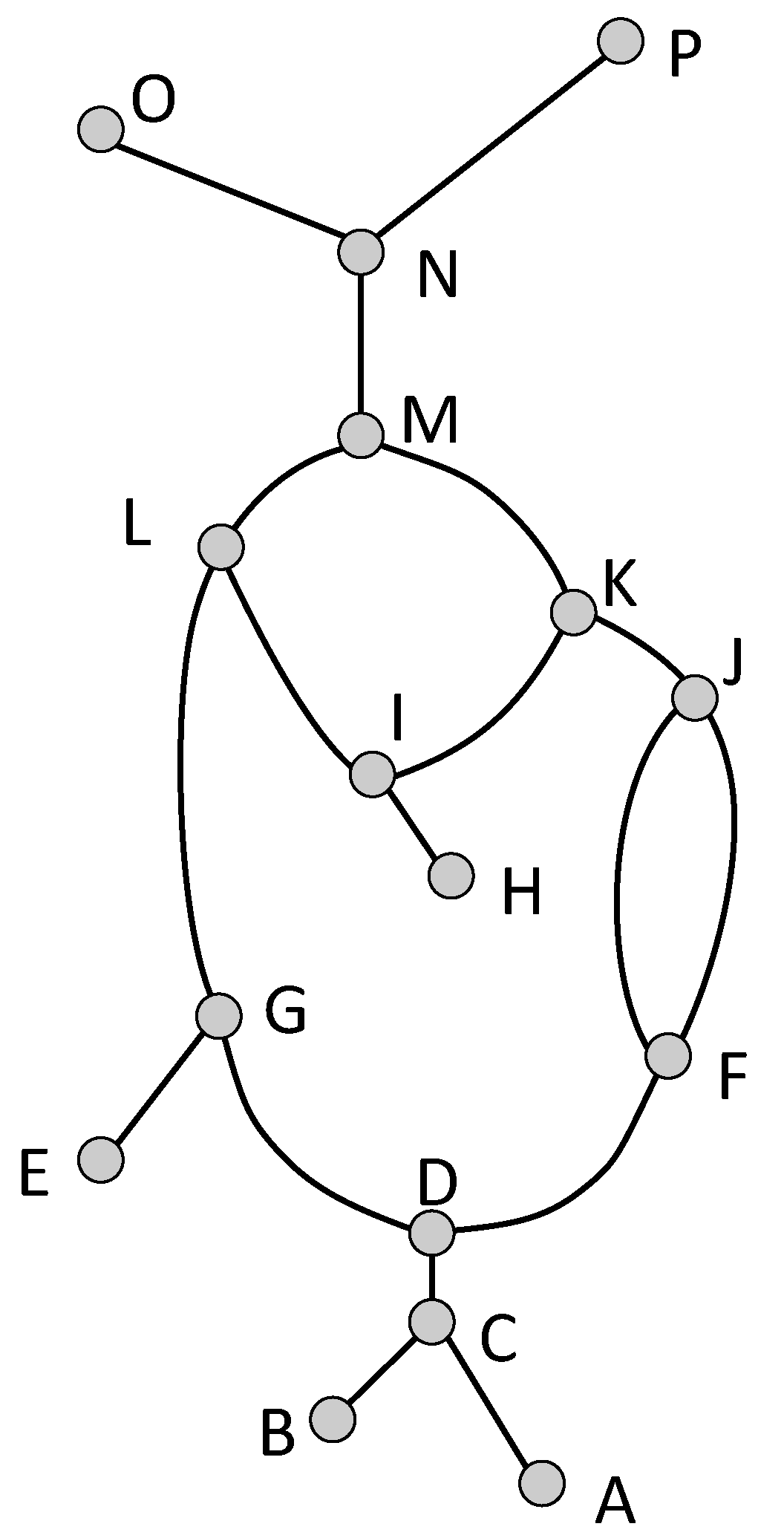}\hspace{0pt}}
    \end{minipage}}
    \hfill
    {\begin{minipage}{0.13\linewidth}
	        \subfigure[Split Tree\label{fig.reebexample.st}]{\hspace{7pt}\includegraphics[trim = 65pt 0 65pt 0, clip, width=33.5pt]{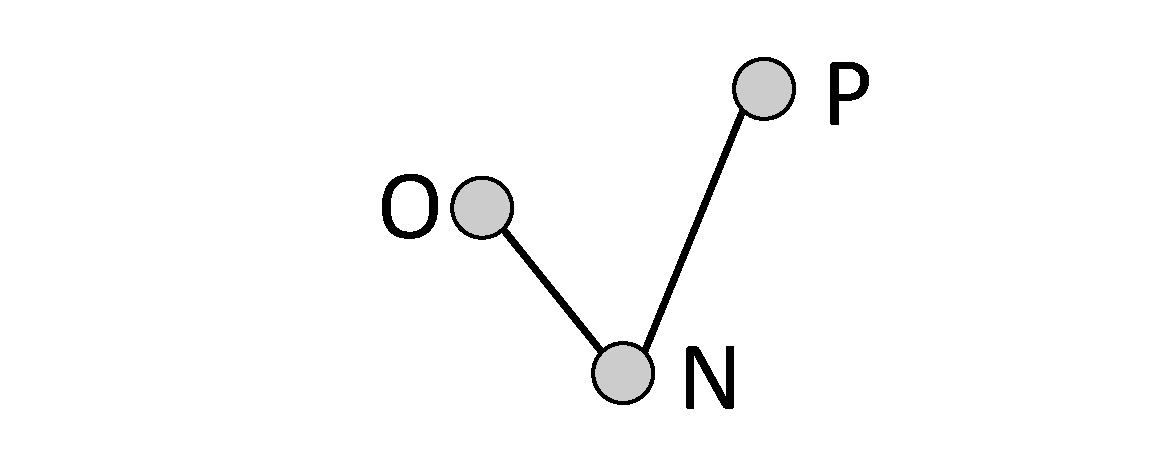}\hspace{7pt}}
	        \subfigure[Join Tree\label{fig.reebexample.mt}]{\hspace{6pt}\includegraphics[trim = 65pt 0 65pt 0, clip, width=33.5pt]{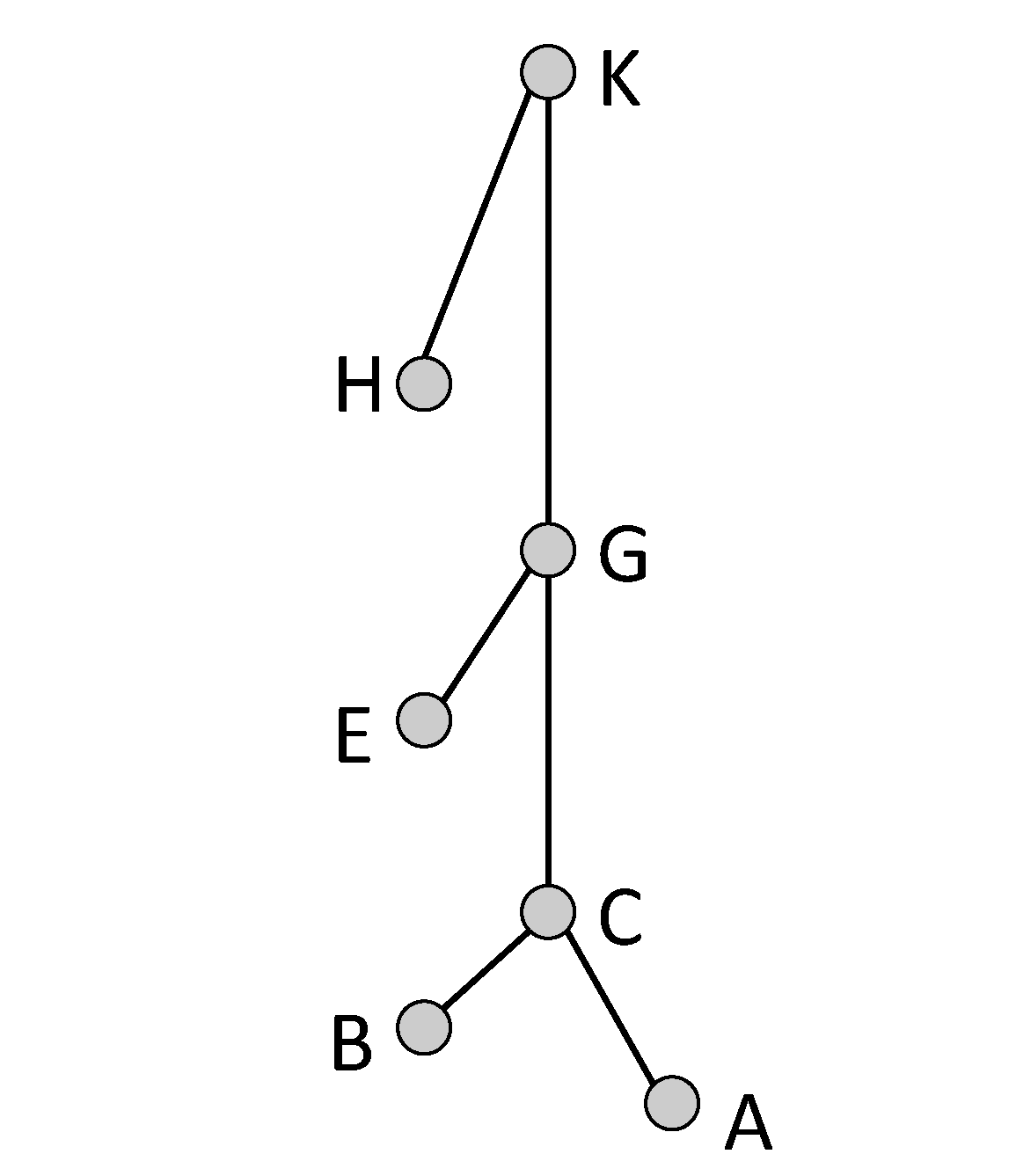}\hspace{6pt}}
    \end{minipage}}
    {\begin{minipage}{0.18\linewidth}
	    \subfigure[Ess.\ Forks\label{fig.reebexample.ess}]{\includegraphics[height=4.35cm]{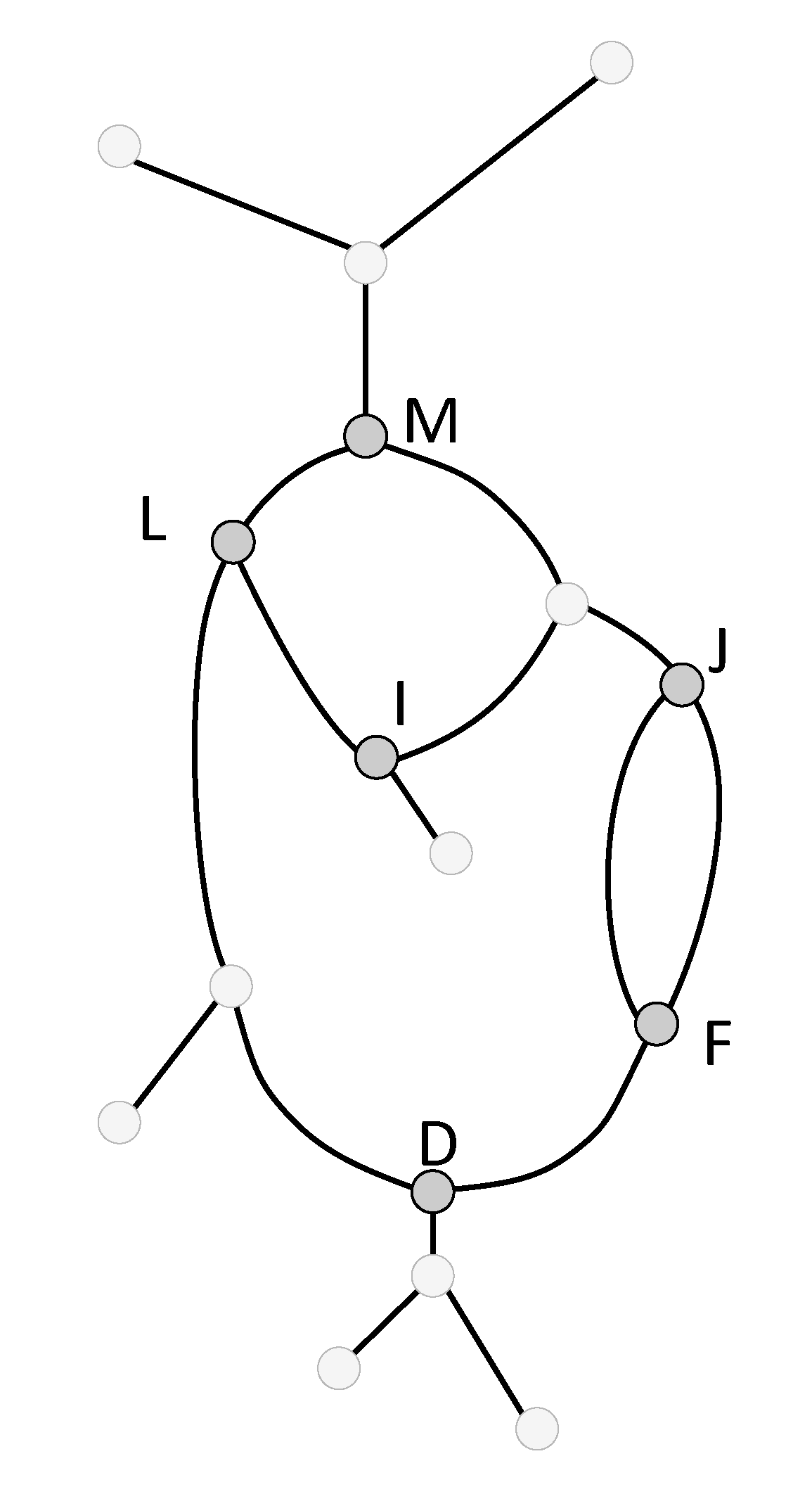}}    
    \end{minipage}}
    \hfill
    {\begin{minipage}{0.15\linewidth}
        \subfigure[$\Dg_0(f)$\label{fig.reebexample.pd0}]{\includegraphics[height=1.75cm]{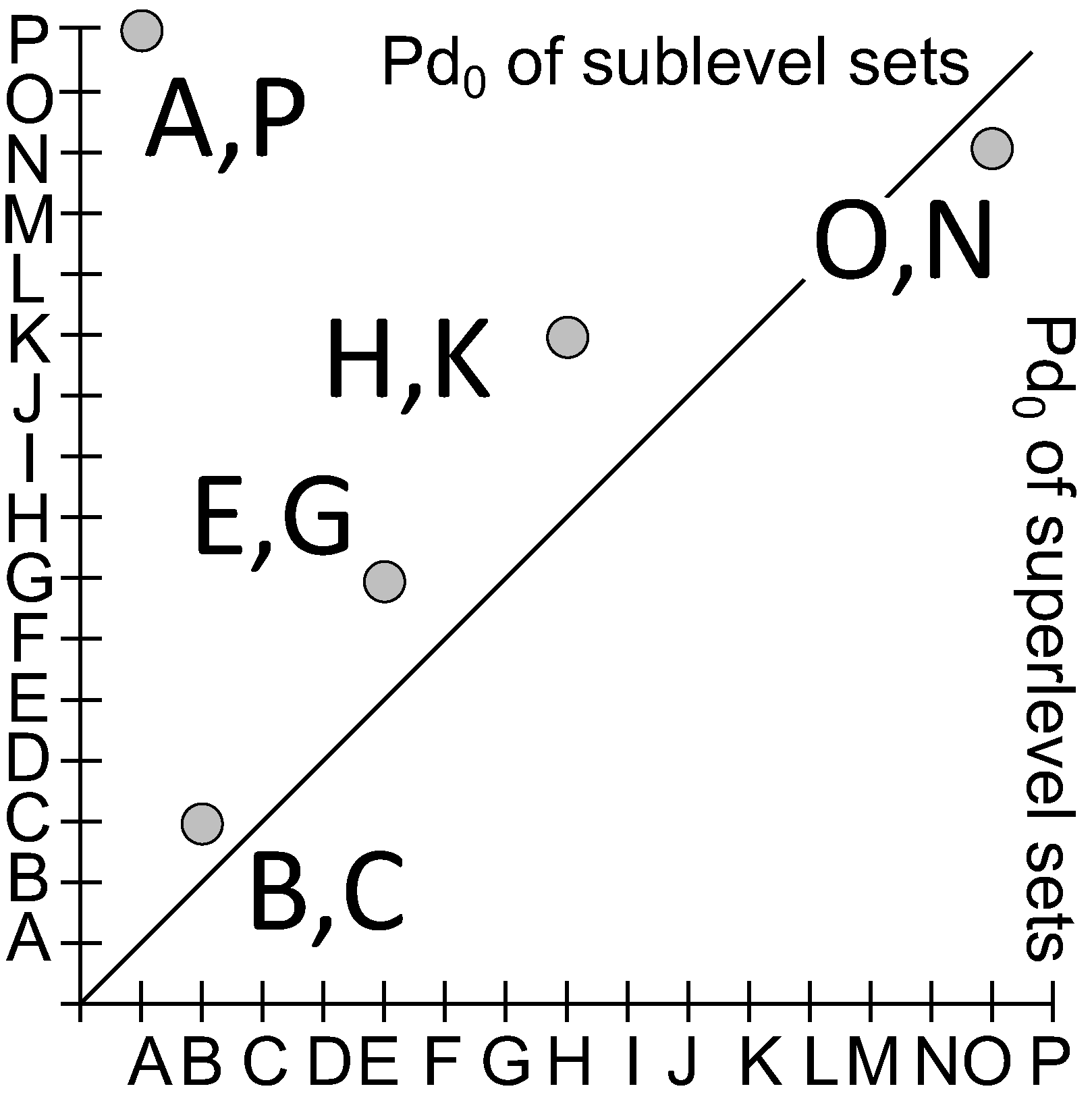}\hspace{1pt}}
        \subfigure[$\eDg_1(f)$\label{fig.reebexample.pd1}]{\hspace{2pt}\includegraphics[height=1.75cm]{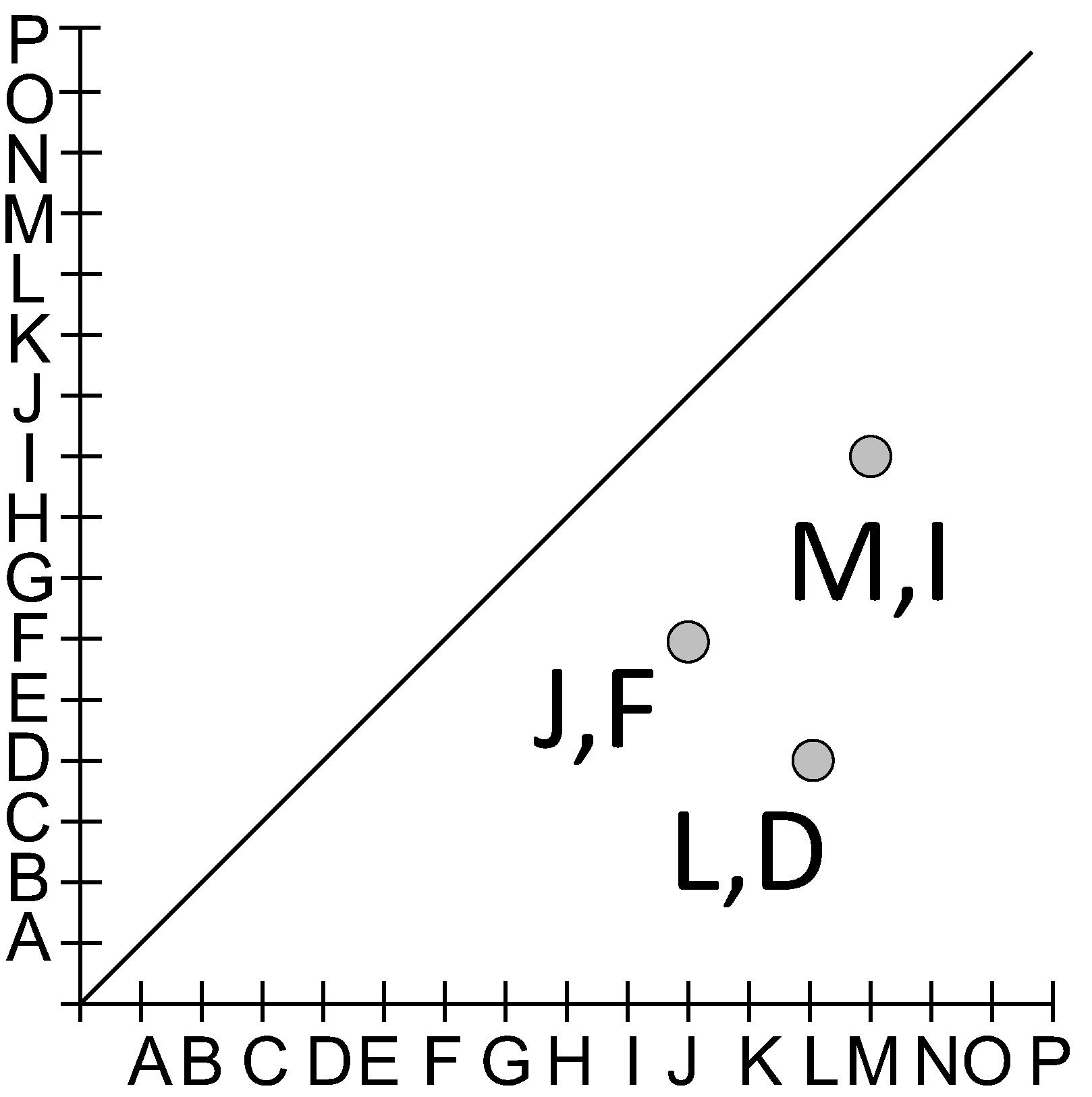}\hspace{2pt}}
    \end{minipage}}

	\caption{(a) A mesh with a scalar function being processed into (b) a Reeb graph, where critical points are paired. In the multipass approach, (c) a split tree and (d) a join tree are first extracted for non-essential pairing. Next, the (e) essential forks are paired, one at a time. (f) The persistence diagram and (g) extended persistence diagram provide a visualization of the pairings.}
	\label{fig.pipeline}
\end{figure}

One challenge with Reeb graphs is that the graph may be too large or complex to directly visualize, therefore requiring further abstraction. Persistent homology~\cite{edelsbrunner2000topological}, parameterizes topological structures by their life-time, providing a topological description called the \textit{persistence diagram}. The notion of persistence can be applied to any act of birth that is paired with an act of death. Since the Reeb graph encodes the birth and the death of the connected components of the level sets of a scalar function, the notion of persistence can be applied to critical points in the Reeb graph~\cite{agarwal2006extreme}. \textit{The advantage of this approach is simplicity and scalability---a large Reeb graph can be reduced to a much easier to interpret scatterplot.} \figref{fig.pipeline} shows an example, where a mesh with a scalar function (\figref{fig.pipeline.mesh}) is converted into a Reeb graph (\figref{fig.pipeline.reeb}). After that, the critical points are paired, and the persistence diagram displays the data (\figref{fig.reebexample.pd0} and~\ref{fig.reebexample.pd1}). This final step can be challenging, particularly when considering \textit{essential critical points}---those critical points associated with cycles in the Reeb graph---that each require an expensive search. While many prior works~\cite{Cole-McLaughlin2004,PascucciScorzelliBremer2007,Harvey2010,Hilaga2001,Tierny2009,Doraiswamy2013,doraiswamy2009efficient, Salman12, doraiswamy2008efficient} have provided efficient algorithms for the calculation of Reeb graph structures, to our knowledge, none have provided a detailed description of an algorithm for pairing critical points.

In this paper, we describe and test 2 algorithms to compute persistence diagrams from Reeb graphs. The first is a multipass approach that pairs non-essential (non-loop) critical points using branch decomposition~\cite{pascucci2004multi} on join and split trees, then pairing essential critical points also using join trees. This leads to our second approach, an algorithm for pairing both non-essential and essential critical points in a single-pass.

\section{Reeb Graphs and Persistence Diagrams}
\label{sec.reebgraph}


\subsection{Reeb graph}

Let $X$ be a triangulable topological space, and let $f: \XX \rightarrow \reals$ be a continuous function defined on it. The Reeb graph, $R_f$, can be thought of as a topological summary of the space $X$ using the information encoded by the scalar function $f$. More precisely, the Reeb graph encodes the changes that occur to connected components of the level sets of $f^{-1}(r)$ as $r$ goes from negative infinity to positive infinity. \figref{fig.pipeline.mesh} and \ref{fig.pipeline.reeb} show an example of a Reeb graph defined on a surface. For the sake of simplicity we plot the Reeb graph using the height function indicated by the vertical coordinate in the Figure.  


The function $f$ can be used to classify points on the Reeb graph as follows. Let $x$ be a point in $R_f$. The \textit{up-degree} of $x$ is the number of branches (1-cells) incident to $x$ that have higher values of $f$ than $x$. The down-degree of $x$ is defined similarly. A point $x$ on $R_f$ is said to be \textit{regular} if its up-degree and down-degree are equal to one.  Otherwise it is a \textit{critical point}. A critical point on the Reeb graph is also a \textit{node} of the Reeb graph. A critical point is called a minimum if its down-degree is equal to $0$. Symmetrically, a critical point is said a maximum if its up-degree is equal to $0$. Finally, a critical point is said to be a down-fork/up-fork if its down-degree/up-degree is larger than $1$.

\subsection{Persistent Homology}

The notion of persistent homology was originally introduced by Edelsbrunner et al.~\cite{edelsbrunner2000topological}. Here we present the theoretical setting for the computation of the persistence diagram associated with a scalar function defined on a triangulated topological space. Consider the $p$-dimensional homology class $\HH_p$ of a space, where $\HH_0$ are components, $\HH_1$ are tunnels/cycles, $\HH_2$ are voids, etc. Persistent homology evaluates a sequence of vector spaces: $0 = \HH_p (X_{0}) \to \HH_p(X_{1}) \to \cdots \to \HH_p(X_{n}) = \HH_p(X)$, where $X_{i}= X_{\leq f_i}$, recording the birth and death events. In particular, the $p$-th \emph{ordinary persistence diagram} of~$f$, denoted as $\Dg_p(f)$, is a multiset of pairs $(b, d)$ corresponding to the birth $b$ and death $d$ values of some $p$-dimensional homology class.

Since the homology $H_p (X)$ may not be trivial in general, any nontrivial homology class of $H_p(X)$, referred to as an \emph{essential homology class}, will never die during the sequence. These events are associated with the cyclic portions of the Reeb graph. We refer to the multiset of points encoding the birth and death time of $p$th homology classes created in the ordinary part and destroyed in the relative part of the sequence as the \emph{$p$-th extended persistence diagram} of $f$, denoted by $\eDg_p(f)$. In particular, for each point $(b, d)$ in $\eDg_p(f)$ there is an essential homology class in $\HH_p(X)$ that is born in $\HH_p(X_{\leq b})$ and dies at $\HH_p(X_{\geq d})$. Observe that for the extended persistence diagram the birth time $b$ for an essential homology class in $\HH_p(X_{\leq b})$ is larger than or equal to death time $d$ for the relative homology class in $\HH_p(X_{\geq d})$ that kills it.

\subsection{Persistence Diagram of Reeb Graph}
\label{sec.reeb.phrg}

Of interest to us are the persistence diagram $\Dg_0(f)$ and extended persistence diagram $\eDg_1(f)$. Pairing critical points can be computed independently of the Reeb graph. However, it is more efficiently computed by considering the Reeb graph $R_f$. We give an intuitive explanation here and refer the reader to \cite{bauer2014measuring} for more details. 

First, we distinguish between 2 types of forks in the Reeb graph, namely the ordinary (non-essential) forks and the essential forks. Let $R_f$ be a Reeb graph and let $s$ be a down-fork such that $a = f(s)$. We say that the down-fork $s$ is an \textit{ordinary fork} if the lower branches of $s$ are contained in disjoint connected components $C_1$ and $C_2$ of $(R_f)_{<a}$. The down-fork $a$ is said to be \textit{essential} if it is not ordinary. The ordinary and essential up-forks are defined similarly.

\paragraph{Ordinary Down-Forks of a Reeb Graph} We first consider pairing down-forks using sublevel set filtration. We track changes that occur in $H_0((R_f)_{\leq a})$ as $a$ increases. A connected component of $(R_f)_{ \leq a }$ is created when $a$ passes through a minimum of $R_f$. Let $C$ be a connected component of $(R_f)_{ \leq a }$. We say that a local minimum $a$ of $R_f$ \textit{creates} $C$ if $a$ is the global minimum of $C$. Every ordinary down-fork is paired with a local minimum to form one point in the persistence diagram $\Dg_0(f)$ as follows. Let $s$ be an ordinary down-fork with $f(a)=s$ and let $C_1$ and $C_2$ be the connected components of $(R_f)_{ <a }$. Let $x_1$ and $x_2$ be the creators of $C_1$ and $C_2$. Without loss of generality we assume that $f(x_1)<f(x_2)$. The homology class $[x_2]$ that is created at $f(x_2)$ and dies at $f(s)$ gives rise to a point $(x_2, s)$ in the ordinary persistence diagram $\Dg_0(f)$. Note, a pair occurs when the minimum is a branch in the Reeb graph, hence we name it a \textit{branching feature}.

\paragraph{Ordinary Up-Forks of a Reeb Graph} Ordinary up-forks are paired similarly using superlevel set filtration, pairing each up-fork with a local maximum to form points in the persistence diagram, $\Dg_0(f)$, with the following variations. For an ordinary up-fork, $s$, with $f(a)=s$, connected components $C_1$ and $C_2$ now come from $(R_f)_{ >a }$. Assuming that $f(x_1)<f(x_2)$, the homology class $[x_1]$ that is created at $f(x_1)$ dies at $f(s)$ and gives rise to a point $(x_1, s)$ in $\Dg_0(f)$.

\paragraph{Cycle Features of a Reeb Graph} Let $s$ be an essential down-fork.  We call the down-fork $s$ a creator of a $1$-cycle in the sublevel set $(R_f)_{ \leq a }$. As shown in \cite{agarwal2006extreme},  $s$ will be paired with an essential up-fork  $s^{\prime}$ to form an \textit{essential pair} $(s^{\prime},s)$, and a point $(s^{\prime}, s)$ in the extended persistence diagram $\eDg_1(f)$. The essential up-fork $s^{\prime}$ is determined as follows. Let $\Gamma_s$ be the set of all cycles born at $s$, each corresponding to a cycle in $R_f$. Let $\gamma_s$ be an element of $\Gamma_s$ with largest minimum value of $f$ among these cycles born at $s$. The point $s^{\prime}$ is the point that the function $f$ achieves this minimum on the cycle $\gamma_s$.

\section{Conditioning the Graph}
\label{sec-graph-prep}

Our approach is restricted to Reeb graphs where all point are either a minimum, maximum, up-fork with up-degree 2, or down-fork with down-degree 2. Fortunately, graphs that do not abide by these requirements can be conditioned to fit them. We define the $J:K$ degree of a node as the $J$ up-degree and $K$ down-degree.

\begin{figure}[!tb]
	\centering
    {\begin{minipage}{0.43\linewidth}
    	\subfigure[Non-Critical\label{fig.condition.noncp}]{\hspace{14pt}\includegraphics[height=1.8cm]{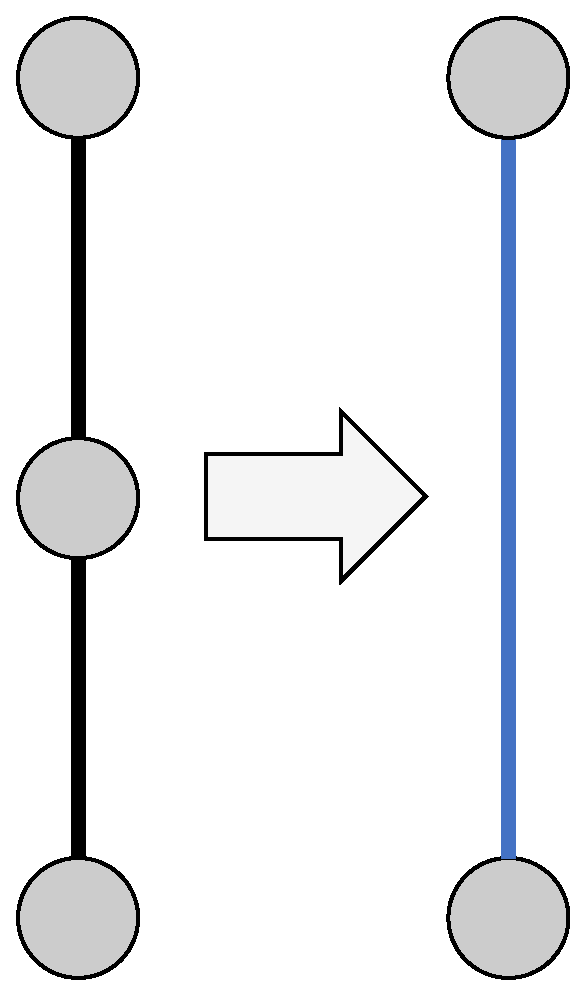}\hspace{14pt}}
    	\hspace{4pt}
    	\subfigure[Degenerate Max \label{fig.condition.maxsaddle}]{\hspace{14pt}\includegraphics[height=1.8cm]{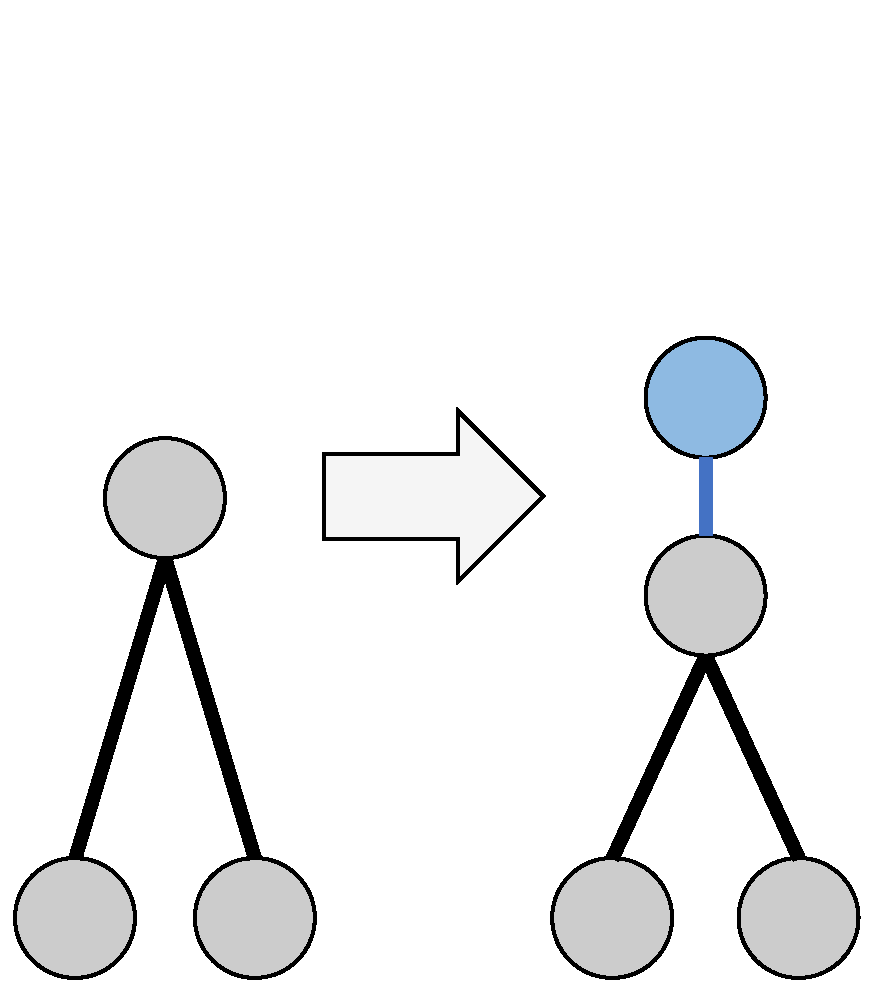}\hspace{14pt}}
    	
        \subfigure[Double Fork\label{fig.condition.doublesaddle}]{\hspace{10pt}\includegraphics[height=1.8cm]{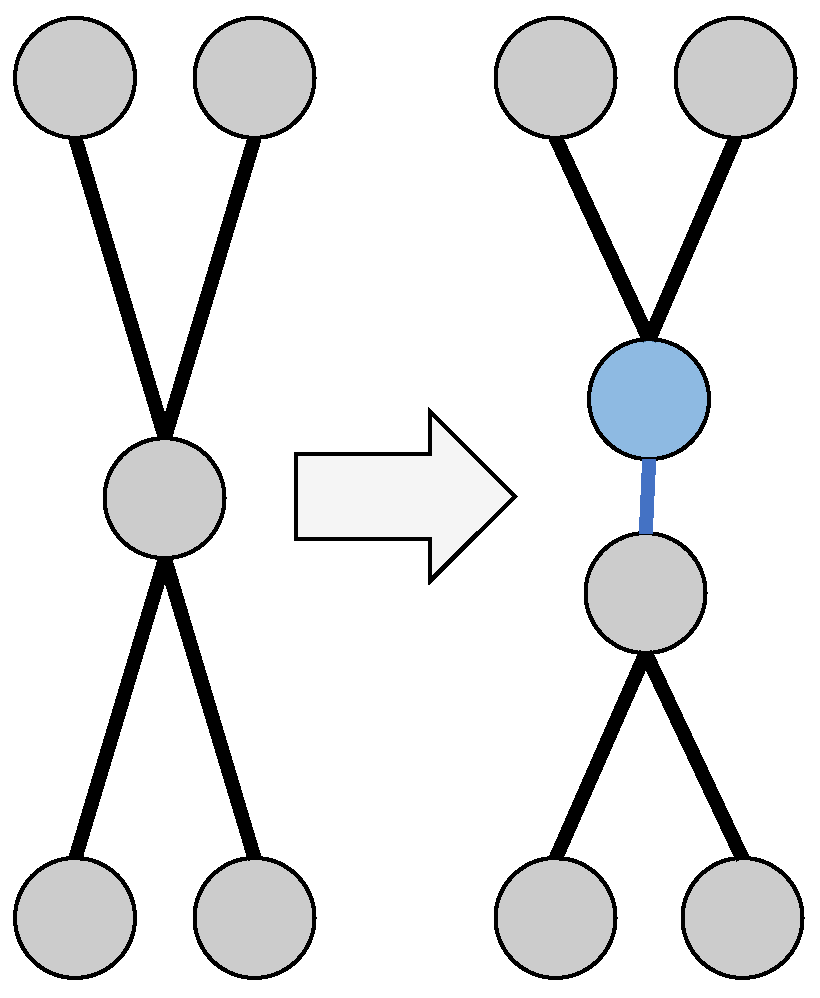}\hspace{10pt}}
    	\hspace{4pt}
    	\subfigure[Complex Fork\label{fig.condition.monkeysaddle}]{\hspace{4pt}\includegraphics[height=1.8cm]{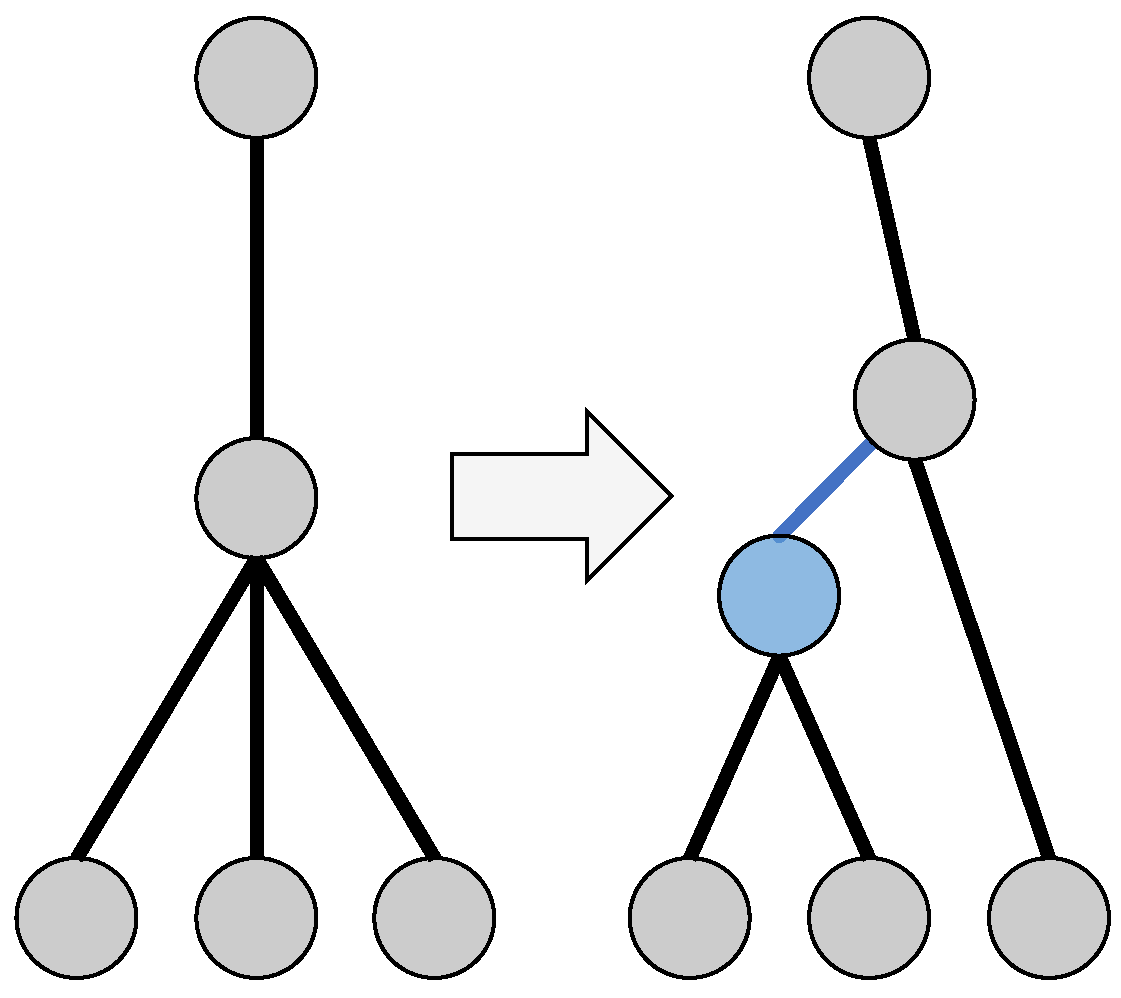}\hspace{4pt}}
    \end{minipage}}
    \hfill
    {\begin{minipage}{0.52\linewidth}
        \subfigure[The 4 cases of stack configurations and their result. Type 1 and 2 involve stack reorganization, while Type 3a and 3b are pairing operations.\label{pairing_ex:ops}]{\hspace{14pt}\includegraphics[width=0.82\linewidth]{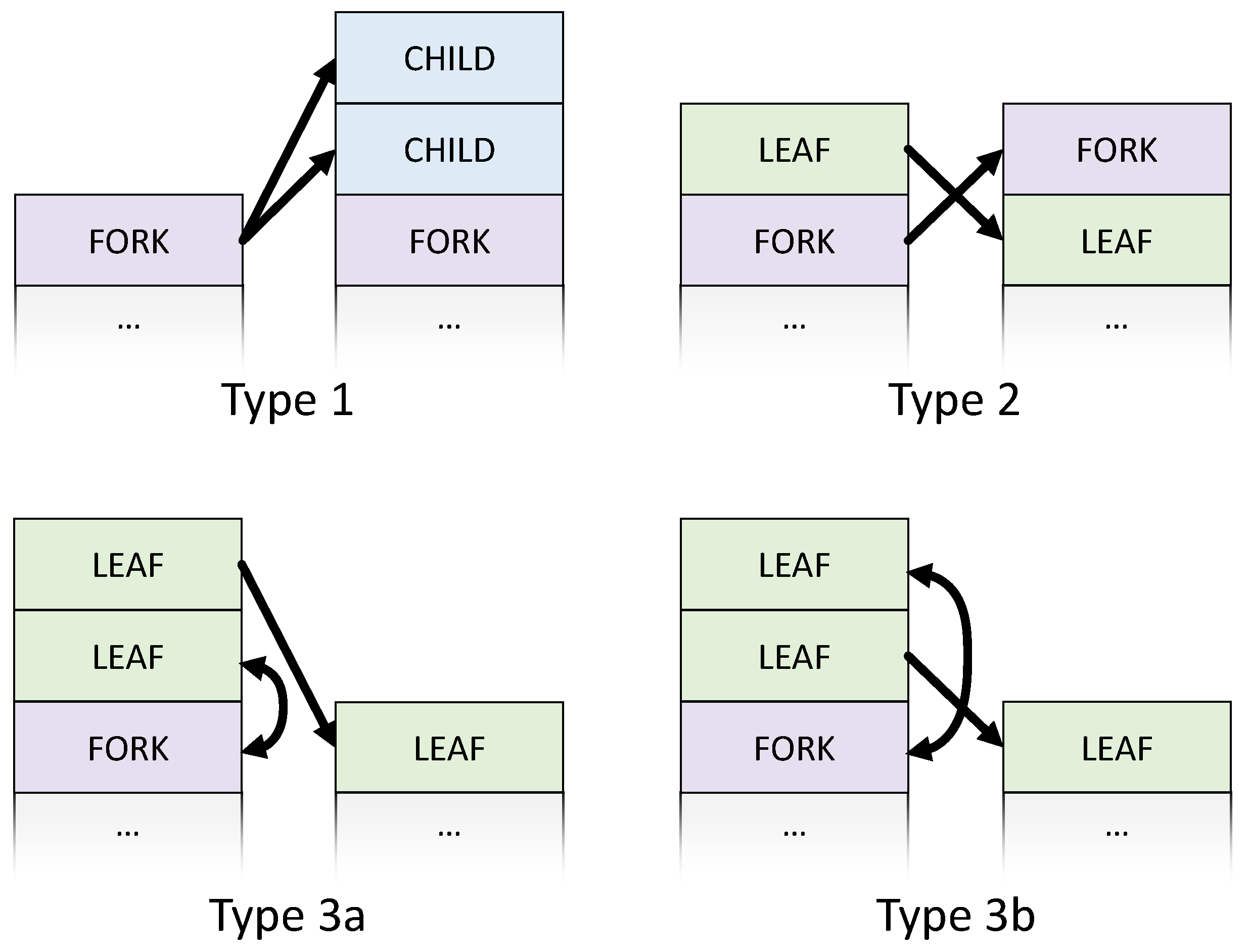}\hspace{14pt}}	
    \end{minipage}}
    
	\caption{(a-d) Before pairing, the nodes of Reeb graph are conditioned considering 4 node configurations. New nodes and edges are shown in blue. (e) For non-essential fork pairing in the multipass algorithm, the 4 cases for stack processing are illustrated with their resulting configurations.}
	\label{fig.condition}
\end{figure}

\vspace{5pt}
\noindent
There are 4 node conditions to be corrected:
\textbf{1:1 nodes}---Nodes with both 1 up- and 1 down-degree are regular. Therefore, they only need to be removed from the graph. This is done by removing the regular point and reconnecting the nodes above and below, as seen in \figref{fig.condition.noncp}.
\textbf{0:2 (and 2:0) nodes}---Nodes with 0 up-degree and 2 down-degree (or vice versa) are degenerate maximum (minimum) nodes, in that they are both down-fork (up-fork) and local maximum (minimum). As shown in \figref{fig.condition.maxsaddle}, this condition is corrected by added a new node for the local maximum $\epsilon$ higher value, where $\epsilon$ is a small number.  This type of degenerate node rarely occurs in Reeb graphs, but it frequently occurs in approximations of a Reeb graph, such as Mapper~\cite{singh2007topological}. 
\textbf{2:2 nodes}---Nodes with both 2 up- and 2 down-degree are degenerate double forks, both down-fork and up-fork. \figref{fig.condition.doublesaddle} shows how double forks can be corrected by splitting into 2 separate forks, one up- and one down-fork, $\epsilon$ distance apart. 
\textbf{1:N\textgreater2 (and N\textgreater2:1) nodes}---Nodes with down-degree (or up-degree) 3 or higher, are difficult forks to pair. These forks correspond to complex saddles in $f$, such as monkey saddles. A single critical point pairing to these forks just reduces the degree of down-fork by 1, requiring complicated tracking of pairs. To simplify this, as seen in \figref{fig.condition.monkeysaddle}, these forks can be split into 2 forks $\epsilon$ apart. The upper down-fork retains 1 of the original down edges. The new down-fork connects with the old and takes the remaining down-edges. For even higher-order forks, the operation can be repeated on the lower down-fork.

Beyond these requirements, the Reeb graph is assumed a single connected component. If not, each connected component can simply be extracted and processed individually. Finally, all nodes on the Reeb graph are assumed to have unique function values. If not, some processing order is arbitrary, and 0-persistence features may result.

\section{Multipass Approach}
\label{sec.approach.multi}

The persistence diagram $\Dg_0(f)$ can be obtained by pairing the non-essential fork nodes of the Reeb graph. The extended persistence diagram $\eDg_1(f)$ can be obtained by pairing of essential fork nodes. We demonstrate these 2 steps using \figref{fig.pipeline} as an example.

\subsection{Non-Essential Fork Pairing}

Identifying the non-essential forks can be reduced to calculating join and split trees on the Reeb graph (see \figref{fig.reebexample.st} and~\ref{fig.reebexample.mt}), in our case, using Carr et al.'s approach~\cite{Carr2003}. Next, a stack-based algorithm, based upon branch decomposition~\cite{pascucci2004multi}, pairs critical points. The algorithm operates as a depth first search that seeks out simply connected forks (i.e., forks connected to 2 leaves) and recursively pairs and collapses the tree.

The algorithm processes the tree using a stack that is initially seeded with the root of the tree. At each iteration, 1 of 3 operation types occurs, as seen in \figref{pairing_ex:ops}. Operation Type 1 occurs when the top of the stack is a fork. In this case, the children of the fork are pushed onto the stack. Operation Type 2 occurs when the top of the stack is a leaf, but the next node is a fork. In this case, the leaf and fork have their orders swapped. Finally, operation Type 3 has 2 variants that occur when 2 leaf nodes sit atop the stack. In both variants, one leaf is paired with the fork, and the other leaf is pushed back onto the stack. The pairing occurs with the leaf that has a value \textit{closer} to the value of the fork. The stack is processed until only a single leaf node remains, the global minimum/maximum of the join/split trees, respectively. The algorithm operates identically on both join and split trees. Finally, the unpaired global minimum and maximum are paired.

\begin{figure}[!t]
    \centering
    
    \subfigure[Example processing the join tree from \figref{fig.reebexample.mt}.\label{pairing_ex:ex}]{\includegraphics[width=0.585\linewidth]{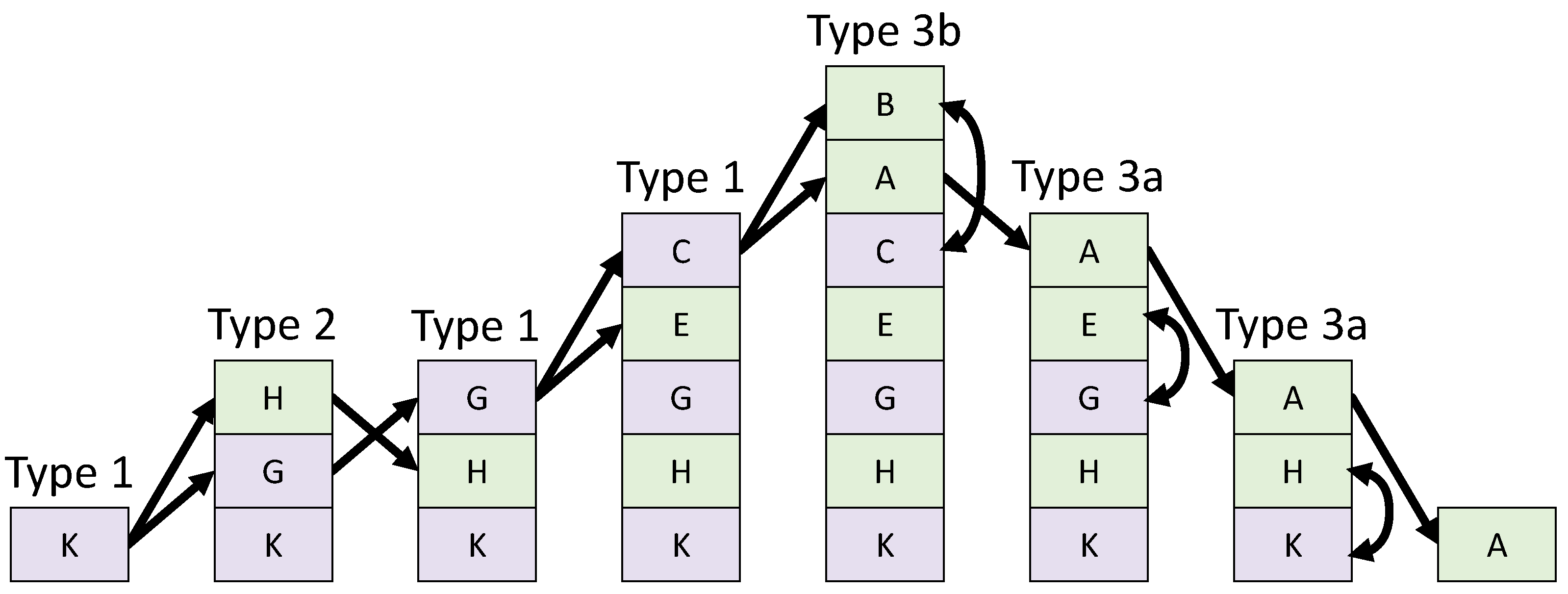}}
    \hfill
    \subfigure[Timestep from scivis\_contest data.\label{fig.result.scivis}]{
        \hspace{2pt}
        \begin{minipage}[b]{0.24\linewidth}
	        {\includegraphics[trim=220pt 210pt 220pt 200pt, clip, width=1\linewidth]{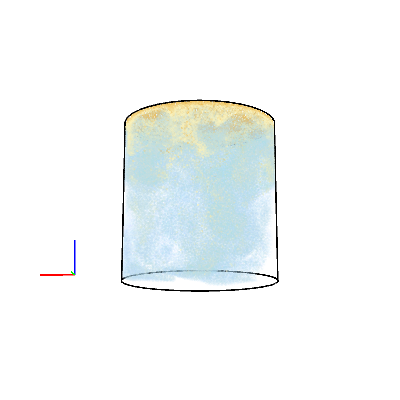}}
	    \end{minipage}    
    	\begin{minipage}[b]{0.12\linewidth}
    		\centering
    		\includegraphics[width=1\linewidth]{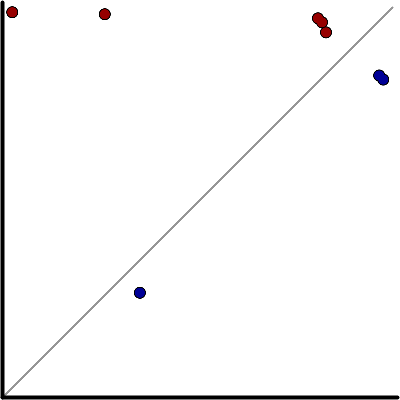}
    		\includegraphics[width=1\linewidth]{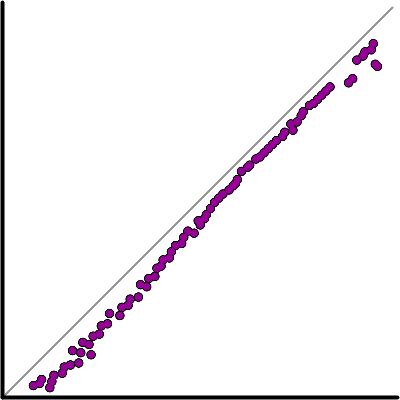}
    	\end{minipage}    
    	\hspace{2pt}
    }
    
    \caption{(a) An example pairing of the join tree from \figref{fig.reebexample.mt} shows the stack at each processing step, from left to right. (b) Timestep (066) from the scivis\_contest data is shown with concentration mapped to color (left), along with $\Dg_0$ (top) showing up-forks in blue and down-forks in red; and the $\eDg_1$ (bottom) showing cycles in purple.}
    \label{pairing_ex}
\end{figure}

\figref{pairing_ex:ex} shows an example for the join tree in \figref{fig.reebexample.mt}. Initially the root $K$ is placed on the stack. A Type 1 operation pushes the children, $G$ and $H$, onto the stack. Next, a Type 2 operation reorders the top of the stack. $G$, a down-fork, in now atop the stack, pushing its 2 children, $E$ and $C$, onto the stack. Another Type 1 pushes $C$'s children, $A$ and $B$ onto the stack. In the next 3 steps, a series of Type 3 operations occur. First $B$ and $C$ are paired, followed by $E$ and $G$, and finally $H$ and $K$. At the end, $A$, the global minimum, is the only point remaining on the stack. The assigned pairs, $B$/$C$, $E$/$G$, and $H$/$K$, appear in the $\Dg_0(f)$ in \figref{fig.reebexample.pd0}, along with the split tree pairing, $O$/$N$, and the global min/max pairing, $A$/$P$.

\subsection{Essential Forks Pairing}

The remaining unpaired forks are essential forks, as seen in \figref{fig.reebexample.ess}. We developed an algorithm from the high-level description of~\cite{bauer2014measuring} to pair them. The essential fork pairing algorithm can be treated as join tree problem, processing forks one at a time. For a given up-fork, $s$, the node can be split into two temporary nodes, $s_L$ and $s_R$. A join tree can be computed by sweeping the superlevel set. At each step of the sweep, the connected components are calculated. The pairing for a selected essential up-fork occurs at the down-fork that merges $s_L$ and $s_R$ into a single connected component.

\begin{figure}[!t]
	\centering
    \begin{minipage}[b]{0.5\linewidth}
    	\subfigure[\label{fig.essMTEX.a}]{\includegraphics[width=0.30\linewidth]{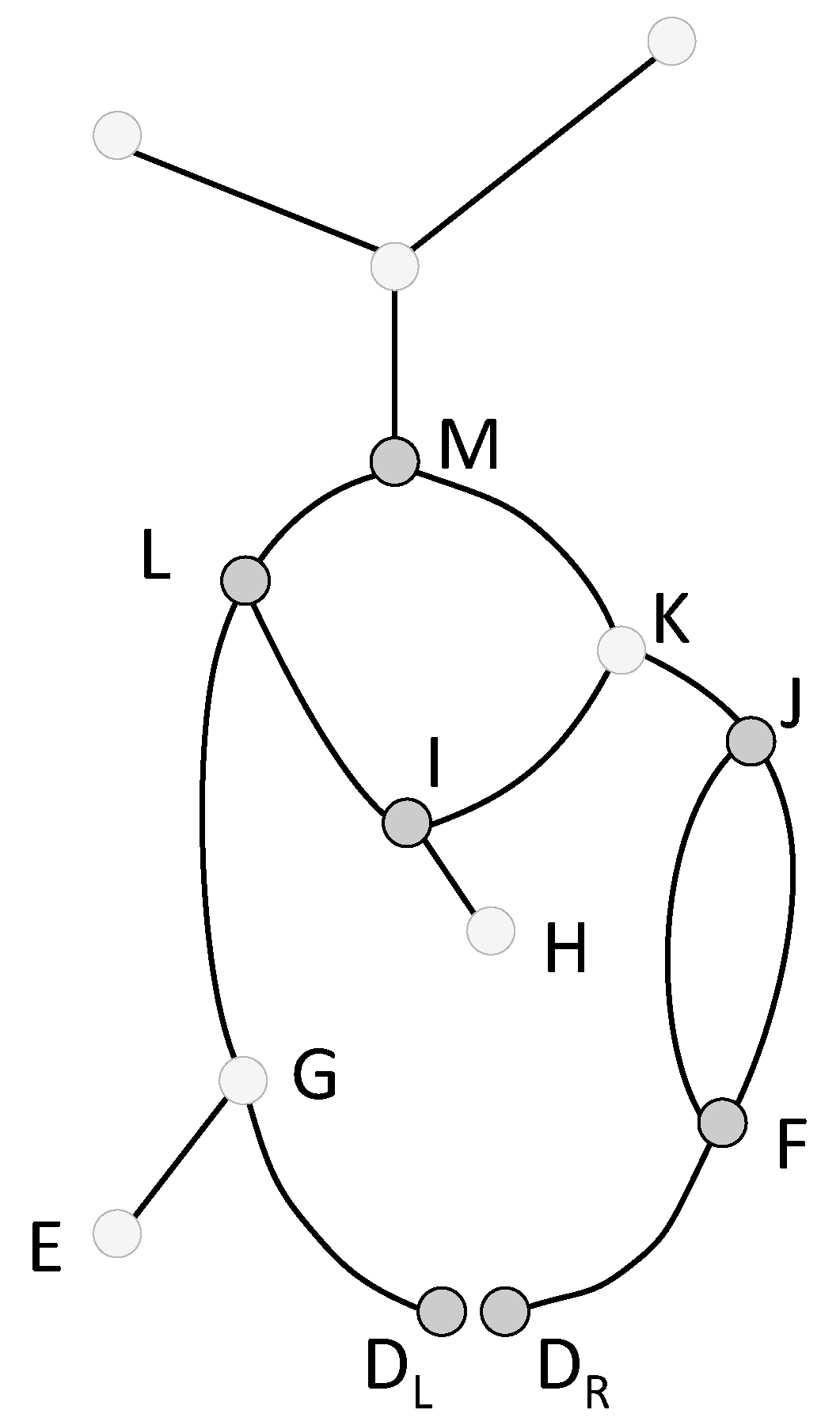}}
    	\hfill
        \subfigure[\label{fig.essMTEX.1}]{\includegraphics[width=0.30\linewidth]{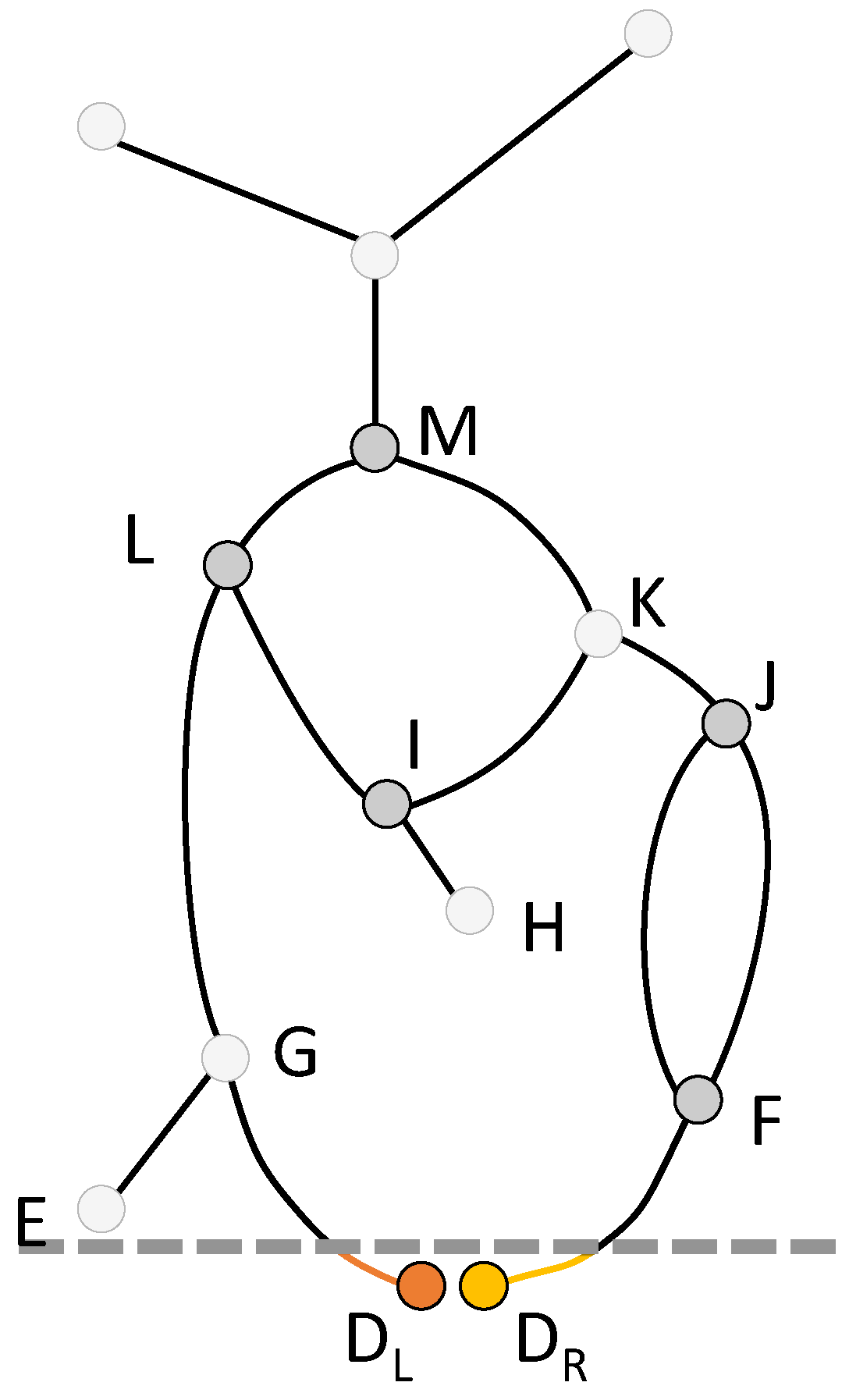}}
        \hfill
        \subfigure[\label{fig.essMTEX.2}]{\includegraphics[width=0.30\linewidth]{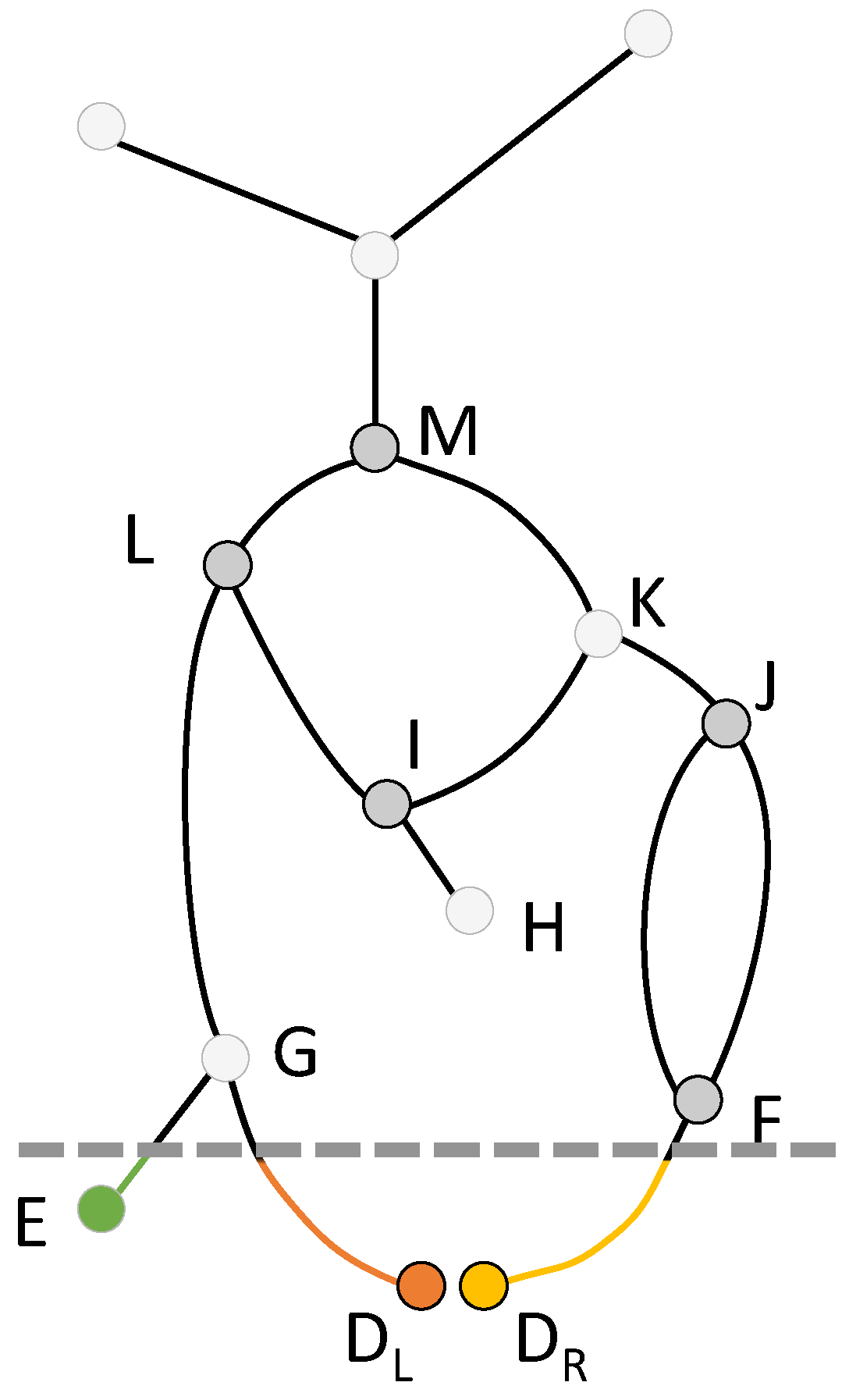}}
        \hfill
        \subfigure[\label{fig.essMTEX.3}]{\includegraphics[width=0.30\linewidth]{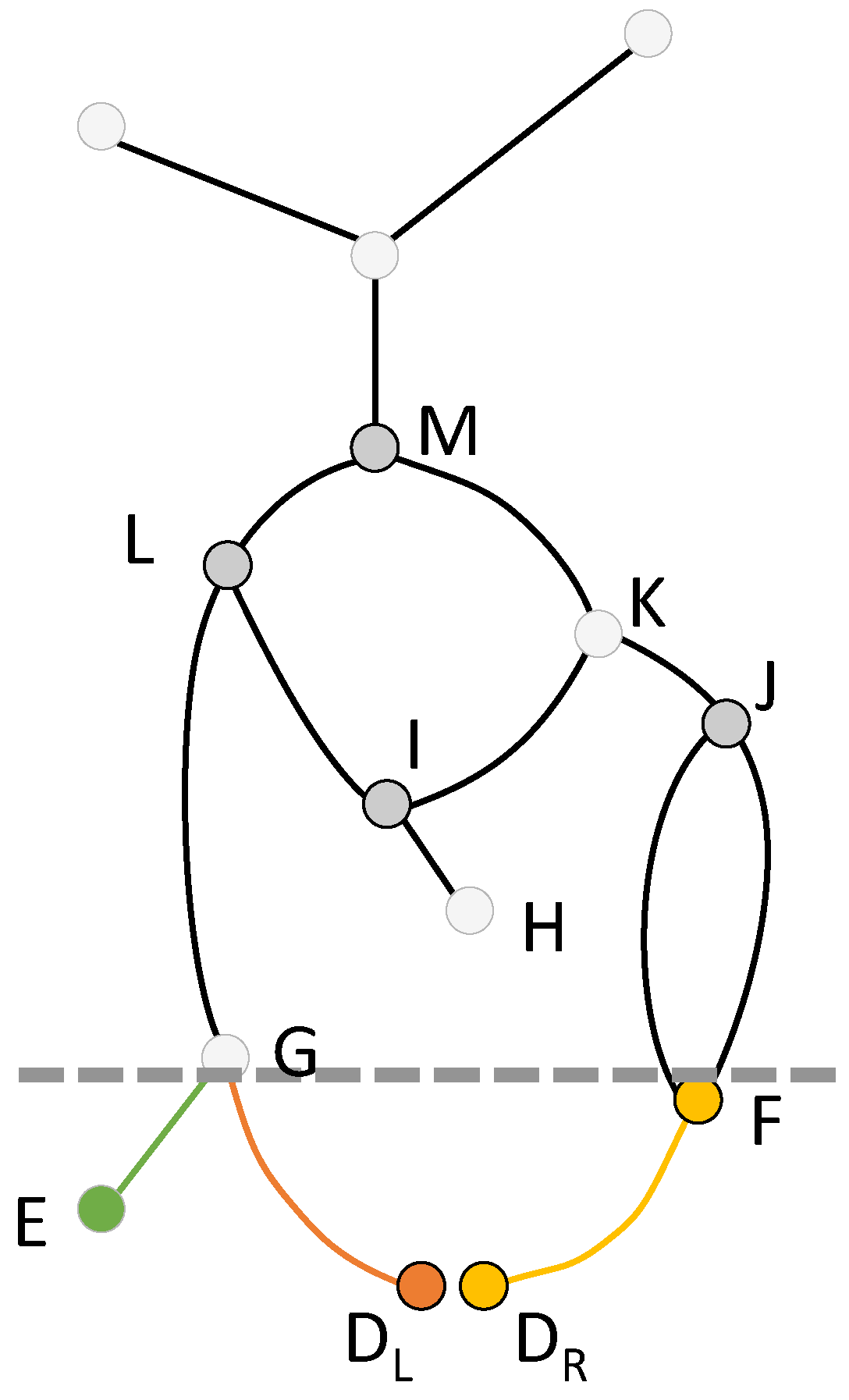}}
        \hfill
        \subfigure[\label{fig.essMTEX.4}]{\includegraphics[width=0.30\linewidth]{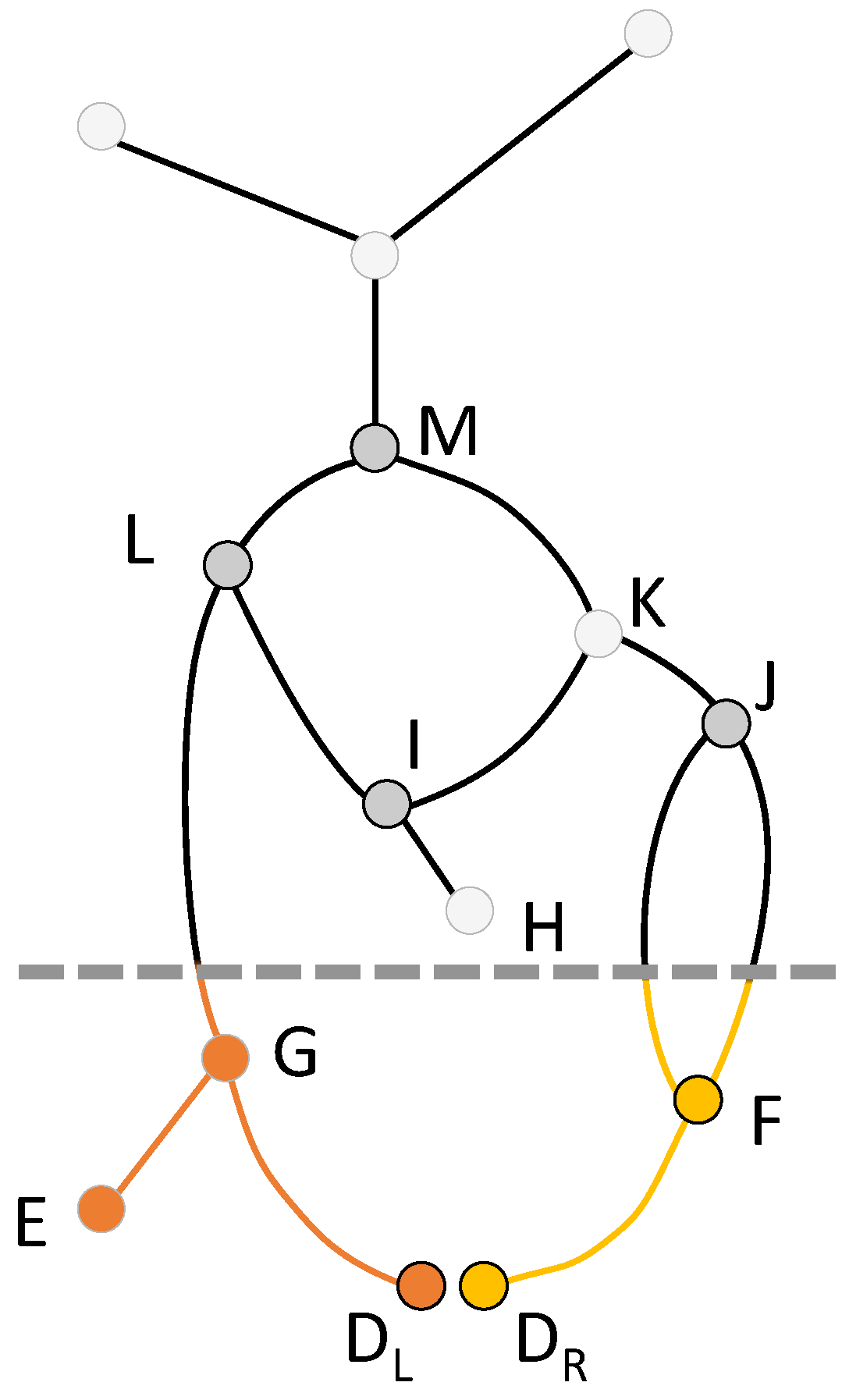}}
        \hfill
        \subfigure[\label{fig.essMTEX.5}]{\includegraphics[width=0.30\linewidth]{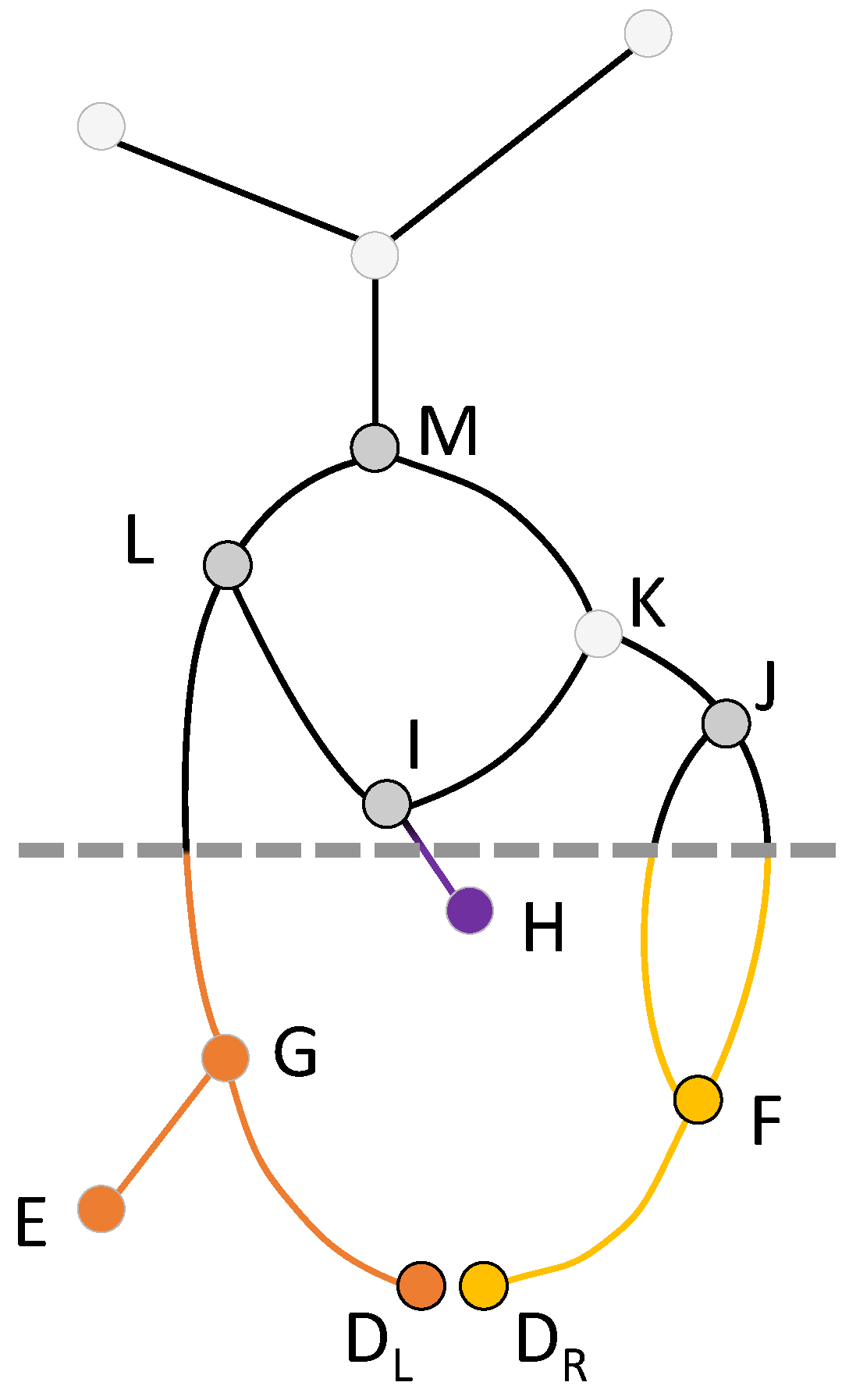}}
        \hfill
        \subfigure[\label{fig.essMTEX.6}]{\includegraphics[width=0.24\linewidth]{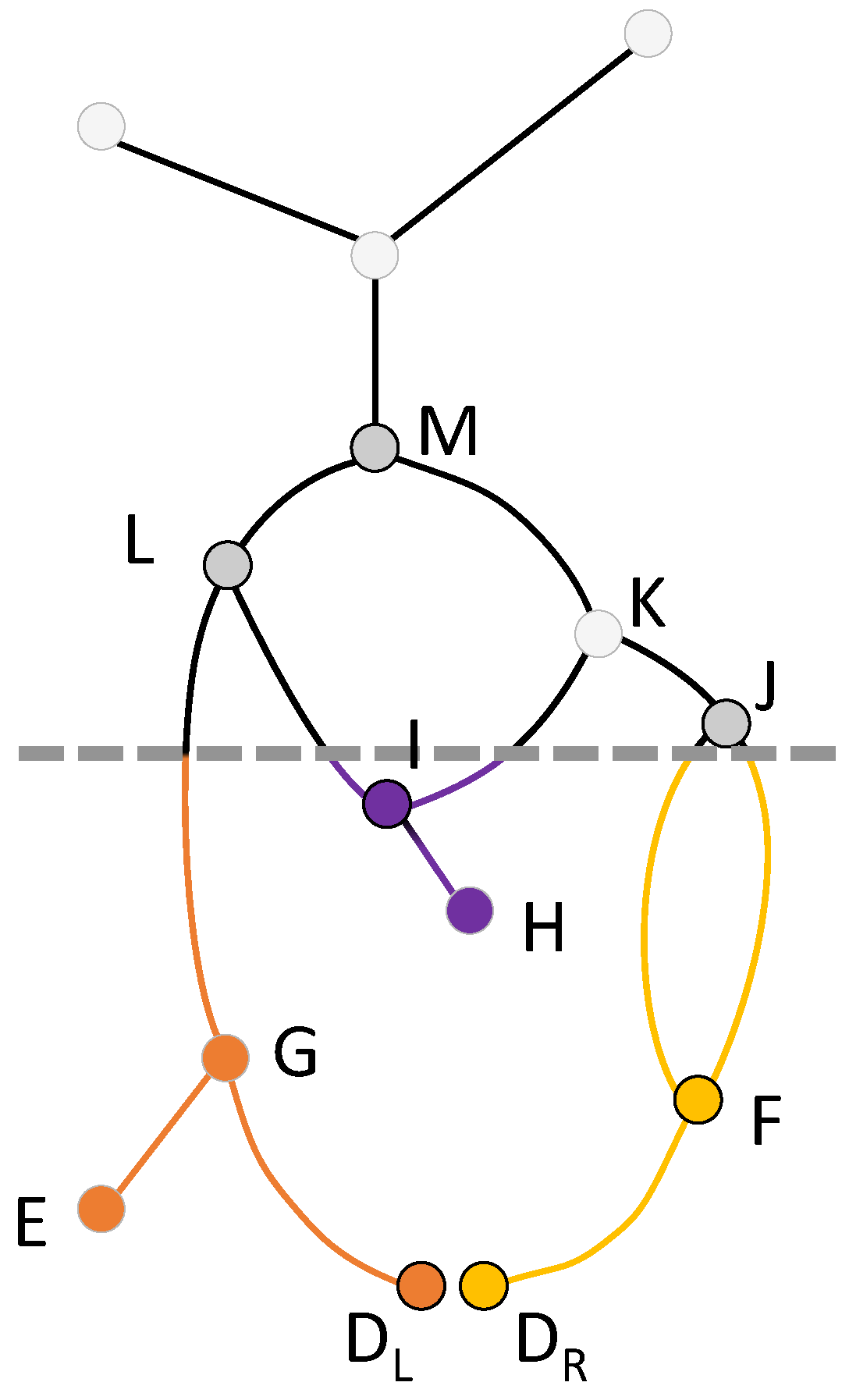}}
        \hfill
        \subfigure[\label{fig.essMTEX.7}]{\includegraphics[width=0.24\linewidth]{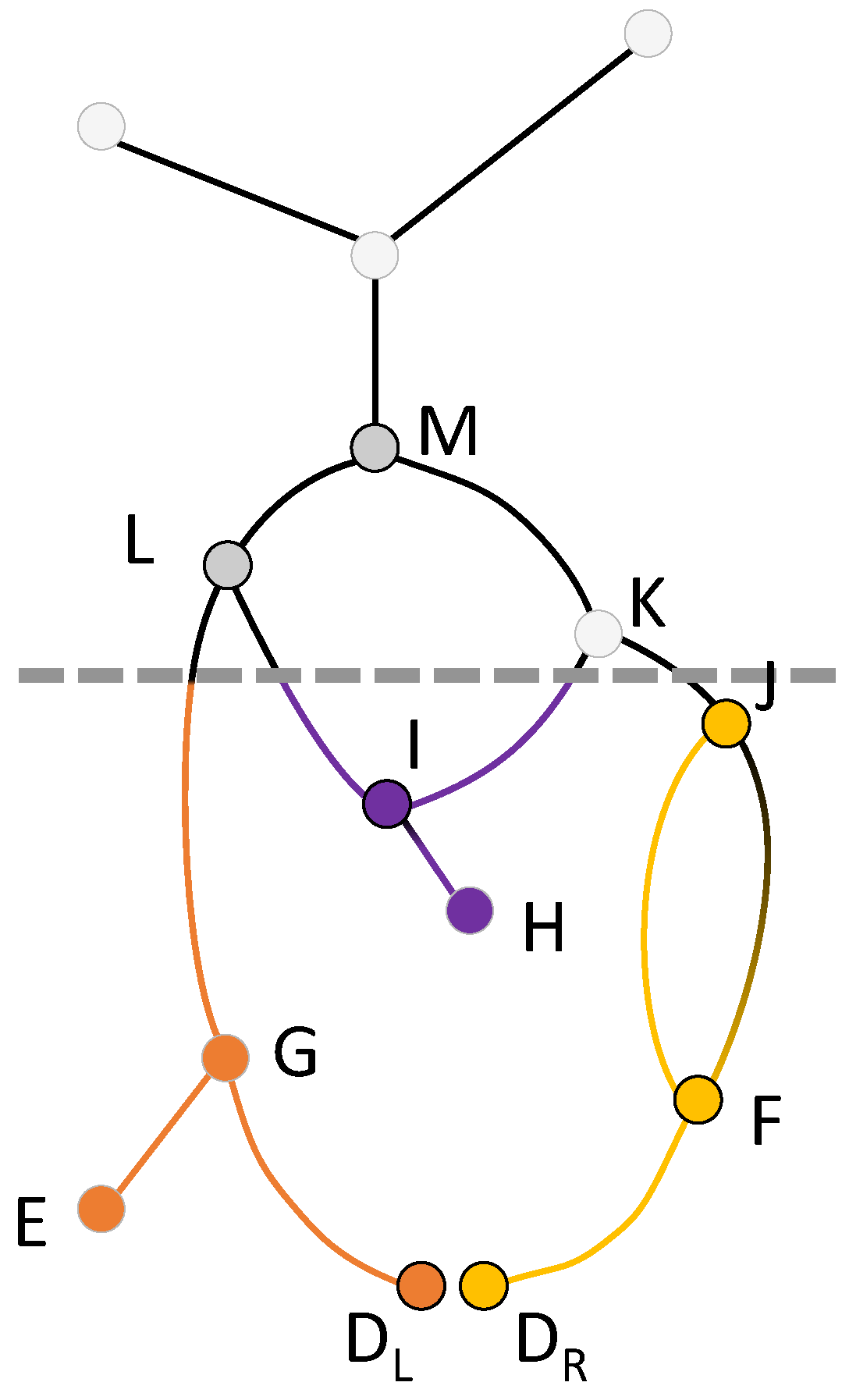}}
        \hfill
        \subfigure[\label{fig.essMTEX.8}]{\includegraphics[width=0.24\linewidth]{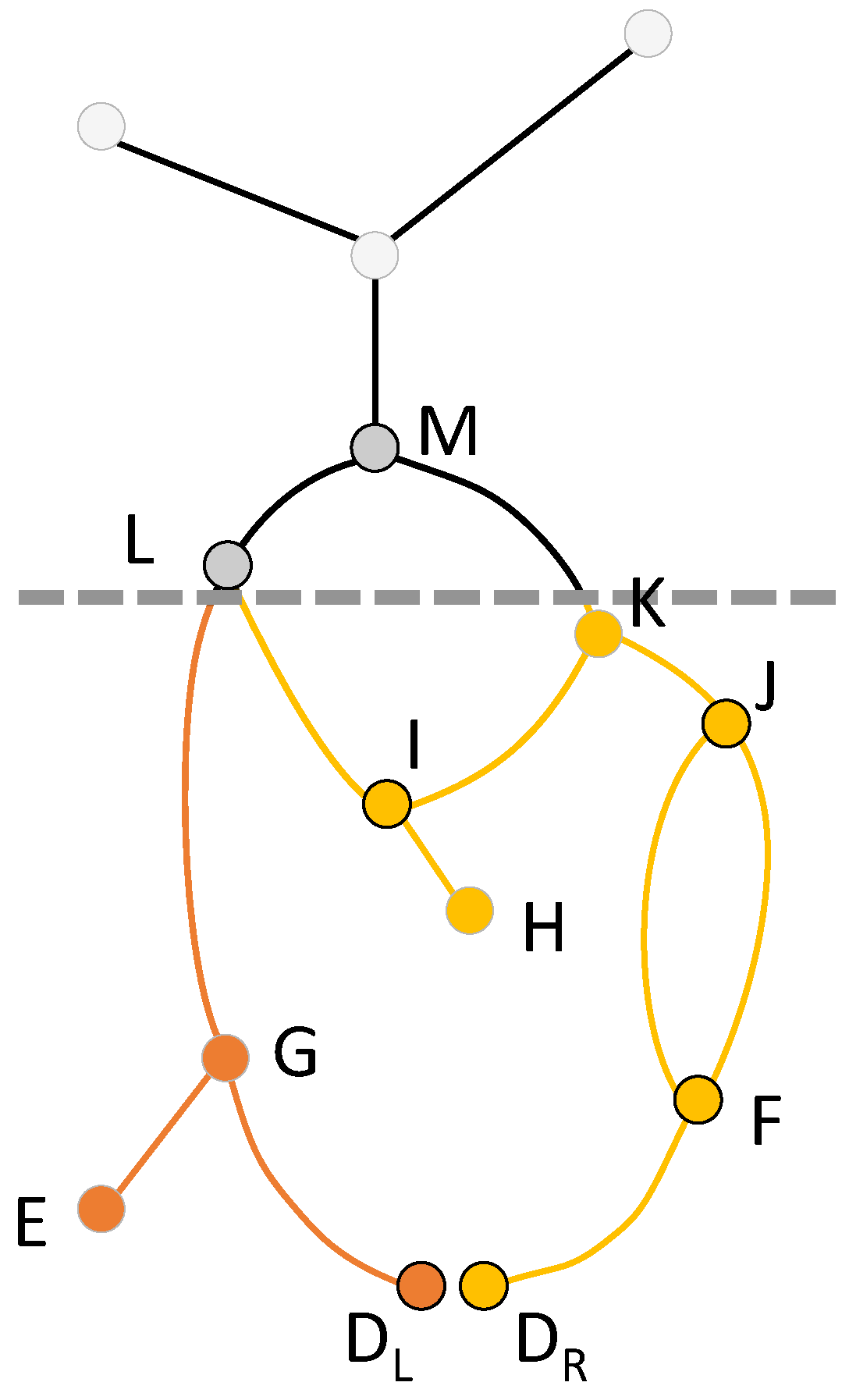}}
        \hfill
        \subfigure[\label{fig.essMTEX.9}]{\includegraphics[width=0.24\linewidth]{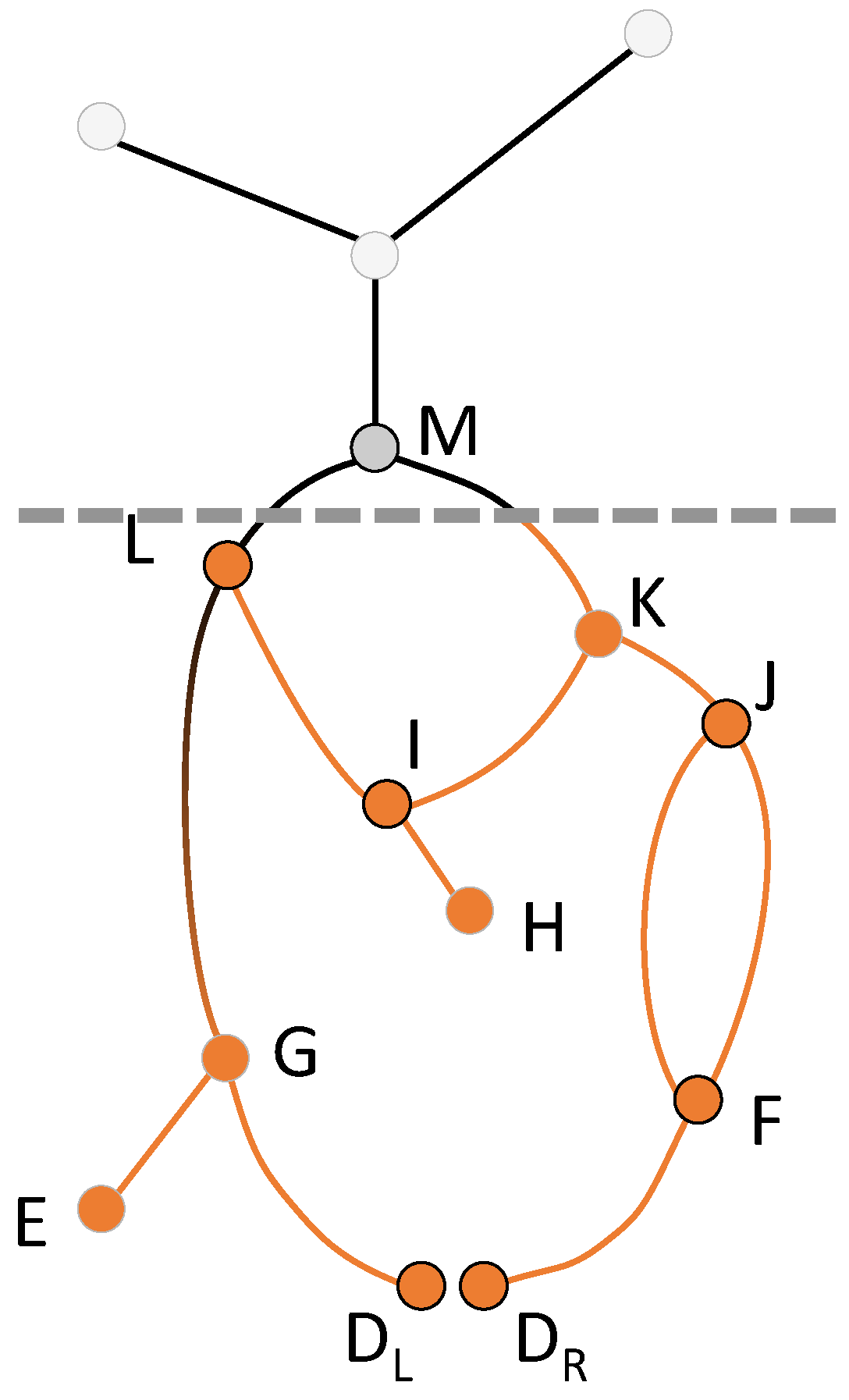}}
    \end{minipage}
    \hfill
    \begin{minipage}[b]{0.45\linewidth}
    	\subfigure[$D$ superlevel set (left) and join tree (right)\label{fig.ess_mt.d_sls}]{
    	    \hspace{30pt}
    	    \includegraphics[height=2.7cm]{figs/example/essMTS_d_1}
    	    \includegraphics[trim=0 0 0 0, clip, height=1.90cm]{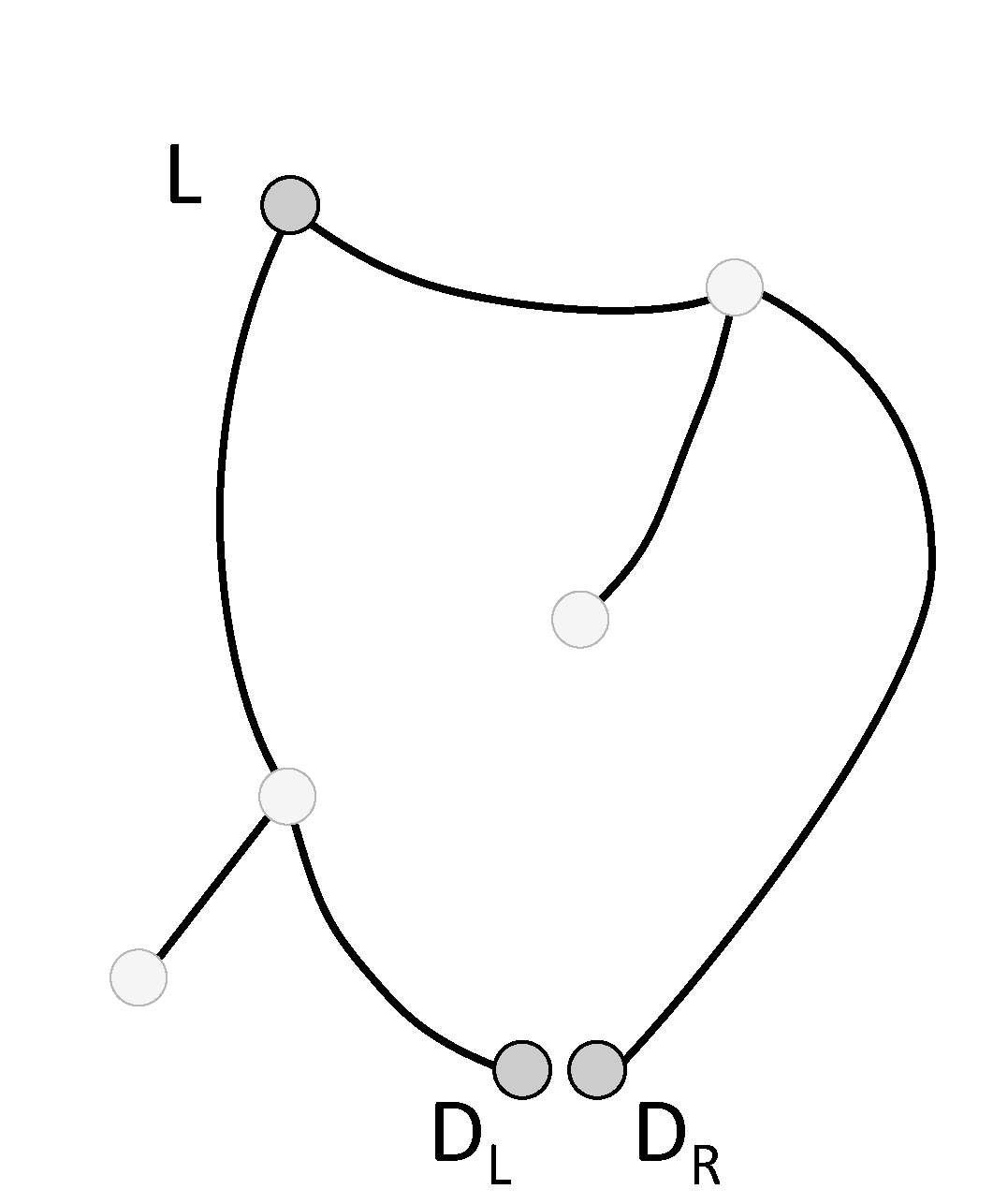}
    	    \hspace{30pt}
    	}
	    \subfigure[$F$ superlevel set (left) and join tree (right)\label{fig.ess_mt.f_sls}]{
	        \hspace{40pt}
	        \includegraphics[height=2.7cm]{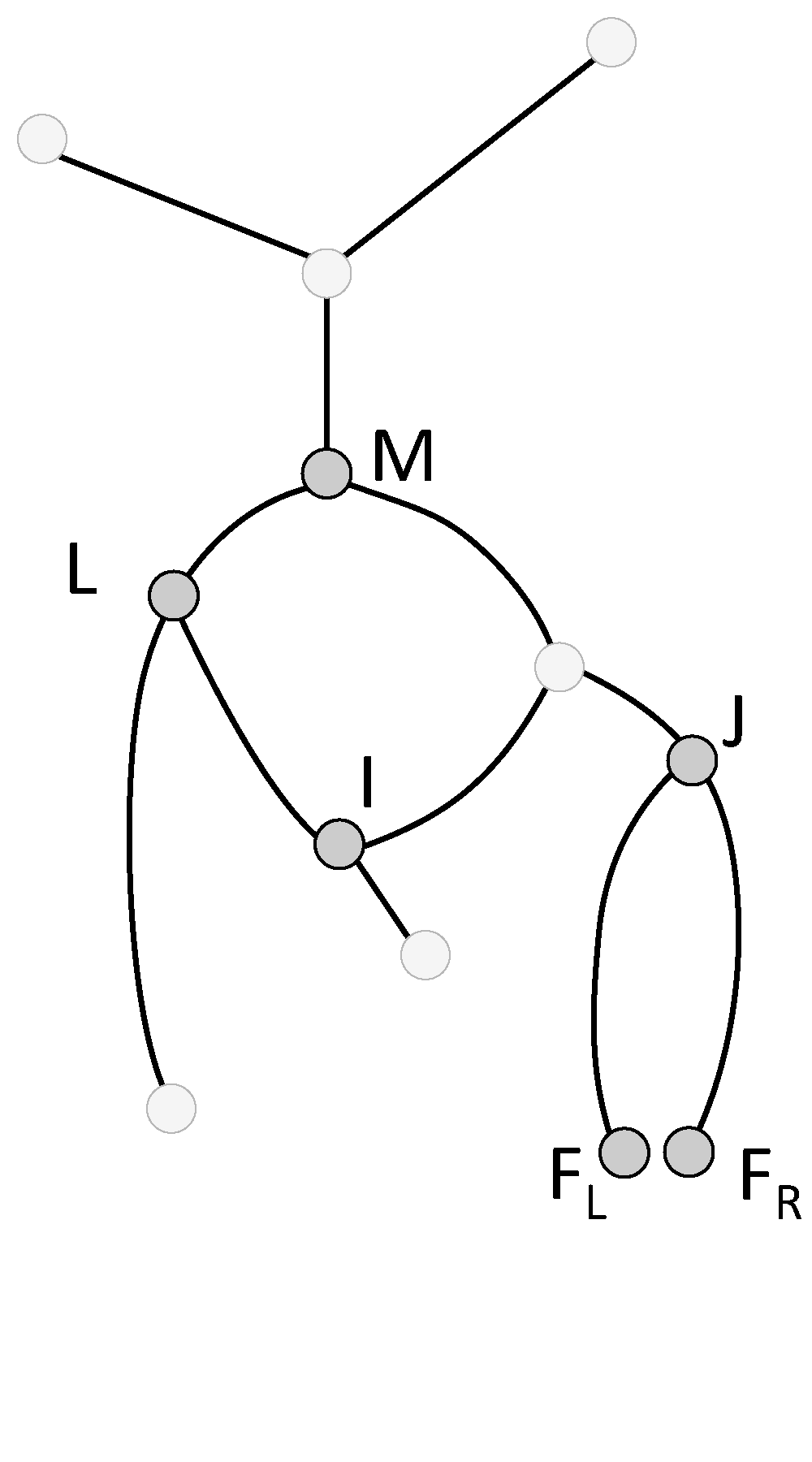}
	        \includegraphics[trim=100pt 0 0 0, clip, height=1.90cm]{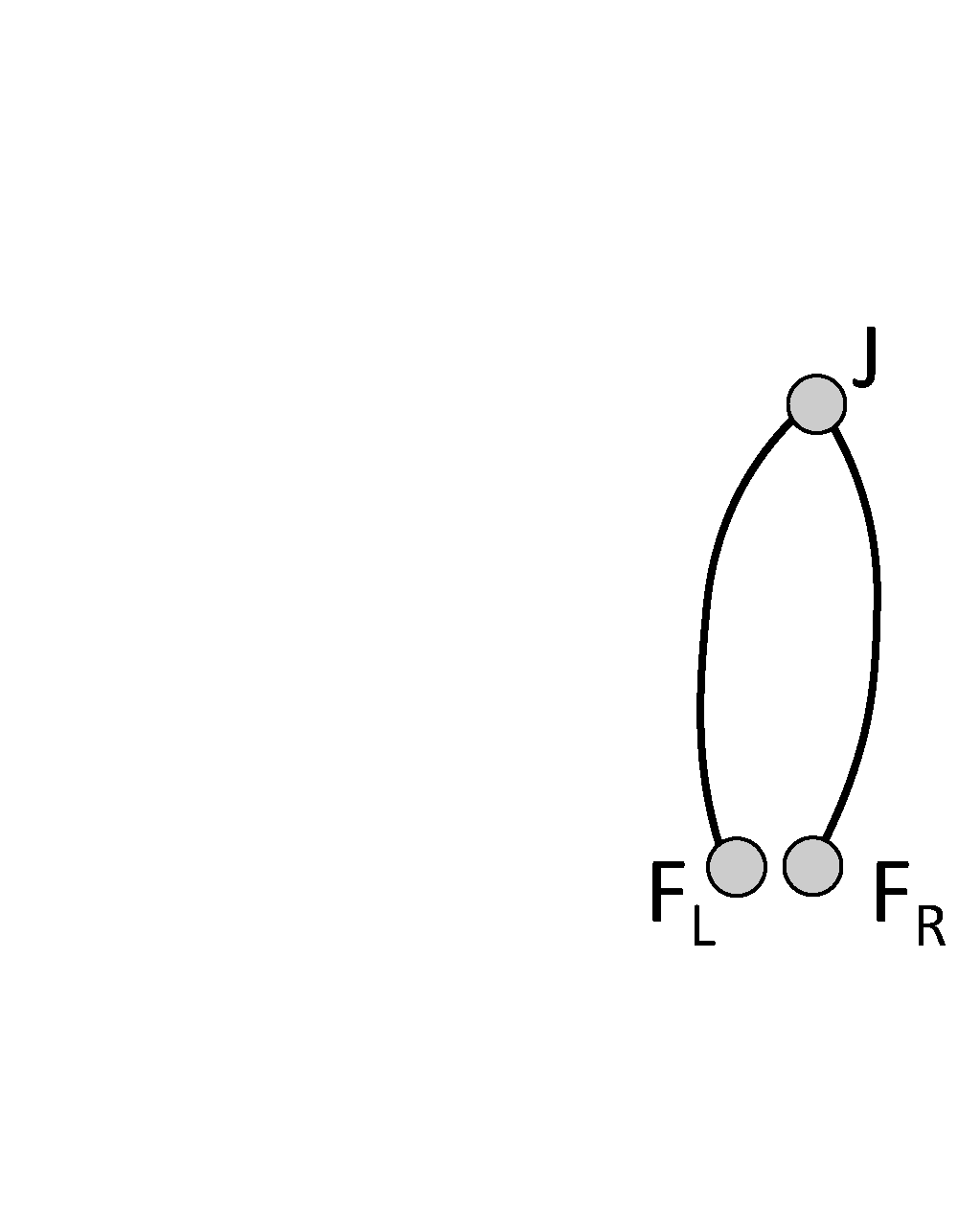}
	        \hspace{40pt}
	    } 
	    \subfigure[$I$ superlevel set (left) and join tree (right)\label{fig.ess_mt.i_sls}]{
	        \hspace{35pt}
	        \includegraphics[height=2.7cm]{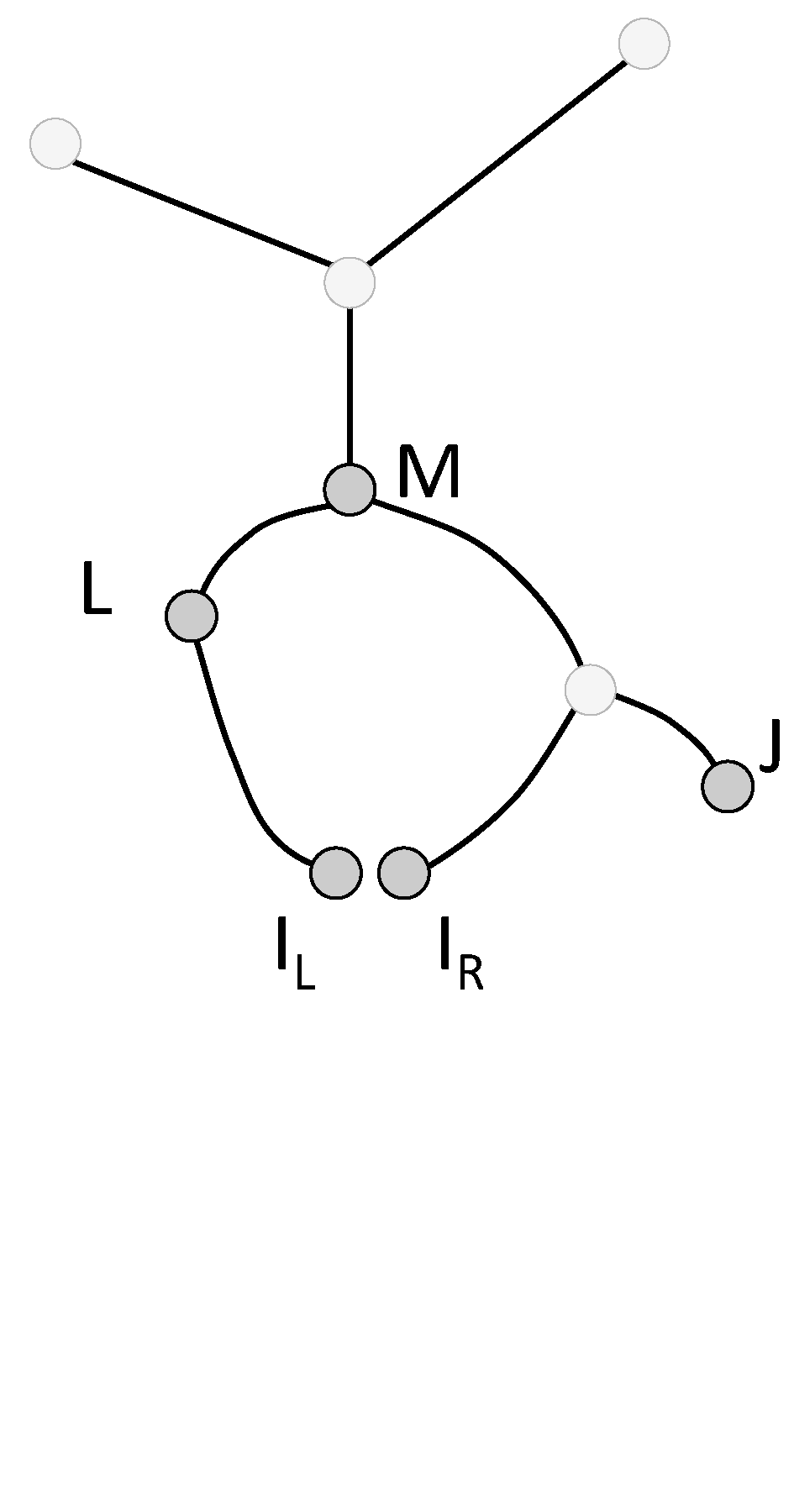}
	        \includegraphics[trim=0 0 0 0, clip, height=1.90cm]{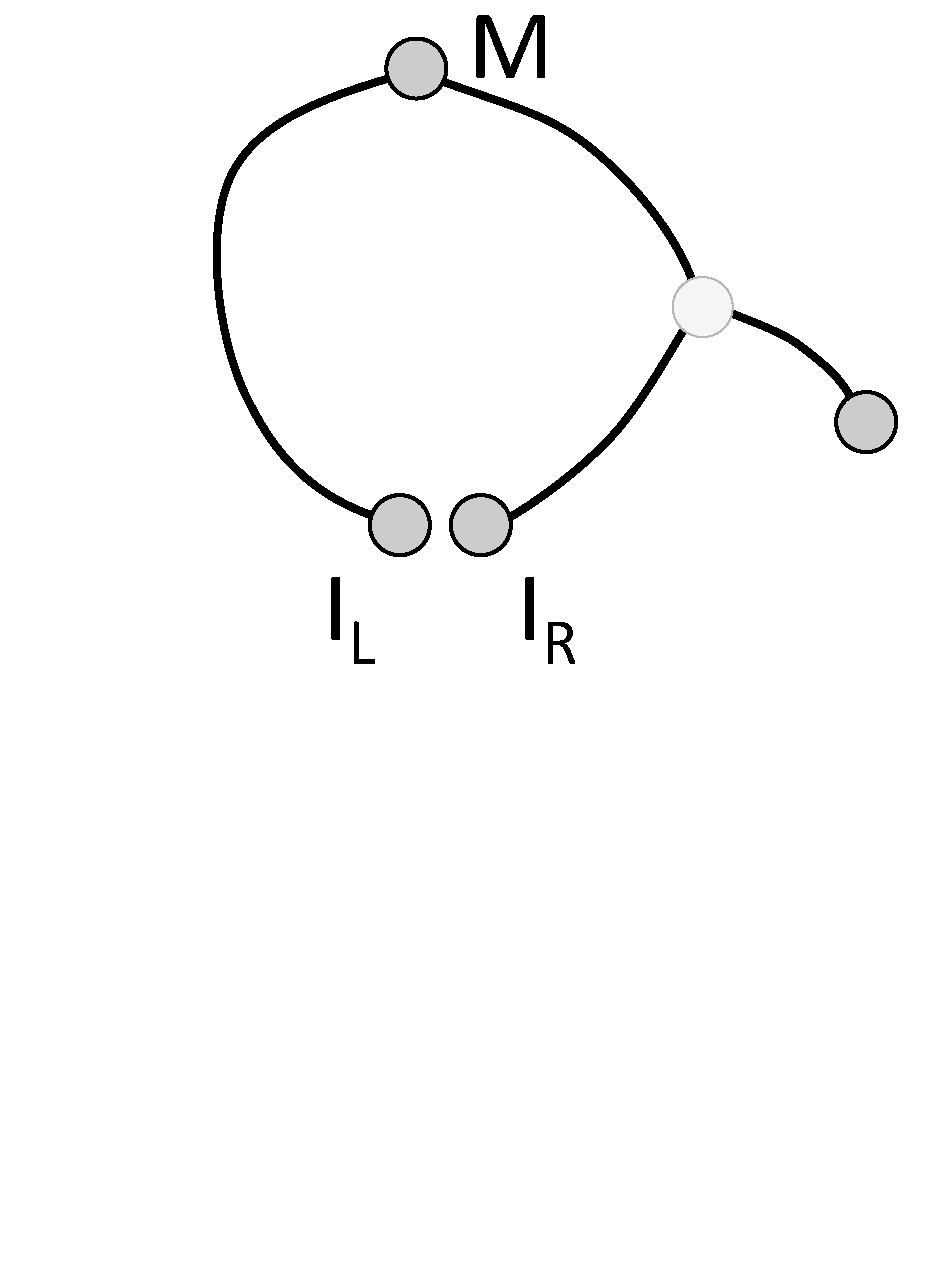}
	        \hspace{35pt}
	   }
    \end{minipage}
	
	\caption{Essential fork pairing in the multipass algorithm for the example Reeb graph from \figref{fig.pipeline}. (a-j) The join tree-based essential fork pairing for up-fork $D$. (a) $D$ is initially split into $D_L$ and $D_R$. (b-i) The colors indicate different connected components as the join tree is swept up the superlevel set. (j) The pairing is found when $D_L$ and $D_R$ are contained in the same connected component. (k-m) Each up-fork ($D$, $F$, and $I$, respectively) is split into 2 pieces and a join tree calculated from the superlevel set to find the partner.}
	\label{fig.essMTEX}
\end{figure}

\figref{fig.essMTEX} shows the sweeping process for the up-fork $D$. Initially (\figref{fig.essMTEX.a}), $D$ is split into $D_L$ and $D_R$, which are each part of separate connected components, denoted by color (\figref{fig.essMTEX.1}). As the join tree is swept past $E$ (\figref{fig.essMTEX.2}), a new connected component is formed. In \figref{fig.essMTEX.3}, $F$ is added to the connected component of $D_R$. As the join tree is swept past $G$ (\figref{fig.essMTEX.4}), the $E$ and $D_L$ connected components join. The process continues until \figref{fig.essMTEX.7}, where 3 connected components exist. The purple and yellow components join at $K$ (\figref{fig.essMTEX.8}). Finally at $L$ (\figref{fig.essMTEX.9}), both $D_L$ and $D_R$ are part of the same connected component. This indicates that $D$ pairs with $L$. \figref{fig.essMTEX}(k-m) shows the superlevel sets and associated join trees for the up-forks $D$, $F$, and $I$. The pairing partner $L$/$D$, $J$/$F$, and $M$/$I$ can all be seen in the $\eDg_1(f)$ in \figref{fig.reebexample.pd1}.

\section{Single-Pass Algorithm: Propagate and Pair}

In the previous section, we showed that the critical point pairing problem could be broken down into a series of merge tree computations. For non-essential forks this was in the form of join and split trees, which are merge trees of the superlevel sets and sublevel sets, respectively. For essential saddles, it came in the form of a special join tree calculation for each essential up-fork. A natural question is whether these merge tree calculations can be combined into a single-pass operation, which is precisely what follows.

\subsection{Basic Propagate and Pair}

The Propagate and Pair algorithm operates by sweeping the Reeb graph from lowest to highest value. At each point, a list of unpaired points from the sublevel set is maintained. When a point is processed in the sweep, 2 possible operations occur on these lists: \textit{propagate} and/or \textit{pair}.

\begin{figure*}[!t]
    \centering

    \subfigure[Local min\label{fig.pp.full.a}]{\hspace{8pt}\includegraphics[height=3.0cm]{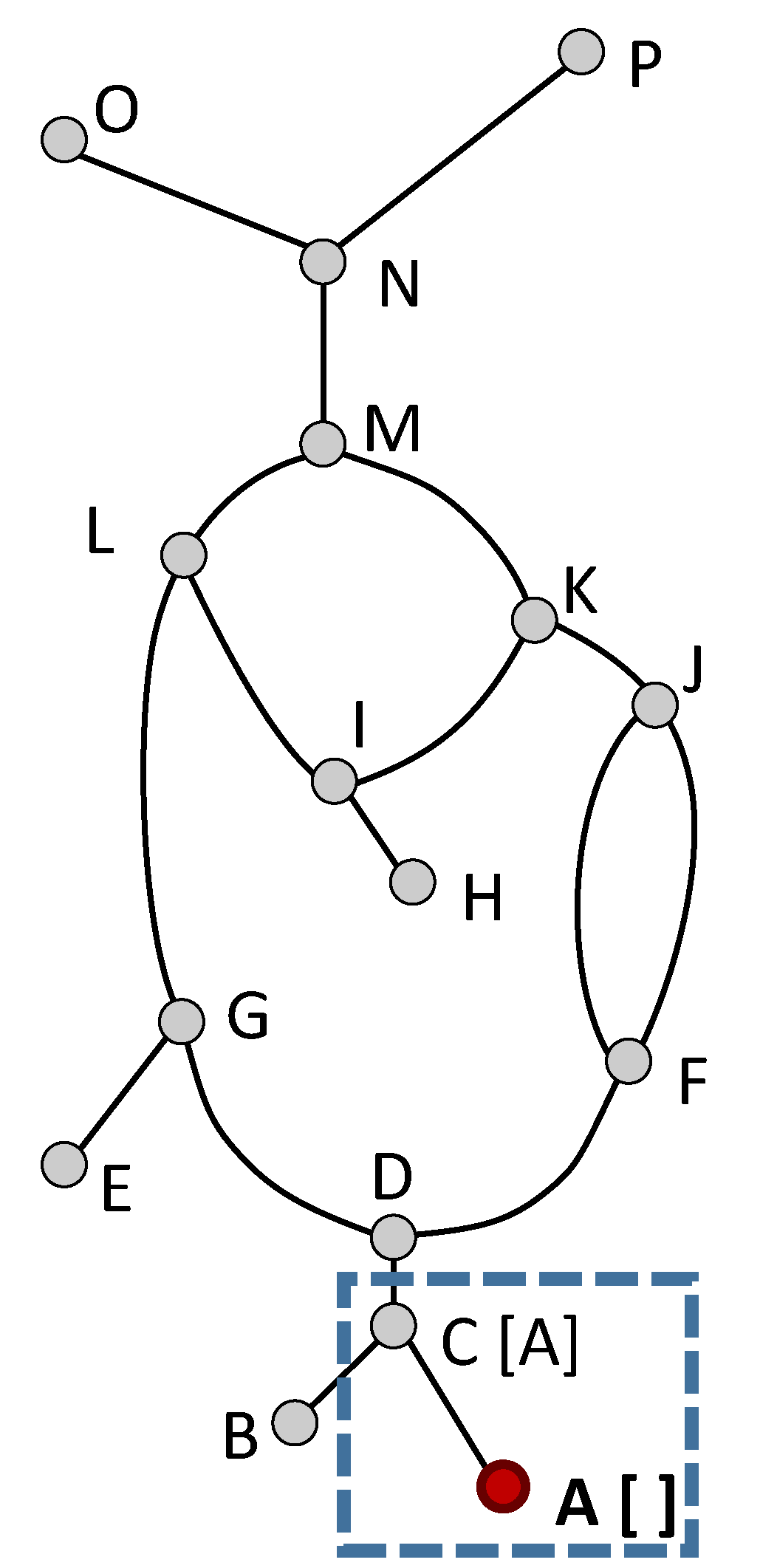}\hspace{8pt}}
    \subfigure[Local min\label{fig.pp.full.b}]{\hspace{8pt}\includegraphics[height=3.0cm]{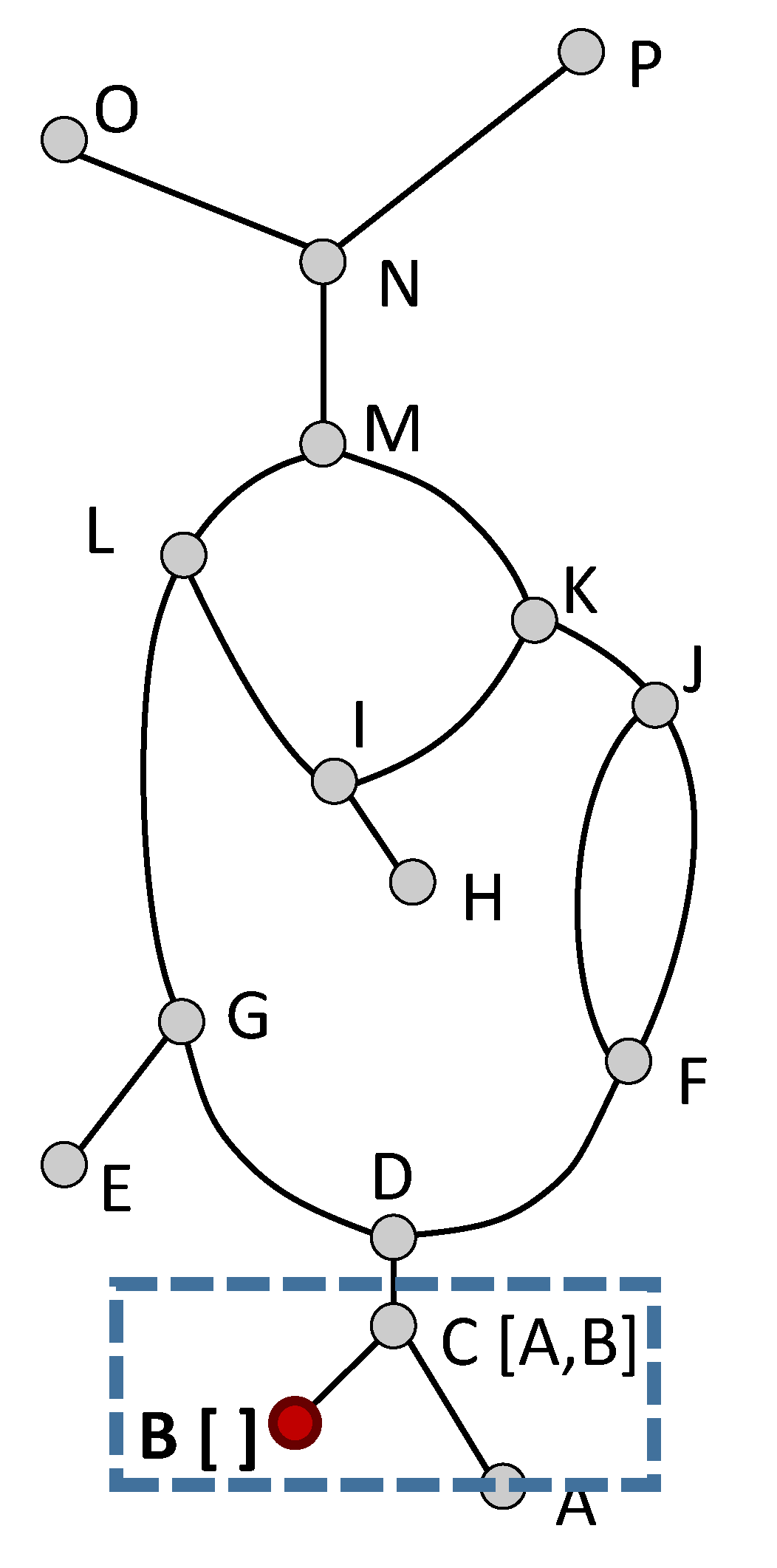}\hspace{8pt}}
    \subfigure[Non-ess.\ d-f\label{fig.pp.full.c}]{\hspace{8pt}\includegraphics[height=3.0cm]{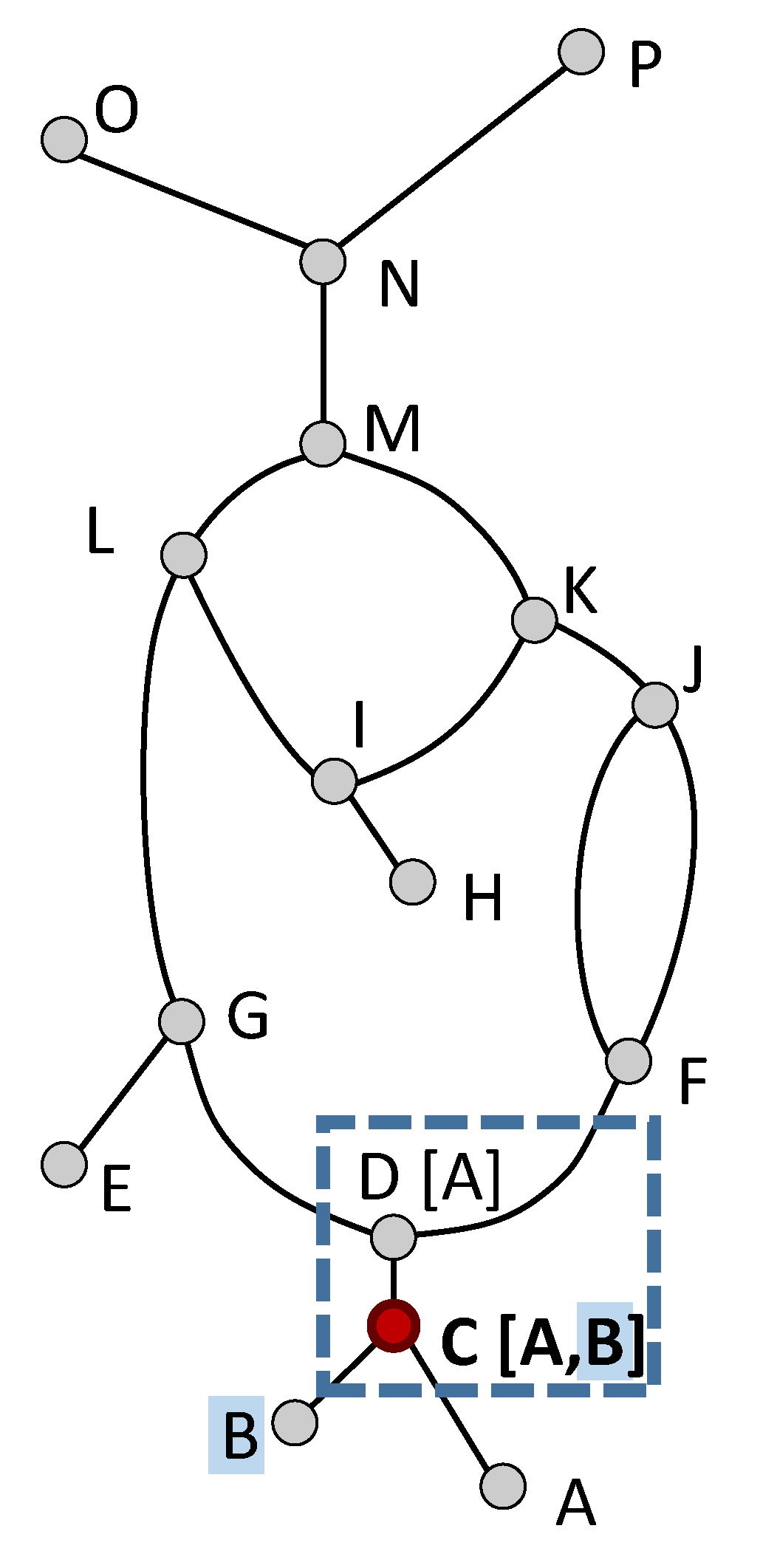}\hspace{8pt}}
    \subfigure[Ess.\ u-f\label{fig.pp.full.d}]{\hspace{8pt}\includegraphics[height=3.0cm]{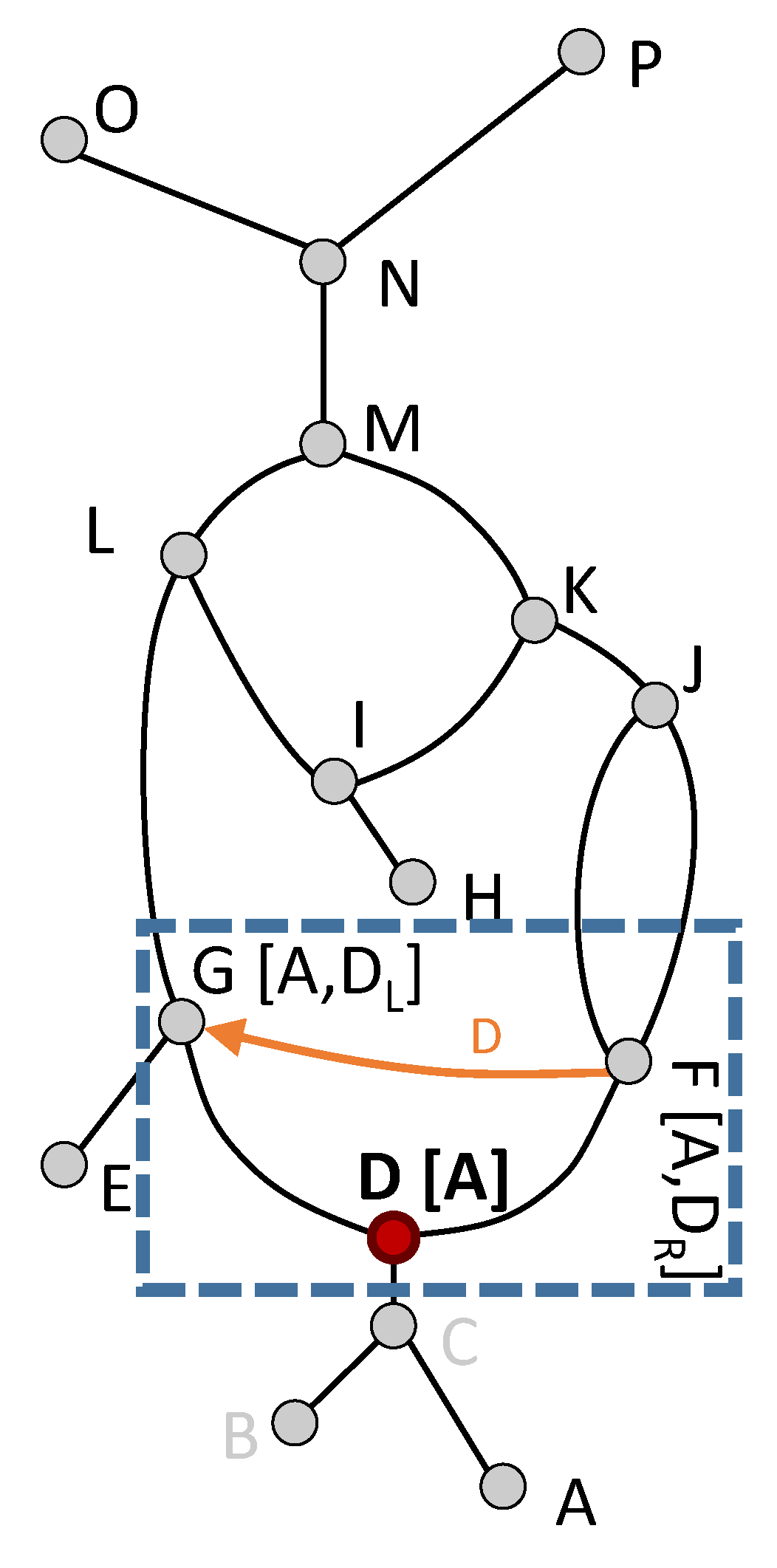}\hspace{8pt}}
    \subfigure[Local min]{\hspace{8pt}\includegraphics[height=3.0cm]{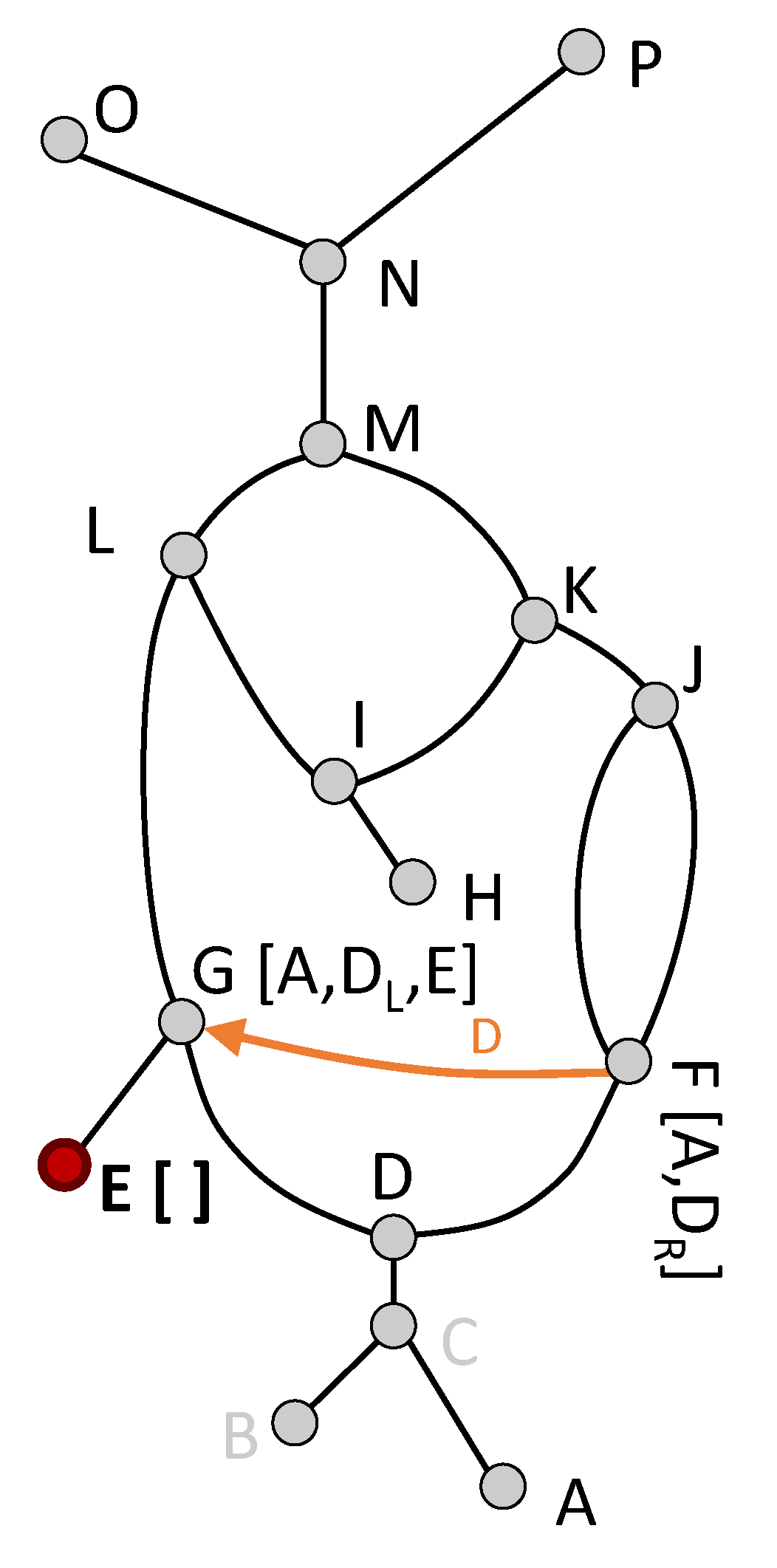}\hspace{8pt}}
    \subfigure[Ess.\ u-f]{\hspace{8pt}\includegraphics[height=3.0cm]{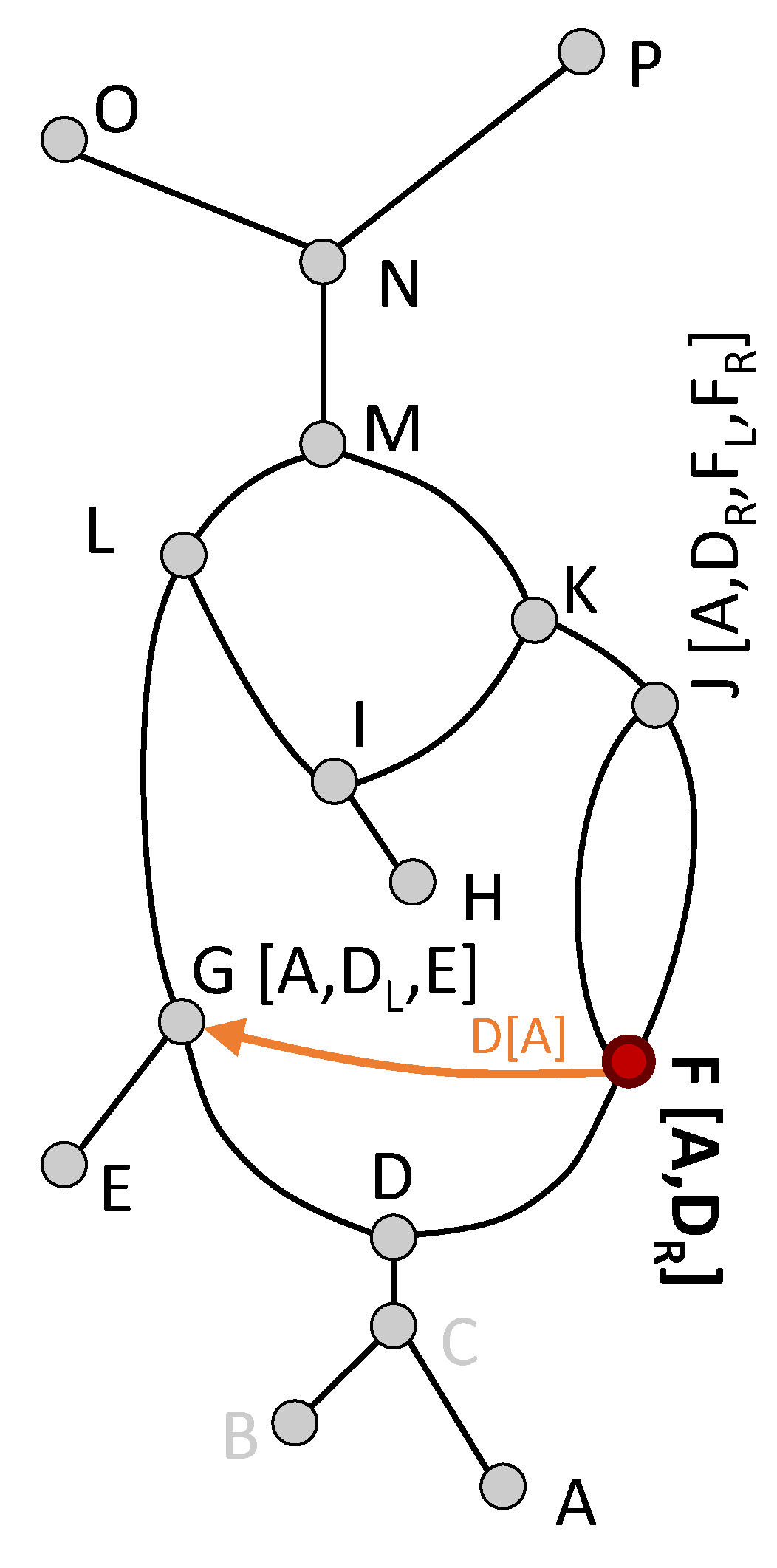}\hspace{8pt}}
    \subfigure[Non-ess.\ d-f]{\hspace{8pt}\includegraphics[height=3.0cm]{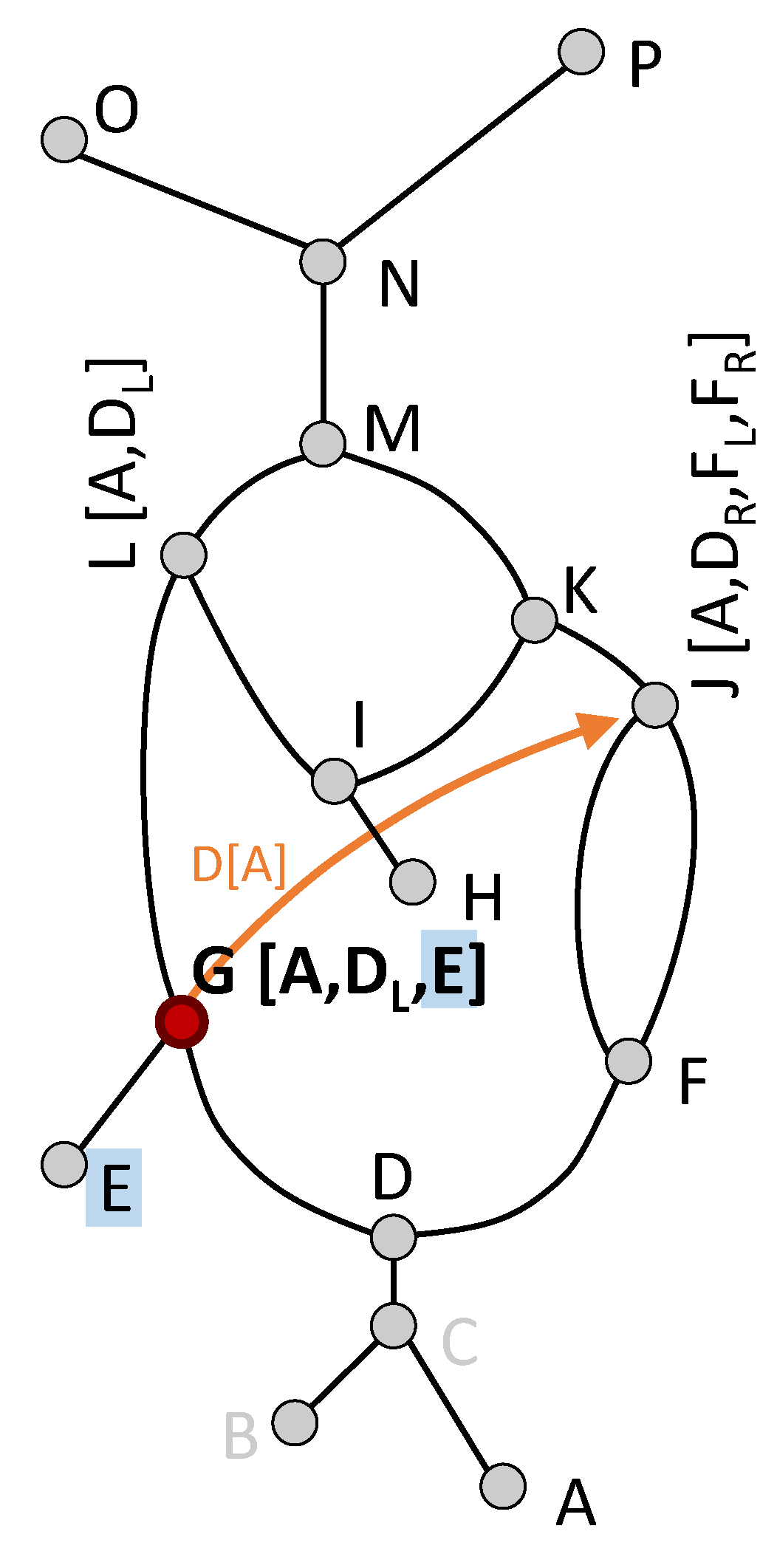}\hspace{8pt}}
    \subfigure[Local min]{\hspace{8pt}\includegraphics[height=3.0cm]{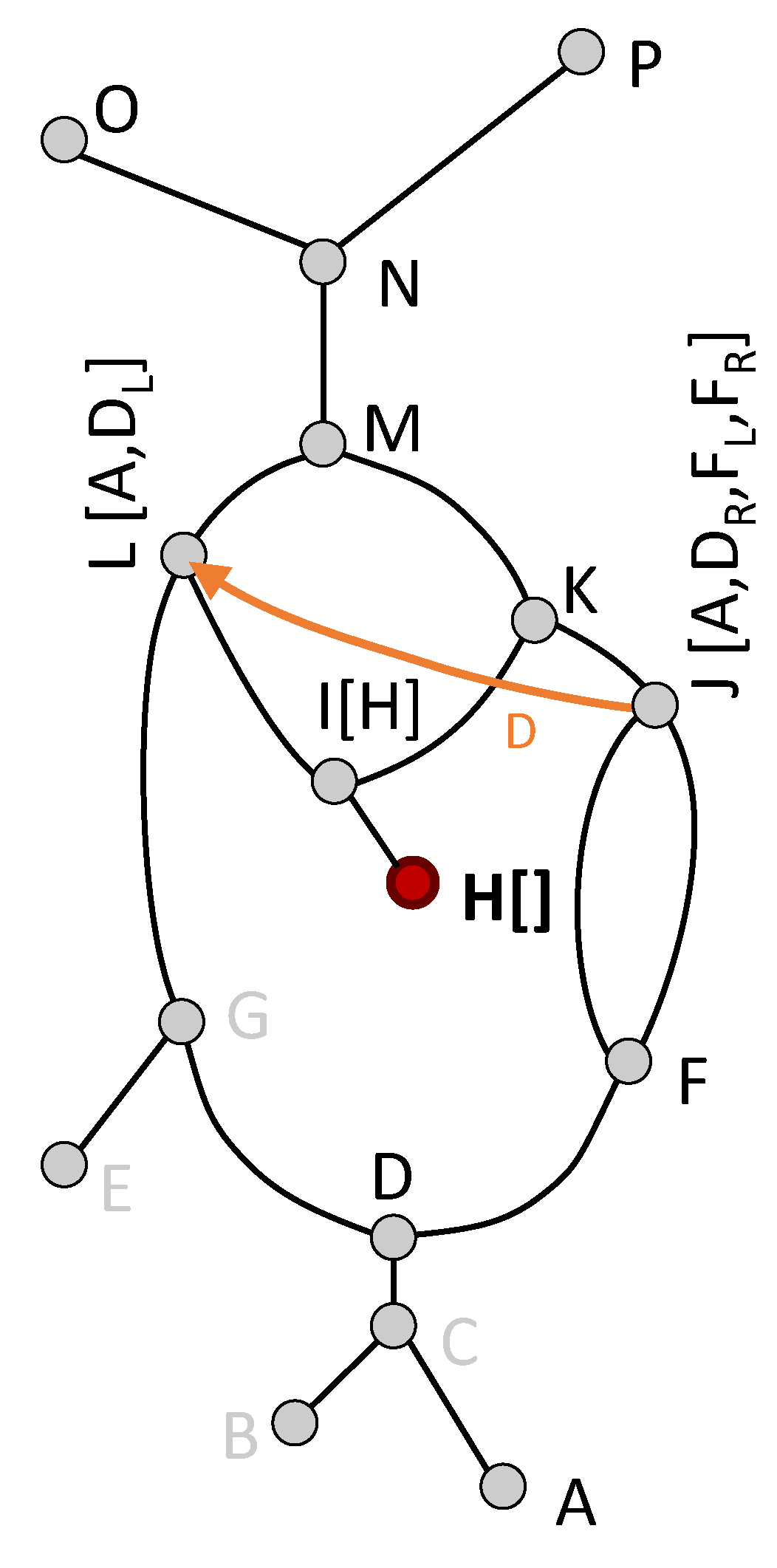}\hspace{8pt}}
    \subfigure[Ess.\ u-f\label{fig.pp.full.i}]{\hspace{8pt}\includegraphics[height=3.0cm]{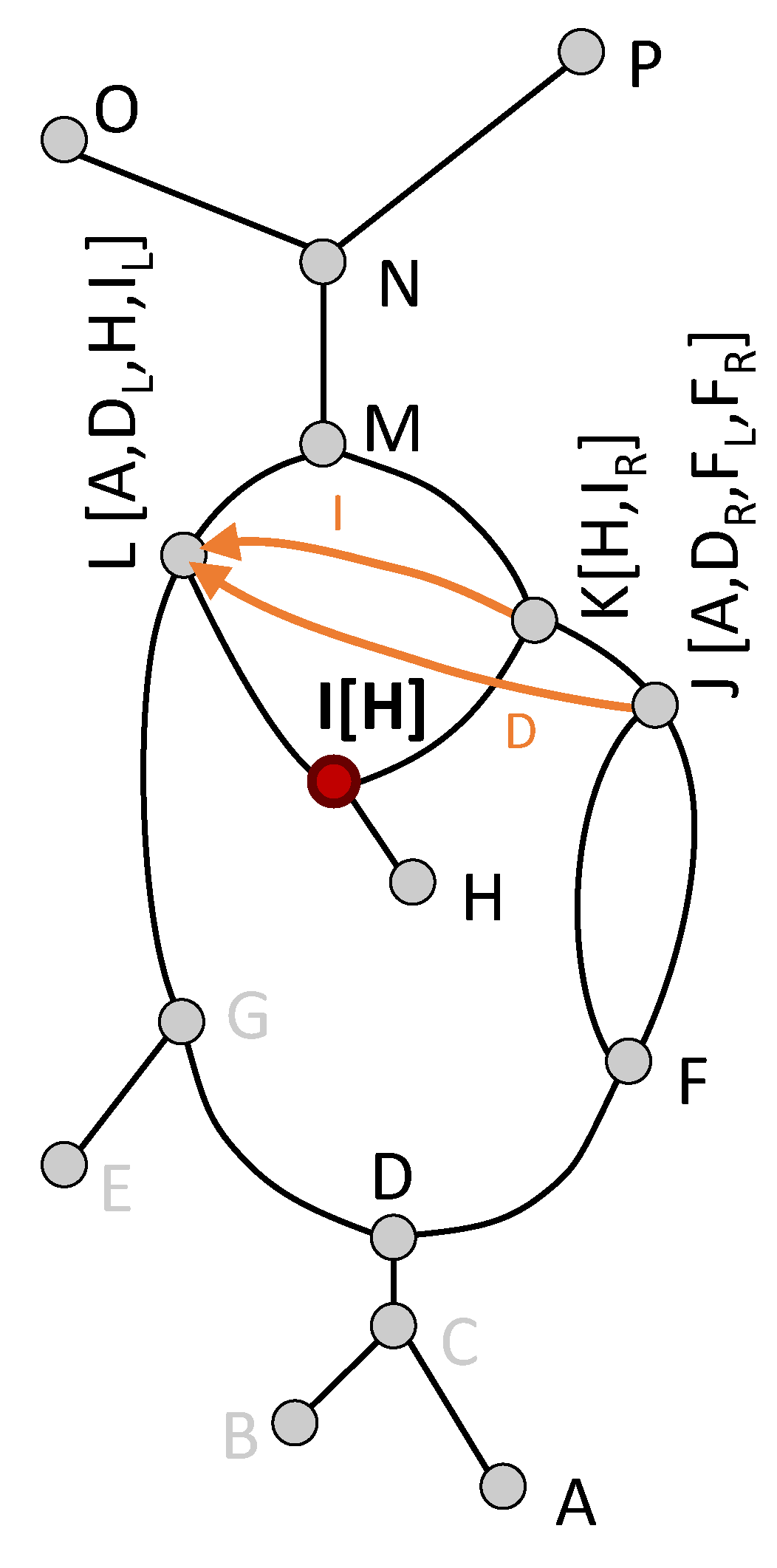}\hspace{8pt}}
    \subfigure[Ess.\ d-f]{\hspace{8pt}\includegraphics[height=3.0cm]{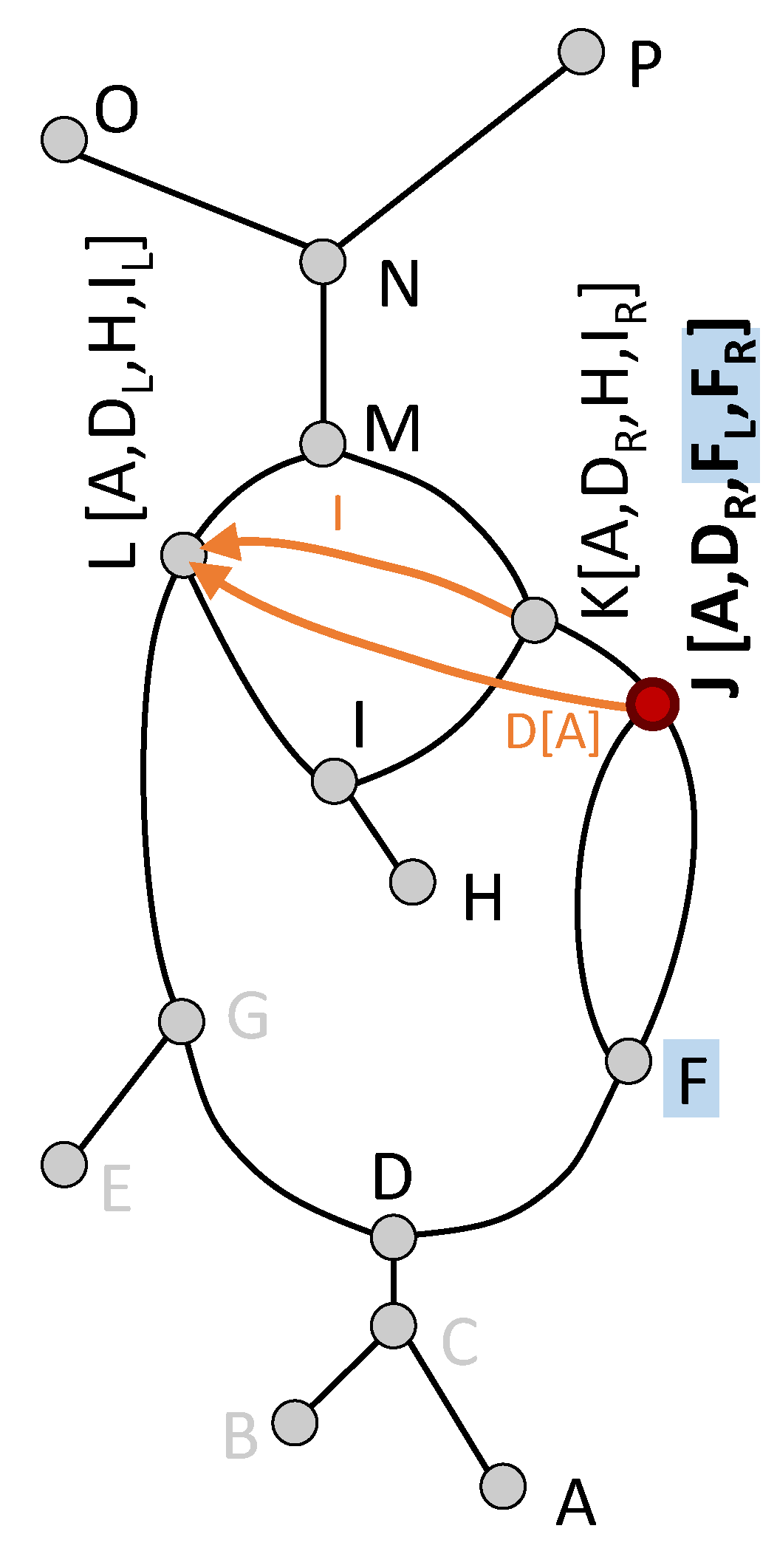}\hspace{8pt}}
    \subfigure[Non-ess.\ d-f]{\hspace{8pt}\includegraphics[height=3.0cm]{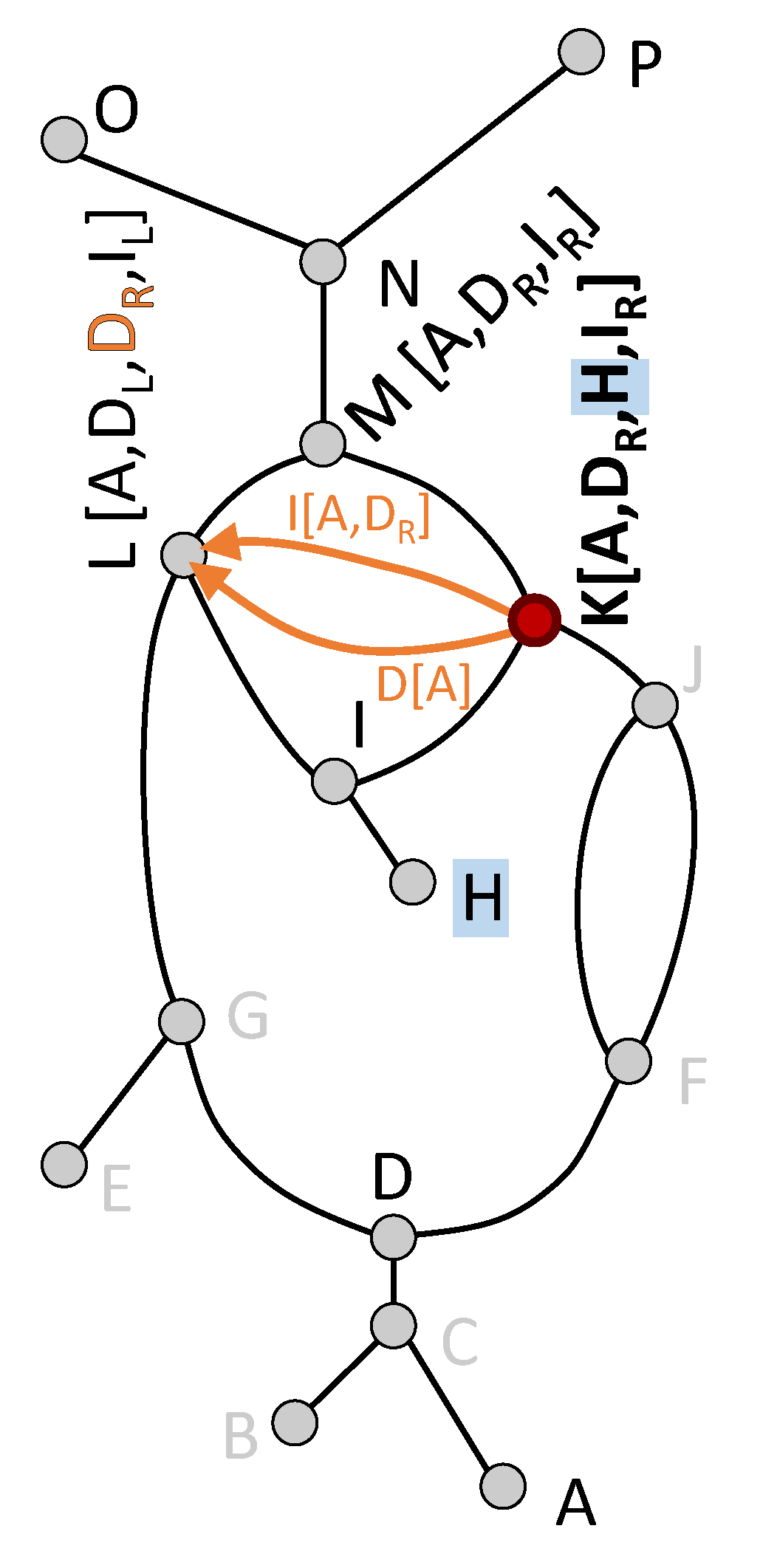}\hspace{8pt}}
    \subfigure[Ess.\ d-f\label{fig.pp.full.l}]{\hspace{8pt}\includegraphics[height=3.0cm]{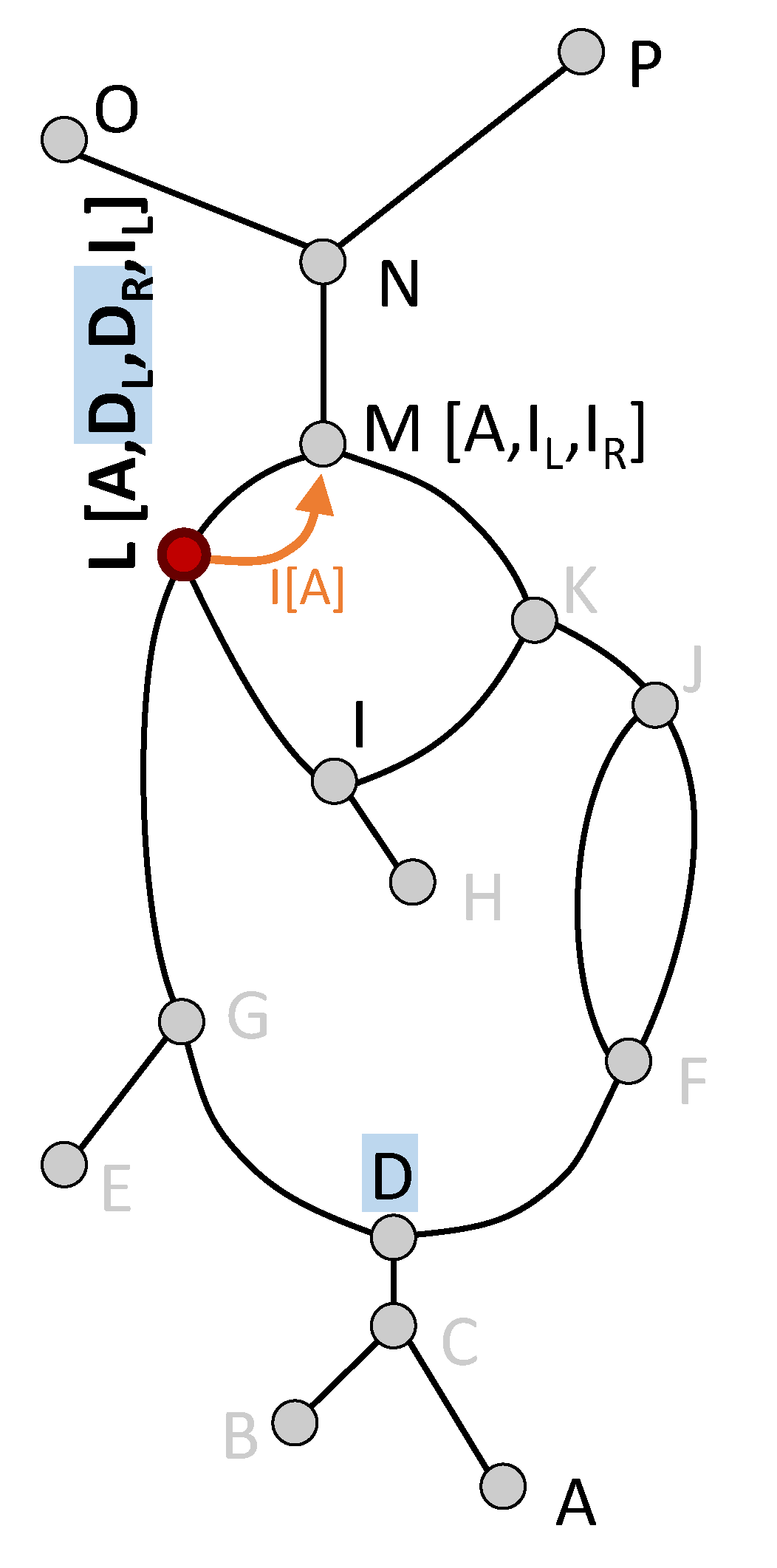}\hspace{8pt}}
    \subfigure[Ess.\ d-f\label{fig.pp.full.m}]{\hspace{8pt}\includegraphics[height=3.0cm]{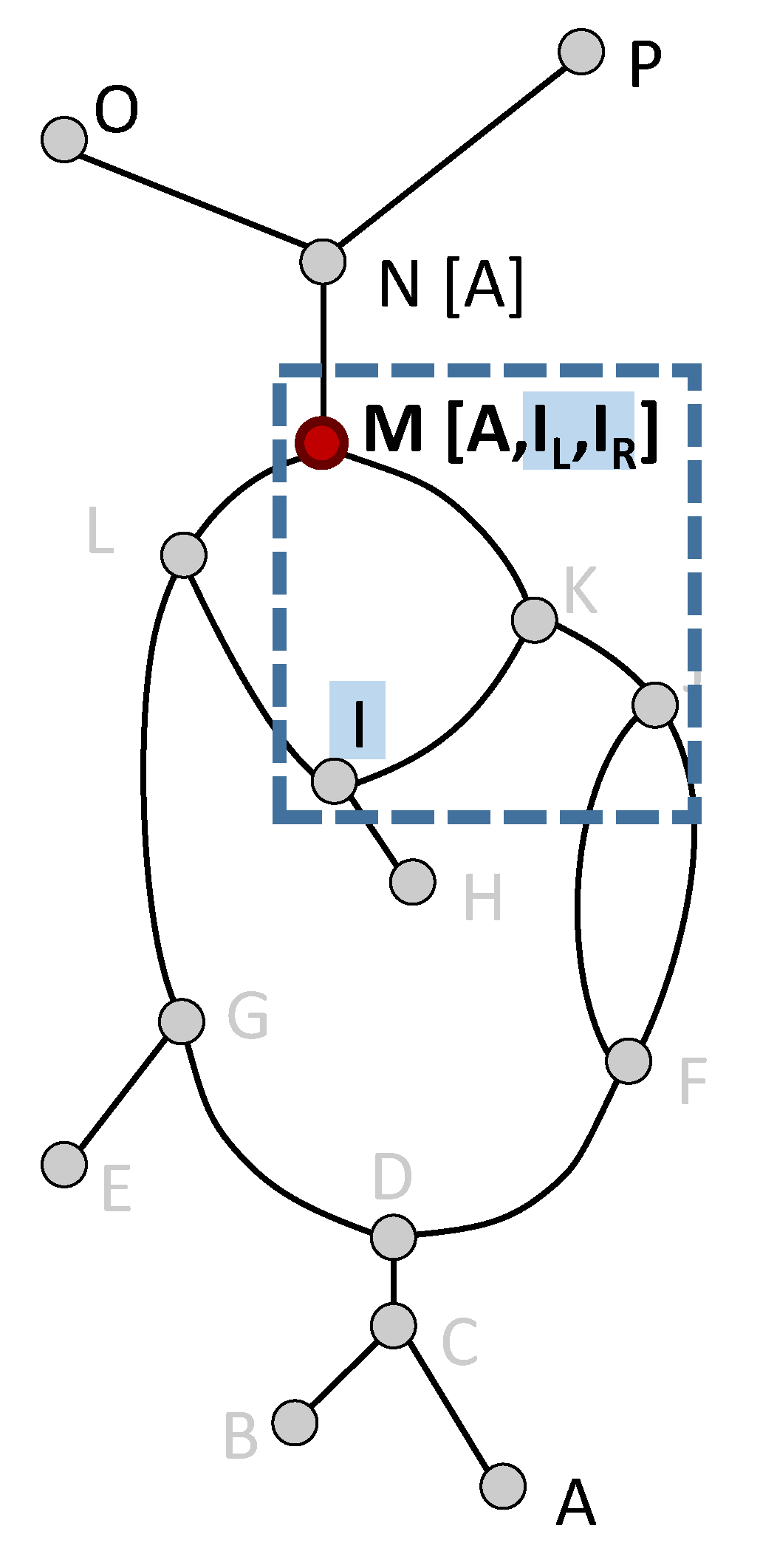}\hspace{8pt}}
    \subfigure[Non-ess.\ u-f]{\hspace{8pt}\includegraphics[height=3.0cm]{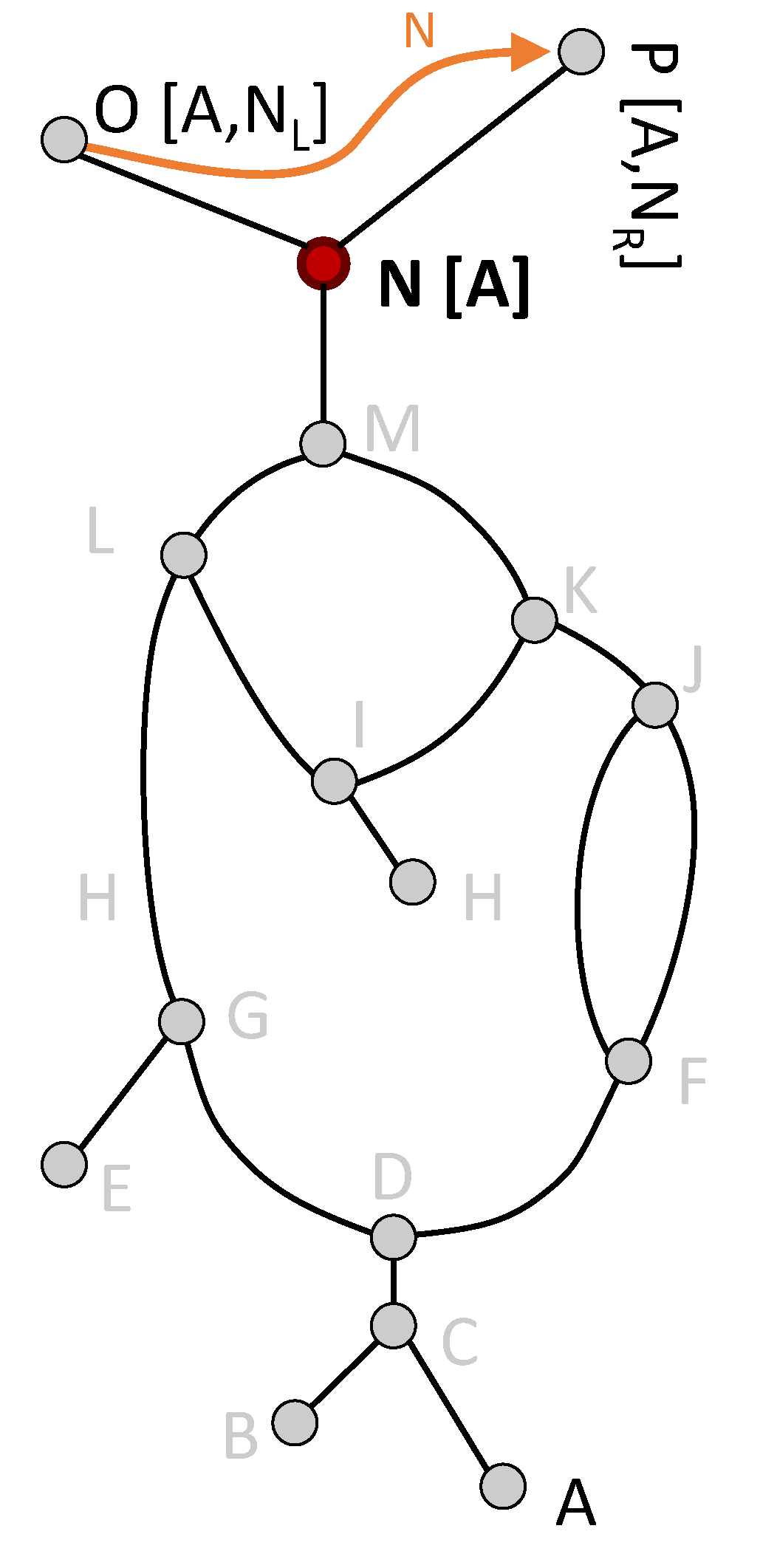}\hspace{8pt}}
    \subfigure[Local max\label{fig.pp.full.o}]{\hspace{8pt}\includegraphics[height=3.0cm]{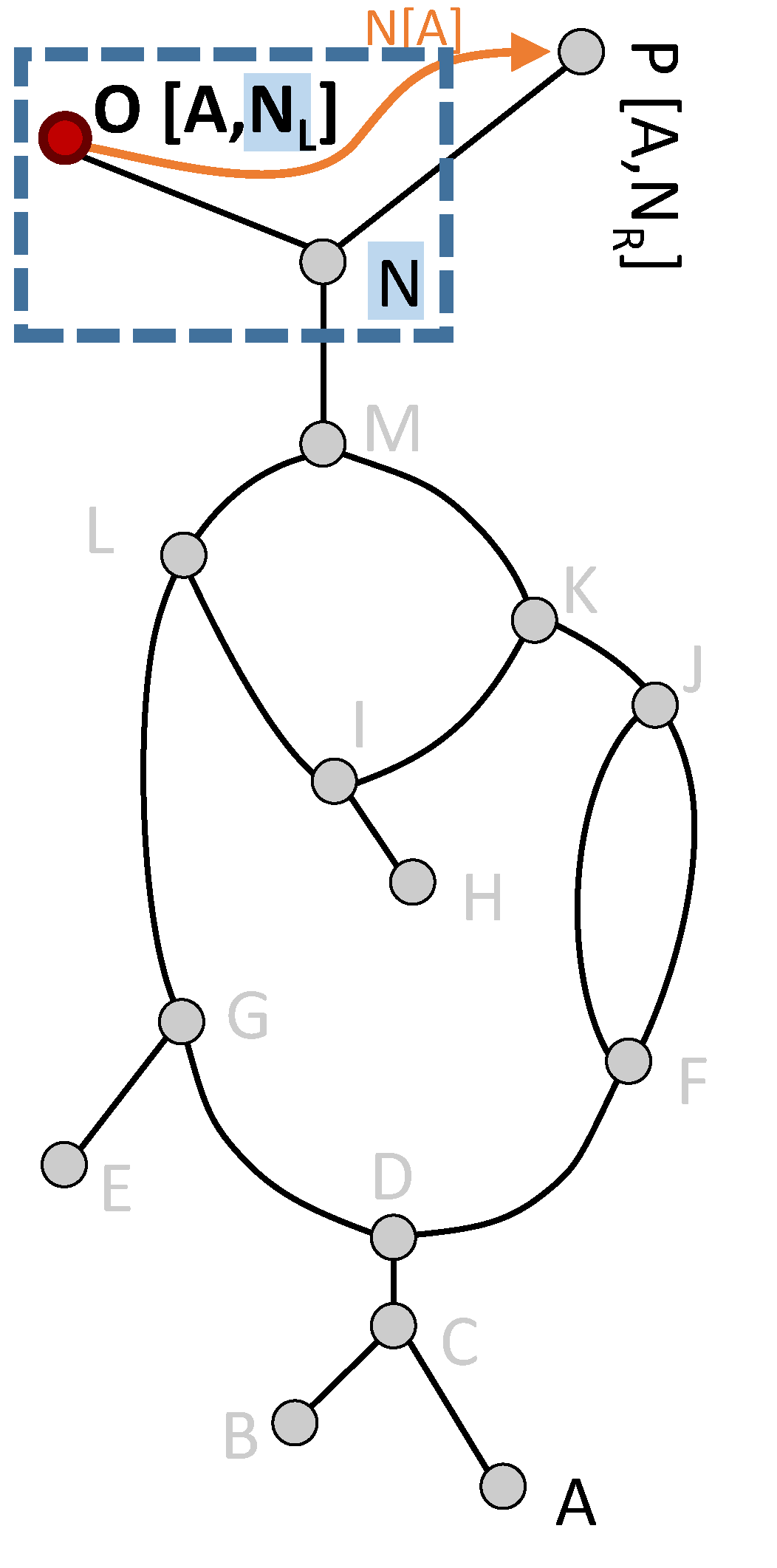}\hspace{8pt}}

    \caption{Propagate and Pair algorithm on the Reeb graph from \figref{fig.pipeline}. At each step, the node being processed is in bold; propagated edges are shown in brackets; pairing is shown in blue; and virtual edges are shown in orange. (ess.: essential; non-ess.: non-essential; d-f: down-fork; u-f: up-fork)}
    \label{fig.pp.full}
\end{figure*}

\paragraph{Propagate} The job of propagate is to push labels from unpaired nodes further up the unprocessed Reeb graph. 4 cases exist.
\begin{itemize}
    \item For \underline{local minima} a label for the current critical point is propagated upward. In the examples of \figref{fig.pp.full.a} and \ref{fig.pp.full.b}, both $A$ and $B$ are propagated to $C$. 
    \item For \underline{local maxima} nothing needs to propagate.
    \item For \underline{down-forks} all unpaired labels are propagated upwards. In the example of \figref{fig.pp.full.c}, the critical points $B$ and $C$ are paired, thus only $A$ is propagated to $D$. 
    \item For \underline{up-forks} all unpaired labels are propagated upwards. Additional labels for the current up-fork are created and tagged with the specific branch of the fork that created them (in the examples with subscripts $L$ and $R$). This tag is critical for closing essential cycles. In the example of \figref{fig.pp.full.d}, the labels $A$ and $D_L$ are propagated to $G$, and labels $A$ and $D_R$ are propagated to $F$. 
\end{itemize}

\paragraph{Pair} The pairing operation searches the list of labels to determine an appropriate pairing partner from the sublevel set. The pairing operation only occurs for local maxima and down-forks. 
\begin{itemize}
    \item For \underline{local maxima} the labels list is searched for the unpaired up-fork with the largest value. Those critical points are then paired. In \figref{fig.pp.full.o}, for local maximum $O$, the list is searched and $N_L$ is determined to be the closest unpaired up-fork. 
    \item For \underline{down-forks} two possible cases exist, essential or non-essential, which can be differentiated by searching the available labels. First, the list is searched for the largest up-fork with both legs. Both legs indicate that the current down-fork closes a cycle with the associated up-fork. In the example, \figref{fig.pp.full.m}, the list of $M$ is searched and labels $I_L$ and $I_R$ found. If no such up-fork exists, then the down-fork is non-essential. In this case, the highest valued local minimum is selected from the list. In the example of \figref{fig.pp.full.c}, no essential up-forks are found for $C$, and the largest local minimum, $B$ is selected instead.
\end{itemize}

\begin{figure}[!t]
\centering
    \begin{minipage}[m]{0.45\linewidth}
        \begin{minipage}[m]{1\linewidth}
            \centering
            \subfigure[Failure case for propagate and pair\label{fig.vedge.failure}]{{\hspace{45pt}\includegraphics[trim=0 0 50pt 0, clip, width=0.3\linewidth]{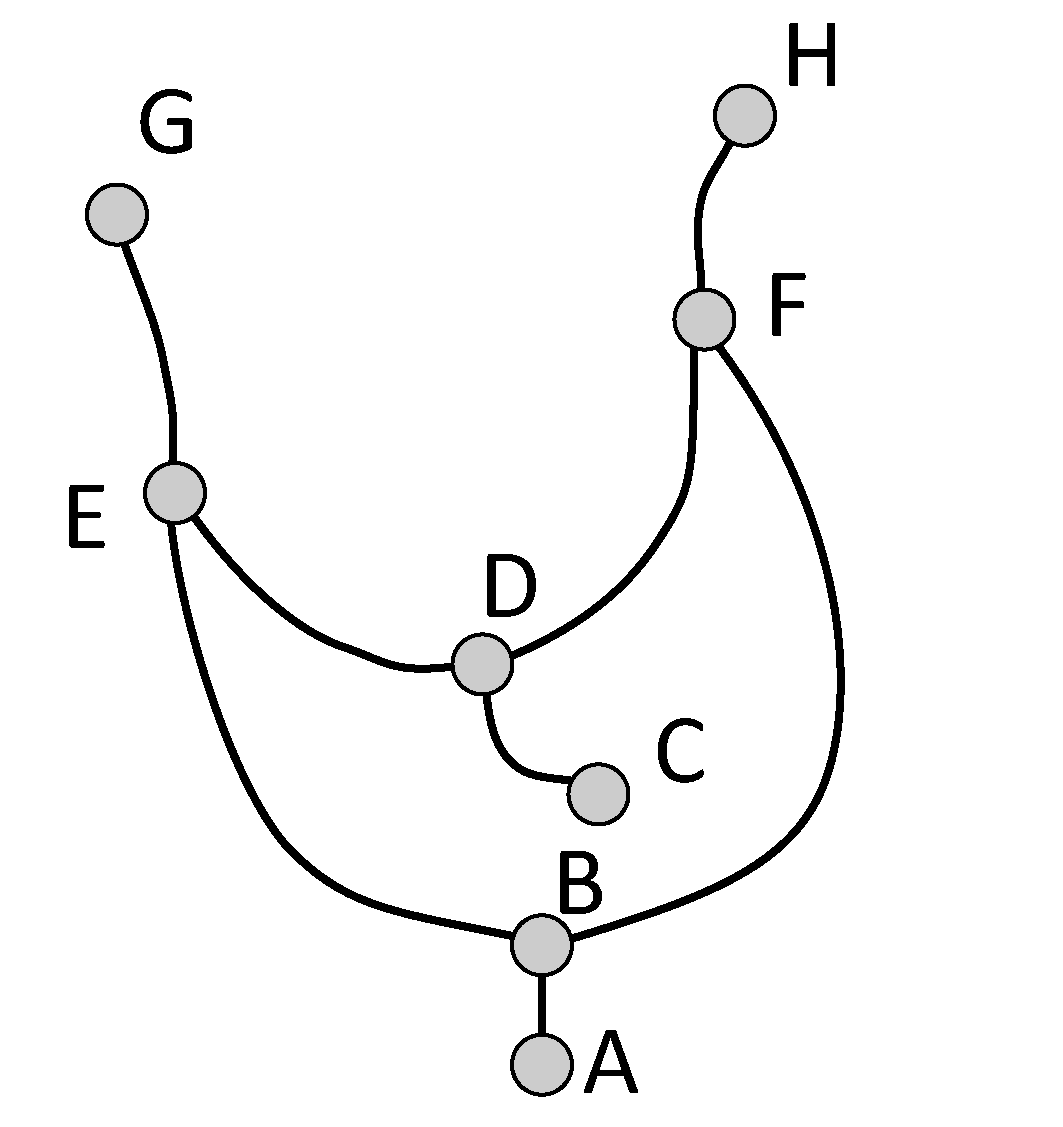}\hspace{45pt}}}
        \end{minipage}
        %
        \begin{minipage}[m]{1\linewidth}
            \subfigure[$B$ created v.\ edge \label{fig.vedge.createB}]{\hspace{12pt}{\includegraphics[width=0.325\linewidth]{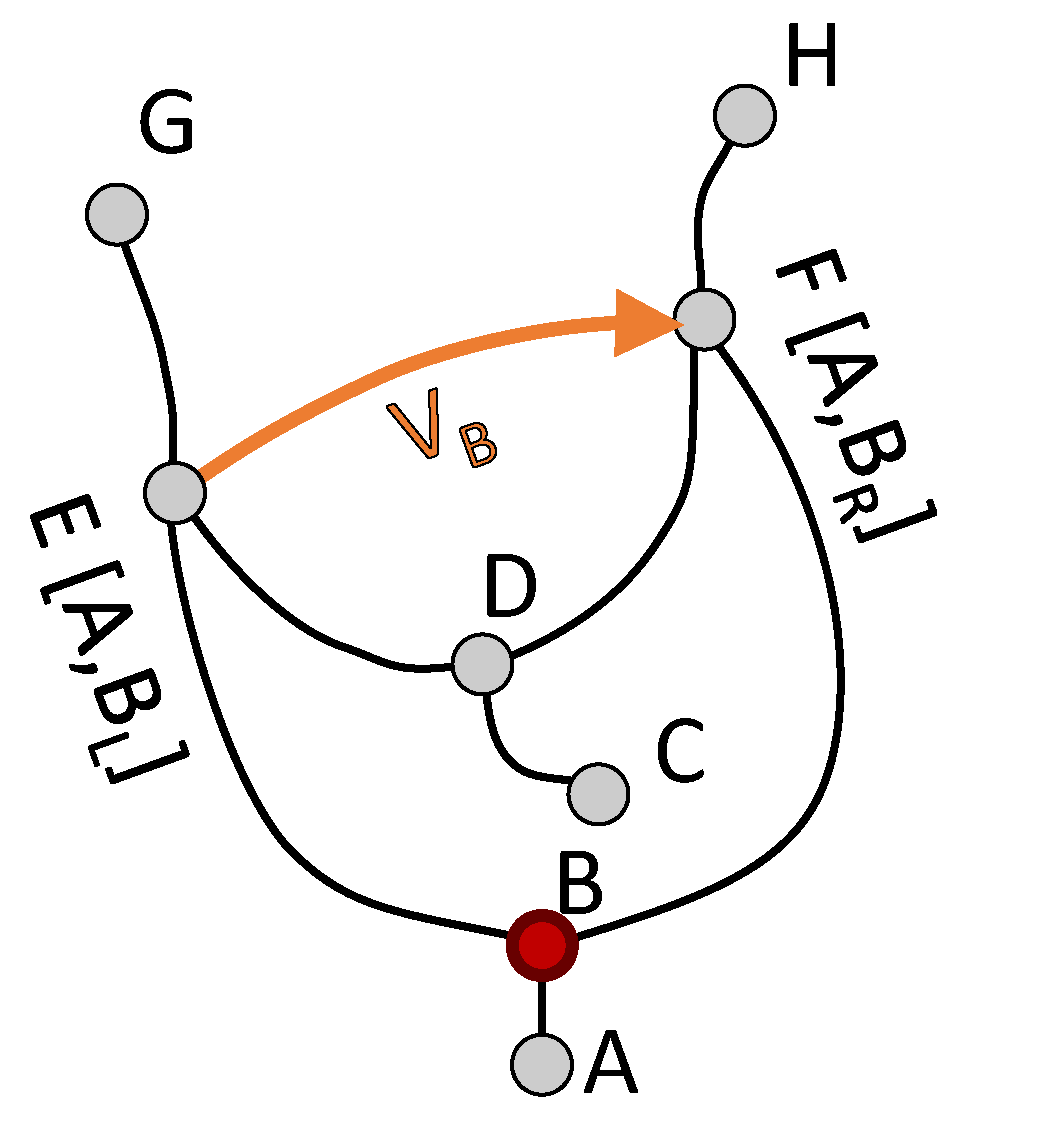}}\hspace{12pt}}
            \hfill
            \subfigure[$D$ created v.\ edge\label{fig.vedge.createD}]{\hspace{12pt}{\includegraphics[width=0.325\linewidth]{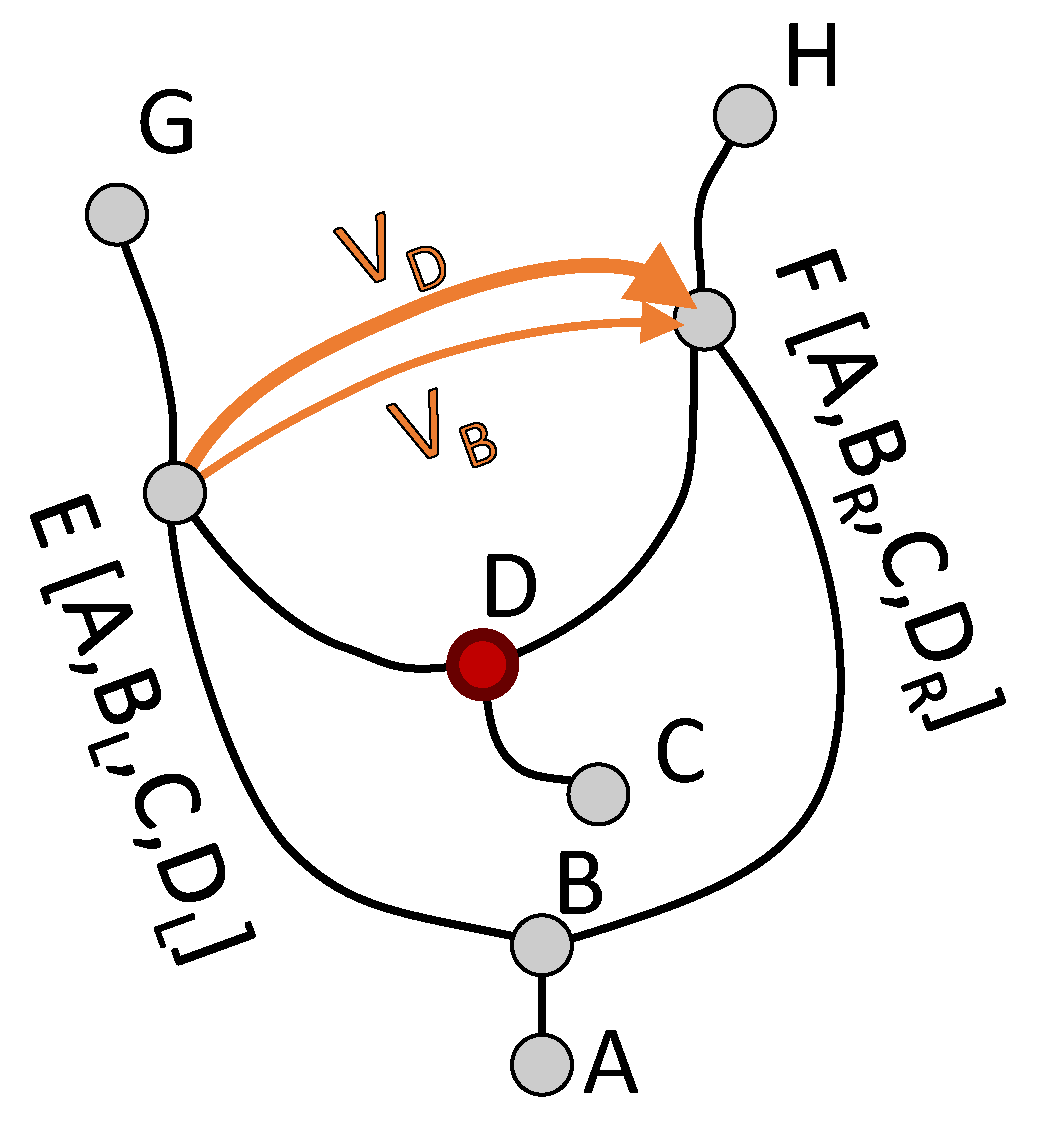}}\hspace{12pt}}
    
            \subfigure[Label prop.\label{fig.vedge.labelP}]{\hspace{12pt}{\includegraphics[width=0.325\linewidth]{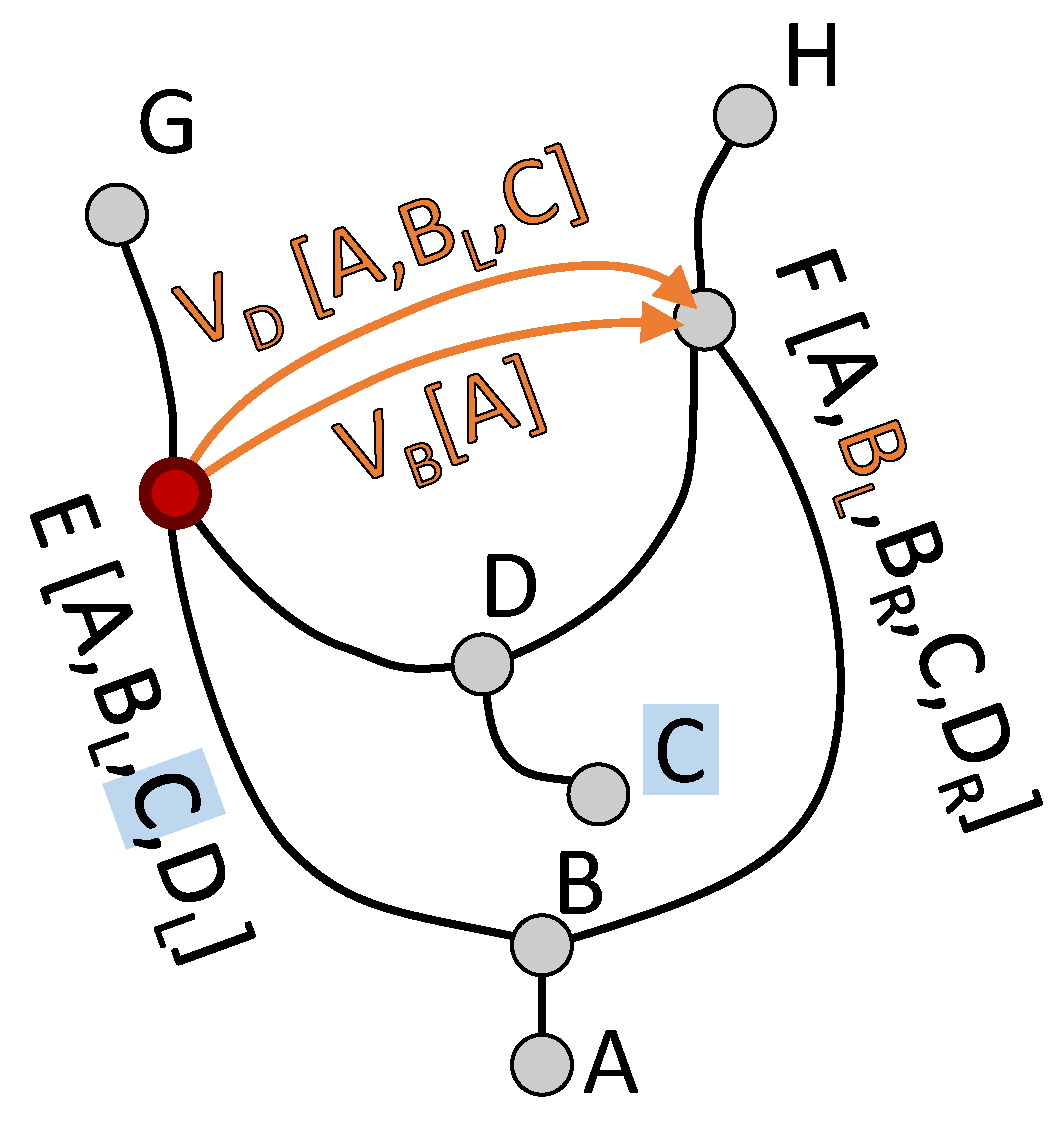}}\hspace{12pt}}
            \hfill
            \subfigure[V.\ edge prop.\label{fig.vedge.edgeP}]{\hspace{12pt}{\includegraphics[width=0.325\linewidth]{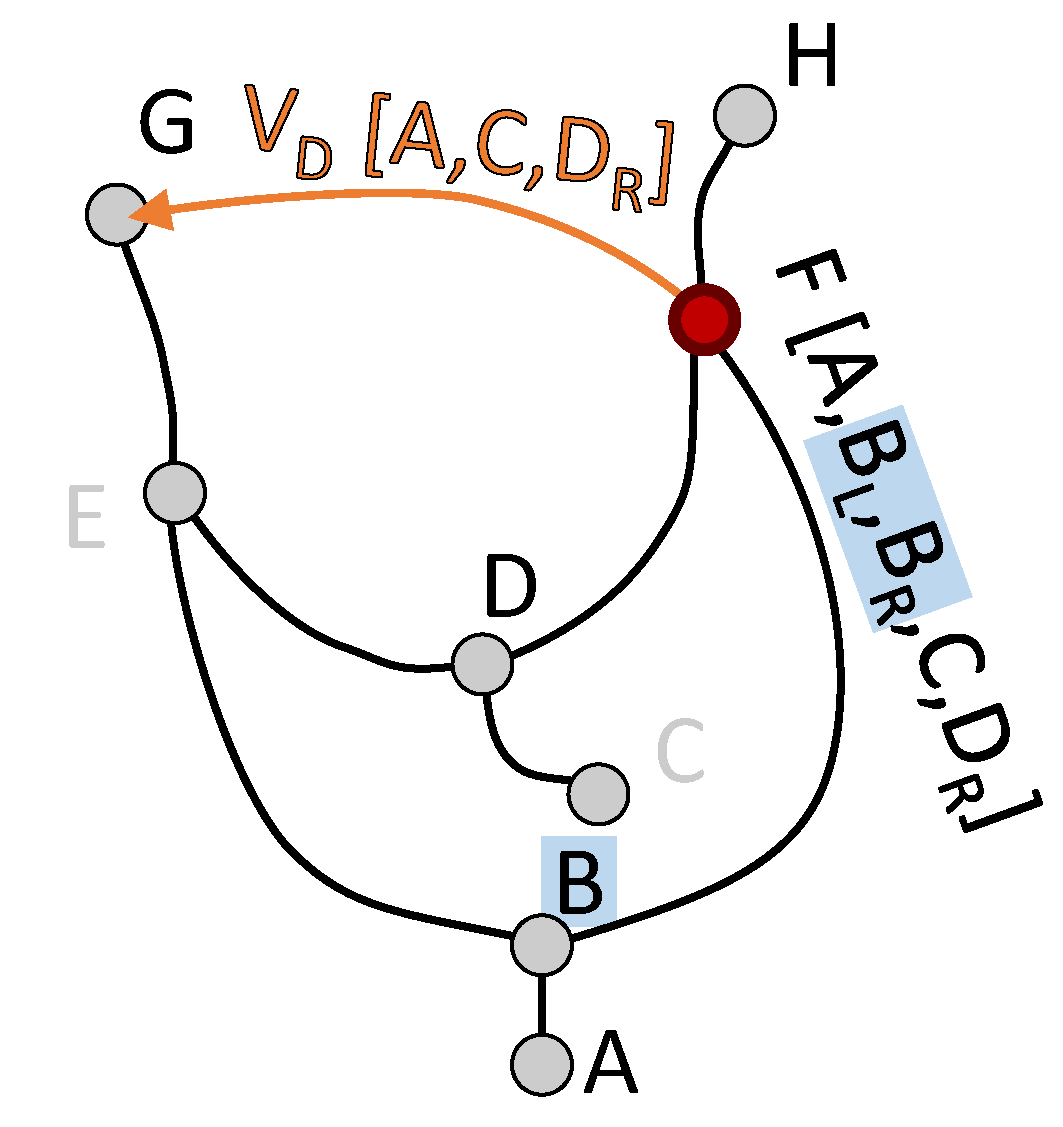}}\hspace{12pt}}
        \end{minipage}
    \end{minipage}
    \hfill
    \begin{minipage}[m]{0.52\linewidth}
        \subfigure[Initial]{\includegraphics[trim=0 0 0pt 0, clip, width=0.485\linewidth]{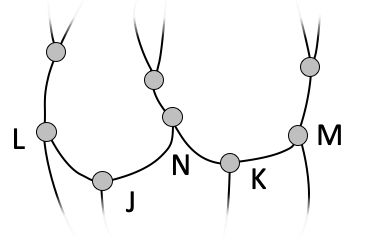}}
        \hfill
        \subfigure[$J$ created v.\ edge]{\includegraphics[trim=0 0 0pt 0, clip, width=0.485\linewidth]{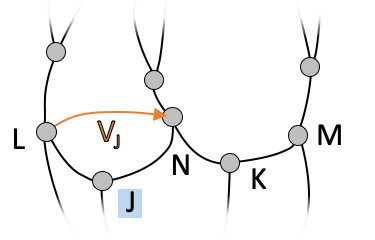}}
        \subfigure[$K$ created v.\ edge]{\includegraphics[trim=0 0 0pt 0, clip, width=0.485\linewidth]{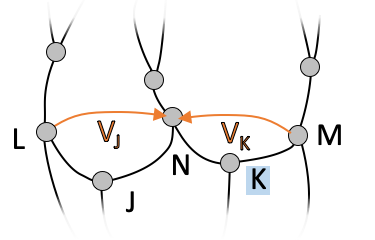}}
        \hfill
        \subfigure[V.\ edge prop.]{\includegraphics[trim=0 0 0pt 0, clip, width=0.485\linewidth]{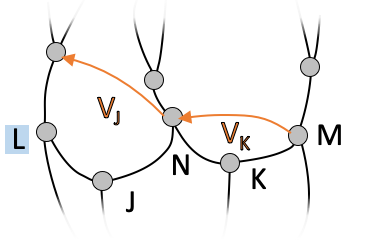}}
        \subfigure[V.\ edge prop.]{\includegraphics[trim=0 0 0pt 0, clip, width=0.485\linewidth]{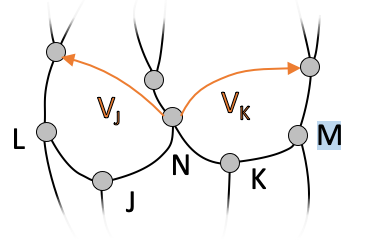}}
        \hfill
        \subfigure[V.\ edge merge to $V_{JK}$\label{fig.vedge.merge.f}]{\includegraphics[trim=0 0 0pt 0, clip, width=0.485\linewidth]{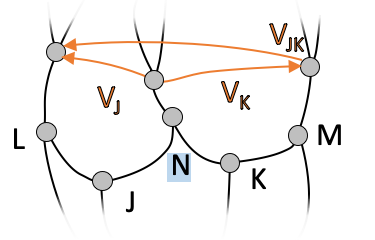}}
    \end{minipage}

\caption{Example where the basic propagate and pair algorithm fails. (a-e) In this case, $B$ and $F$ should pair but do not.  To overcome this limitation, (b-c) virtual edges are created as up-forks are processed. (d) Labels can then be propagated across virtual edges. (e) The virtual edges themselves are propagated and redundant edges removed. (f-k) An example requiring virtual edge merging. (g-h) Virtual edges are created. (i-j) Virtual edges are propagated. (k) At the down-fork $N$, virtual edges $V_J$ and $V_K$ are propagated and merged into $V_{JK}$.}
\label{fig.vedge}
\end{figure}

\subsection{Virtual Edges for Propagate and Pair}

The basic propagate and pair will fail in certain cases, such as in \figref{fig.vedge.failure}. The failure arises from the assumption that the superlevel set is the only thing needed to propagate labels. In this case, label information needs to be communicated between $E$ and $F$, which are connected by the node $D$ in the sublevel set. To resolve this communication issue, virtual edges are used. Virtual edges have 4 associated operations.

\paragraph{Virtual Edge Creation} Virtual edges are created on all up-fork operations. For example in \figref{fig.vedge.createB}, when processing $B$, the endpoints of the fork, $E$ and $F$ are connected with virtual edge $V_B$. Similarly, in \figref{fig.vedge.createD}, when processing up-fork $D$, another virtual edge $V_D$ is created connecting the endpoint, $E$ and $F$.

\paragraph{Label Propagation} Propagating labels across virtual edges is similar to standard propagation with one additional condition. A label can only be propagated if its value is less than that of the up-fork that generated the virtual edge. In other words, for a given label $X$ and a virtual edge $V_Y$, $X$ is only propagated if $f(X)<f(Y)$. Looking at the example in \figref{fig.vedge.labelP}, for the virtual edge $V_B$, only $A$ is propagated because $f(A)<f(B)$. For the virtual edge $V_D$: $A$, $B_L$, and $C$ are all propagated, since they all have values smaller than $D$.

\paragraph{Virtual Edge Merging} When processing down-forks, all incoming virtual edges need to be pairwise merged. \figref{fig.vedge.merge.f} shows an example. When processing down-fork $N$, the virtual edges $V_J$ and $V_K$ are merged into a new virtual edge $V_{JK}$. For the purpose of label propagation, the virtual edge uses its minimum saddle, in this case $J$.

\paragraph{Virtual Edge Propagation} Finally, virtual edges themselves need to be propagated. 
For up-forks, all virtual edges are propagated up to both neighboring nodes. In the case of down-forks, all virtual edges are similarly propagated, as we see in \figref{fig.vedge.edgeP}. During the virtual edge propagation phase, redundant virtual edges can also be culled. For example, the virtual edge $V_D$ is a superset of $V_B$. Therefore, $V_B$ can be discarded. 
The necessity of the virtual edge process can also be seen in \figref{fig.pp.full}. In Figures~\ref{fig.pp.full.i}-\ref{fig.pp.full.l}, the pairing of $L$ with D is only possible because of the virtual edge created by $I$ in \figref{fig.pp.full.i}.

\newcolumntype{C}[1]{>{\centering\let\newline\\\arraybackslash\hspace{0pt}}m{#1}}

\begin{table}[!b]
    \caption{Performance for all datasets tested. Bold indicates the faster algorithm.}
    \label{table.results}
    \resizebox{1.0\textwidth}{!}{%
        \begin{tabular}{|@{ \ }p{3.2cm}||C{1.5cm}||C{1.6cm}|C{1.6cm}|C{1.6cm}|C{1.6cm}|C{1.6cm}||C{1.9cm}|C{2.6cm}| }
             \hline
                \multirow{2}{*}{Data} & \multirow{2}{*}{Figure} &\multicolumn{2}{c|}{Mesh} & \multicolumn{2}{c|}{Reeb Graph Nodes} & \multirow{2}{*}{Cycles} & Multipass & Single-pass \\
                 & & Vertices & Faces & Initial & Cond. & & Time (ms) & Time (ms) \\
             \hline
            \hline
            \multirow{2}{*}{random\_tree\_100}  & & & & \multirow{2}{*}{401} & \multirow{2}{*}{204} & \multirow{2}{*}{0} & \multirow{2}{*}{2.45e-02} & 2.71e-02 (split) \\ 
              & & & & & & & & \textbf{9.06e-03} (join) \\ 
            \hline
            \multirow{2}{*}{random\_tree\_500}  & & & & \multirow{2}{*}{2001} & \multirow{2}{*}{1004} & \multirow{2}{*}{0} & \multirow{2}{*}{0.13} & 0.18 (split)\\ 
              & & & & & & & & \textbf{4.90e-02} (join) \\ 
            \hline
            \multirow{2}{*}{random\_tree\_1000}  & & & & \multirow{2}{*}{4001} & \multirow{2}{*}{2004} & \multirow{2}{*}{0} & \multirow{2}{*}{0.42} & 0.30 (split) \\ 
              & & & & & & & & \textbf{0.11} (join) \\ 
            \hline
            \multirow{2}{*}{random\_tree\_3000}  & & & & \multirow{2}{*}{12001} & \multirow{2}{*}{6004} & \multirow{2}{*}{0} & \multirow{2}{*}{1.10} & 1.98 (split) \\ 
              & & & & & & & & \textbf{0.39} (join) \\ 
            \hline
            \multirow{2}{*}{random\_tree\_5000}  & & & & \multirow{2}{*}{20001} & \multirow{2}{*}{10004} & \multirow{2}{*}{0} & \multirow{2}{*}{2.11} & 3.39 (split) \\ 
              & & & & & & & & \textbf{0.75} (join) \\ 
            \hline
            \hline
            random\_graph\_100 & & & & 401 & 112 & 46 & 1.90e-02 & \textbf{1.76e-02}  \\ 
            \hline
            random\_graph\_500 & & & & 2001 & 542 & 231 & \textbf{0.48} & 0.57  \\ 
            \hline
            random\_graph\_1000 & & & & 4001 & 1010 & 497 & \textbf{0.55} & 0.59  \\ 
            \hline
            random\_graph\_3000 & & & & 12001 & 3014 & 1495 & \textbf{1.71} & 1.91  \\ 
            \hline
            random\_graph\_5000 & & & & 20001 & 5204 & 2400 & \textbf{14.35} & 24.45  \\ 
            \hline
            \hline
            4\_torus & \ref{fig.result.mesh.4torus} &10401 &20814 & 23 & 10 & 4 & 2.06e-03 & \textbf{1.47e-03} \\ 
            \hline
            buddah & \ref{fig.result.mesh.buddha} &10098 &20216& 33 & 14 & 6 & 1.61e-03 & \textbf{1.16e-03} \\ 
            \hline
            david & \ref{fig.result.mesh.david}  & 26138 &52284& 8 & 8 & 3 & 7.82e-04 & \textbf{4.17e-04} \\ 
            \hline
            double\_torus & \ref{fig.result.mesh.doubletorus}  & 3070 & 6144 & 13 & 6 & 2 & 5.29e-04 & \textbf{2.80e-04} \\ 
            \hline
            female & \ref{fig.result.mesh.female}  & 8410 & 16816 & 15 & 8 & 0 & 7.82e-04 & \textbf{3.45e-04} \\ 
            \hline
            flower & \ref{fig.result.mesh.flower} & 4000 & 8256 & 132 & 132 & 65 & 2.80e-02 & \textbf{2.43e-02} \\ 
            \hline
            greek & \ref{fig.result.mesh.greek}  & 39994 & 80000 & 23 & 10 & 4 & 8.62e-04 & \textbf{4.81e-04} \\ 
            \hline
            topology & \ref{fig.result.mesh.topology}  & 6616 & 13280 & 28 & 28 & 13 & 4.34e-03 & \textbf{4.02e-03} \\ 
            \hline
            \hline
            scivis\_contest & \ref{fig.result.scivis} & 194k (avg) & --- & 117 (avg) & 178.2 (avg) & 81.3 (avg) & \textbf{3.82} (total) & 4.18 (total) \\
            \hline
        \end{tabular}%
    }
\end{table}

\section{Evaluation}

We have implemented the described algorithms using Java. Performance reported in Table~\ref{table.results} was calculated on a 2017 MacBook Pro, 3.1 Ghz i5 CPU, 8 GB RAM.

\begin{figure}[!t]
    \centering
    
    \subfigure[double torus\label{fig.result.mesh.doubletorus}]{
    {\begin{minipage}[m]{0.095\linewidth}\centering
    {\includegraphics[height=2.2cm]{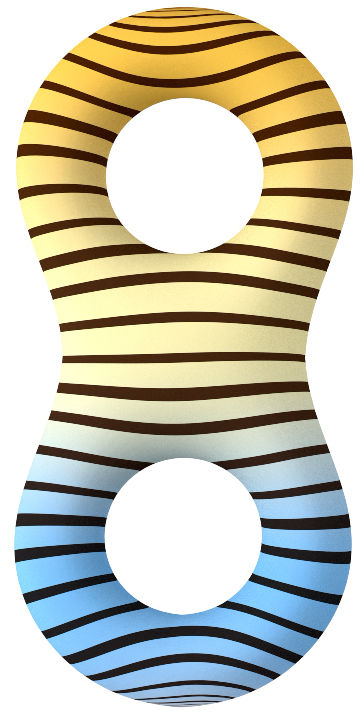}}
    \end{minipage}}
    {\begin{minipage}[m]{0.085\linewidth}\centering
    {\includegraphics[height=2.2cm]{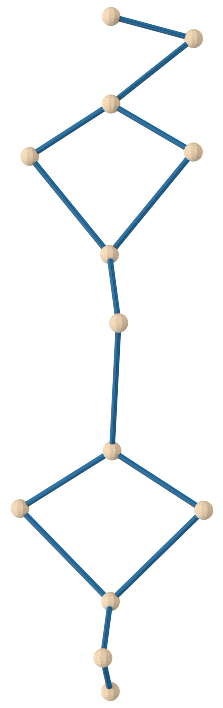}}
    \end{minipage}}
    {\begin{minipage}[m]{0.095\linewidth}\centering
    \includegraphics[height=1.1cm]{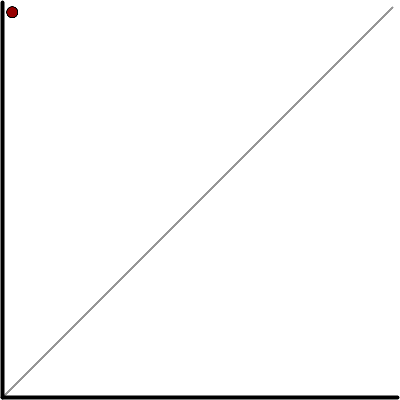}\hspace{10pt}
    \includegraphics[height=1.1cm]{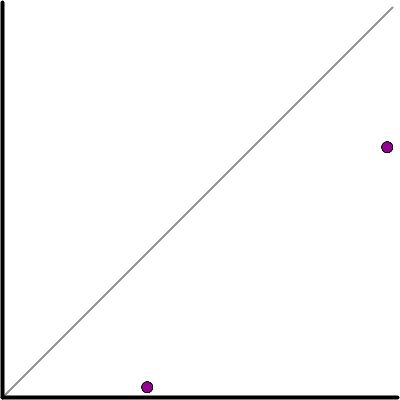}
    \end{minipage}}
    }
    \subfigure[female\label{fig.result.mesh.female}]{
    {\begin{minipage}[m]{0.105\linewidth}\centering
    {\includegraphics[height=2.2cm]{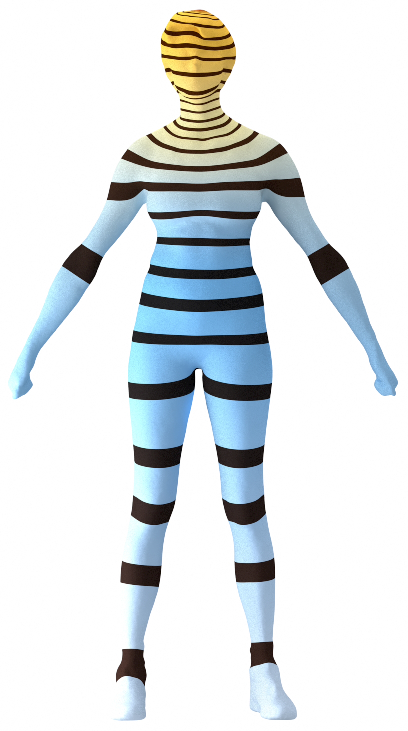}}
    \end{minipage}}
    {\begin{minipage}[m]{0.11\linewidth}\centering
    {\includegraphics[trim= 0 0 1pt 0, clip, height=2.2cm]{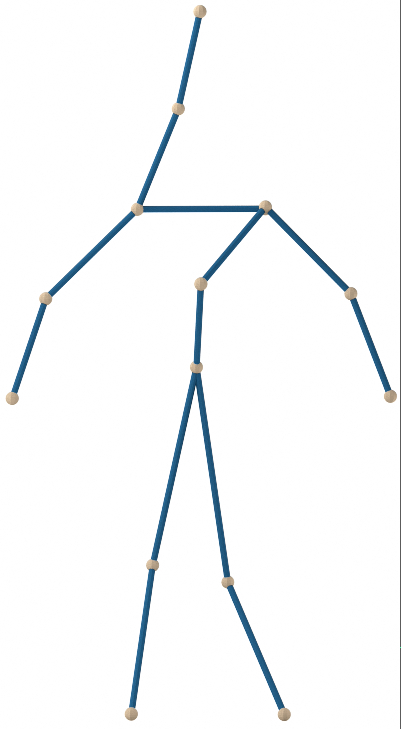}}
    \end{minipage}}
    {\begin{minipage}[m]{0.095\linewidth}\centering
    \includegraphics[height=1.1cm]{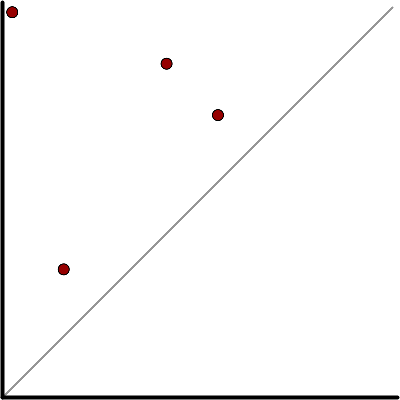}\hspace{10pt}
    \includegraphics[height=1.1cm]{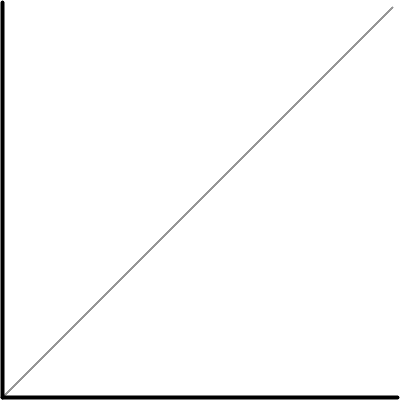}
    \end{minipage}}
    }
    \subfigure[buddah\label{fig.result.mesh.buddha}]{
    {\begin{minipage}[m]{0.095\linewidth}\centering
    {\includegraphics[height=2.2cm]{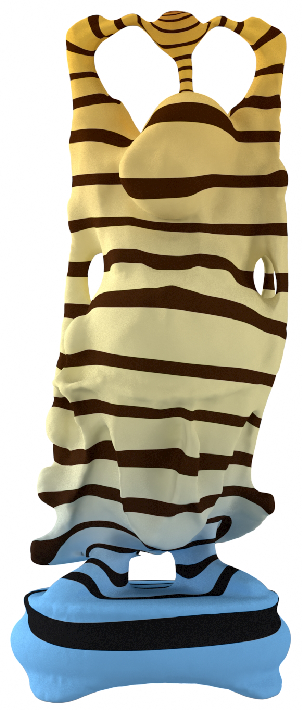}}
    \end{minipage}}
    {\begin{minipage}[m]{0.075\linewidth}\centering
    {\includegraphics[height=2.2cm]{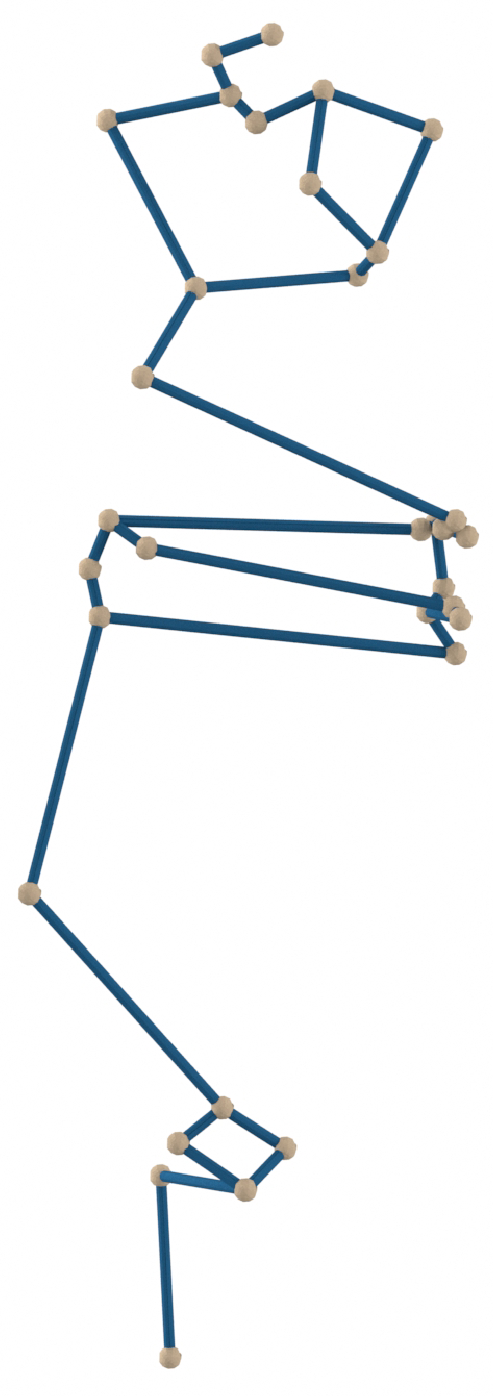}}
    \end{minipage}}
    {\begin{minipage}[m]{0.095\linewidth}\centering
    \includegraphics[height=1.1cm]{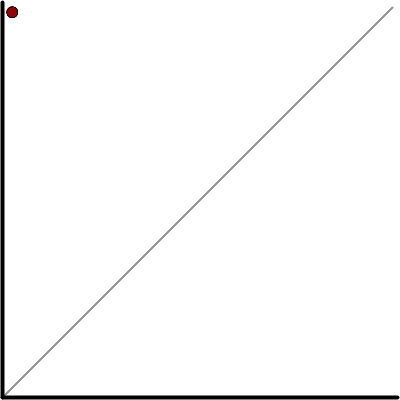}\hspace{10pt}
    \includegraphics[height=1.1cm]{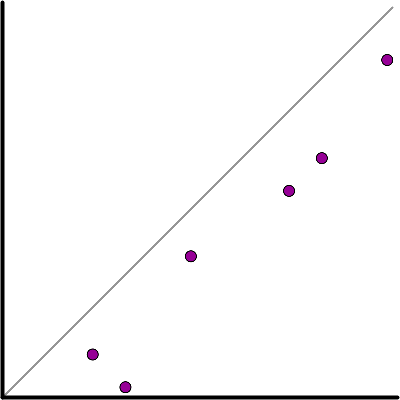}	
    \end{minipage}}
    }

    \subfigure[4 torus\label{fig.result.mesh.4torus}]{
    {\begin{minipage}[m]{0.170\linewidth}\centering
    \includegraphics[height=2.2cm]{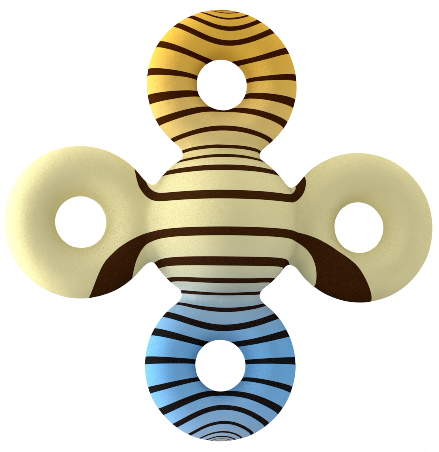}
    \end{minipage}}
    {\begin{minipage}[m]{0.145\linewidth}\centering
    \includegraphics[height=2.2cm]{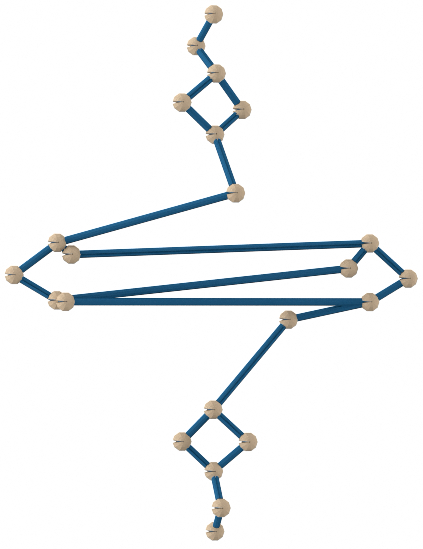}
    \end{minipage}}
    \begin{minipage}[m]{0.095\linewidth}\centering
    \includegraphics[height=1.1cm]{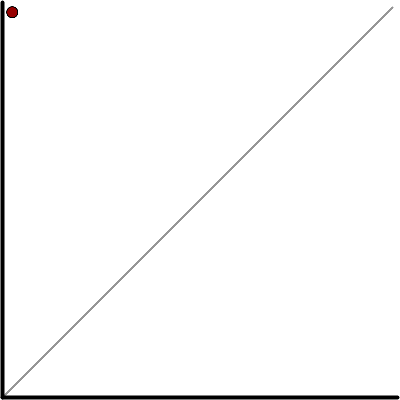}\hspace{10pt}
    \includegraphics[height=1.1cm]{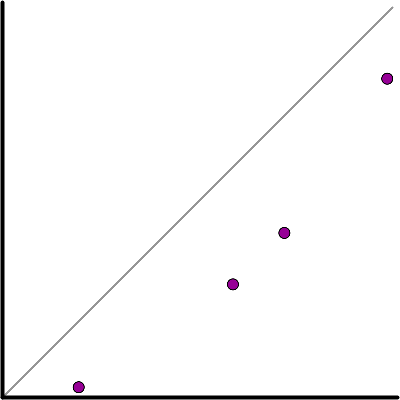}
    \end{minipage}
    }
    \subfigure[david\label{fig.result.mesh.david}]{
    {\begin{minipage}[m]{0.075\linewidth}\centering
    {\includegraphics[height=2.2cm]{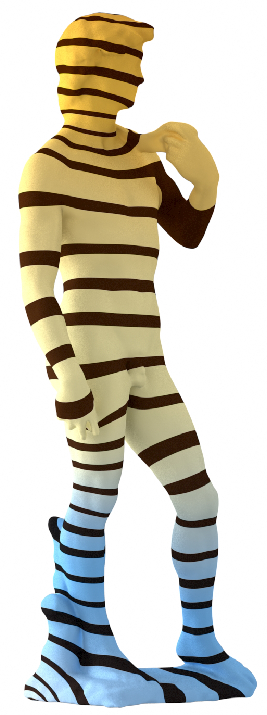}}
    \end{minipage}}
    {\begin{minipage}[m]{0.075\linewidth}\centering
    {\includegraphics[height=2.2cm]{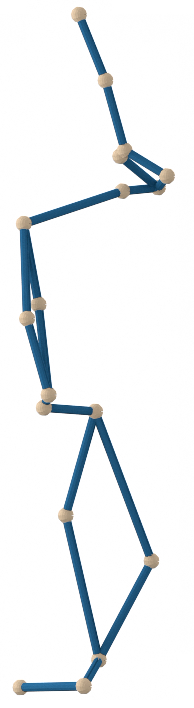}}
    \end{minipage}}
    {\begin{minipage}[m]{0.095\linewidth}\centering
    \includegraphics[height=1.1cm]{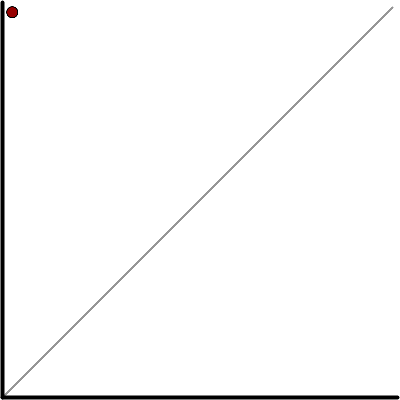}\hspace{10pt}
    \includegraphics[height=1.1cm]{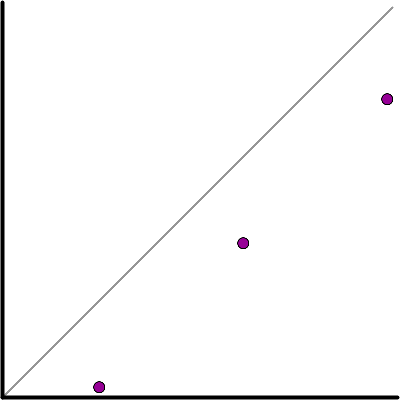}
    \end{minipage}}
    }
    \subfigure[greek\label{fig.result.mesh.greek}]{
    {\begin{minipage}[m]{0.0775\linewidth}\centering
    {\includegraphics[height=2.2cm]{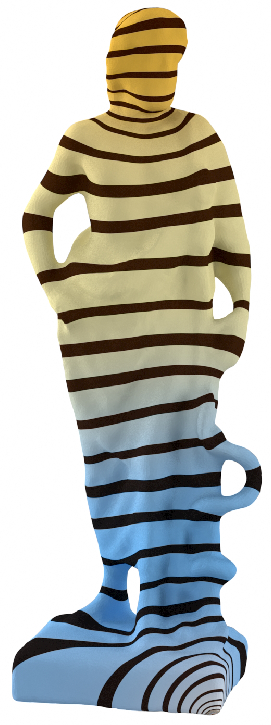}}
    \end{minipage}}
    {\begin{minipage}[m]{0.0625\linewidth}\centering
    {\includegraphics[height=2.2cm]{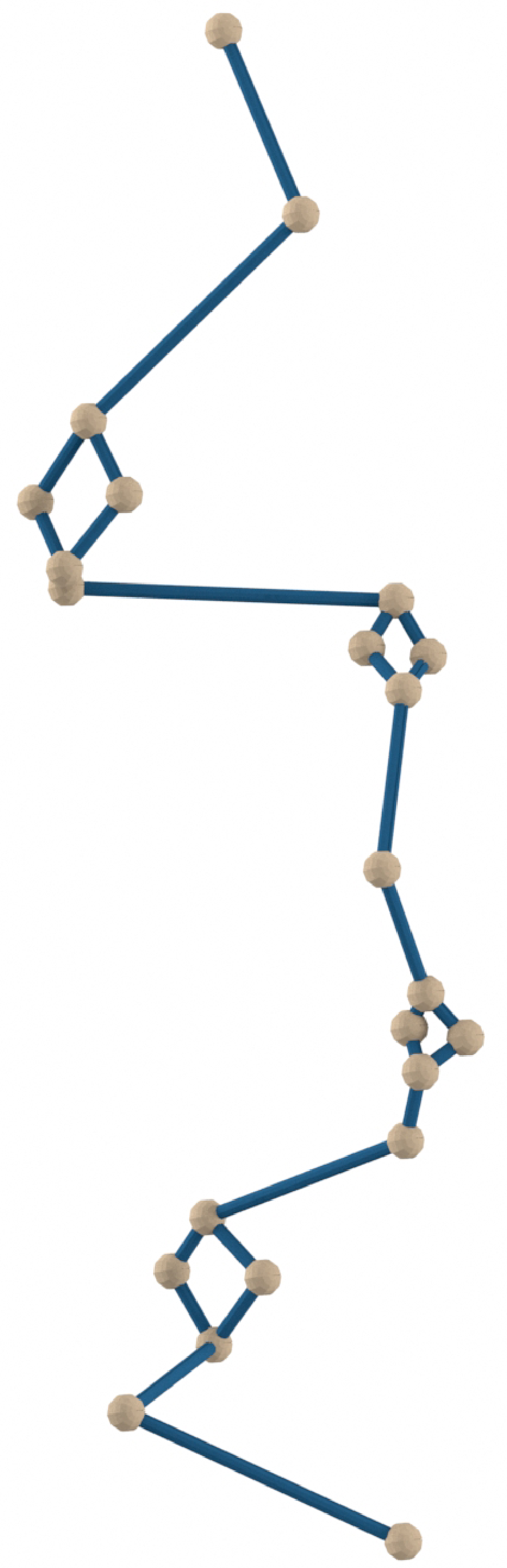}}
    \end{minipage}}
    {\begin{minipage}[m]{0.095\linewidth}\centering
    \includegraphics[height=1.1cm]{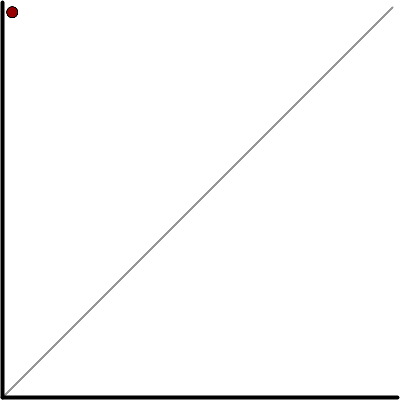}\hspace{10pt}
    \includegraphics[height=1.1cm]{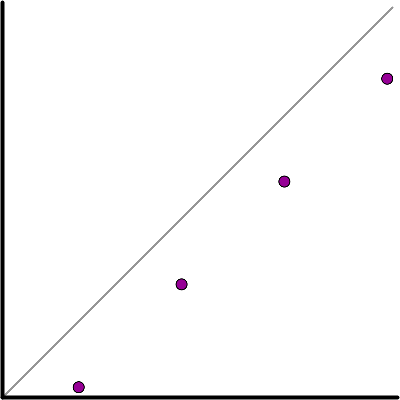}
    \end{minipage}}
    }

    \subfigure[topology\label{fig.result.mesh.topology}]{
    {\begin{minipage}[m]{0.165\linewidth}\centering
    \includegraphics[height=2.2cm]{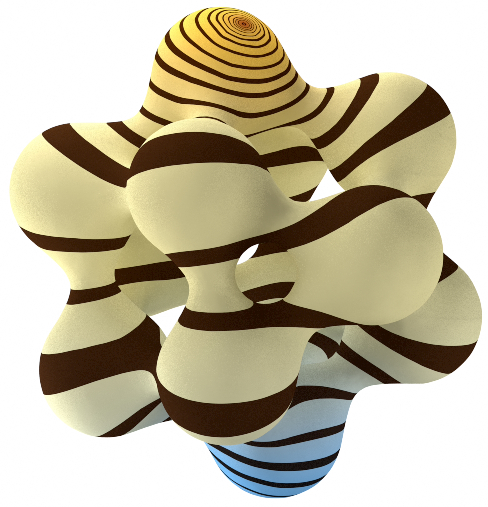}
    \end{minipage}}
    {\begin{minipage}[m]{0.145\linewidth}\centering
    \includegraphics[height=2.2cm]{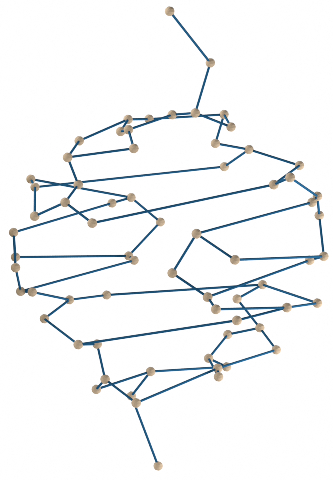}
    \end{minipage}}
    \begin{minipage}[m]{0.095\linewidth}\centering
    \includegraphics[height=1.1cm]{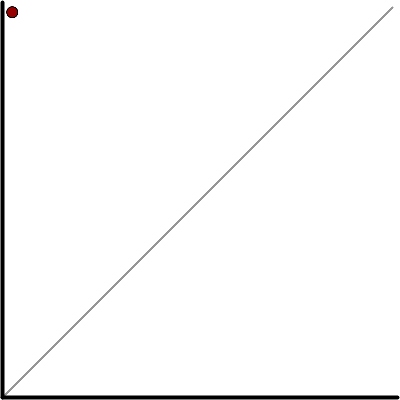}\hspace{10pt}
    \includegraphics[height=1.1cm]{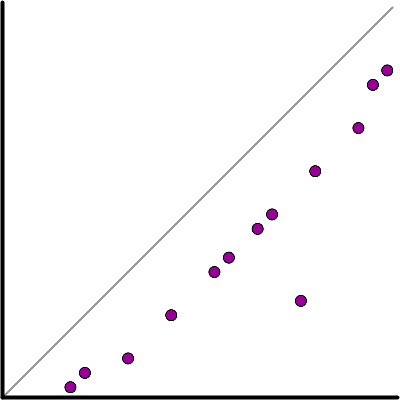}
    \end{minipage}
    }
    \subfigure[flower\label{fig.result.mesh.flower}]{
    {\begin{minipage}[m]{0.175\linewidth}\centering
    \includegraphics[height=2.2cm]{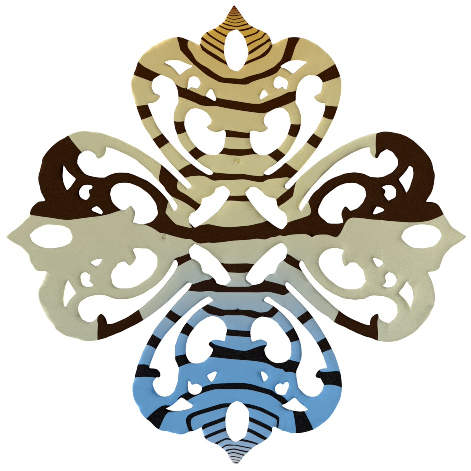}
    \end{minipage}}
    {\begin{minipage}[m]{0.165\linewidth}\centering
    \includegraphics[height=2.2cm]{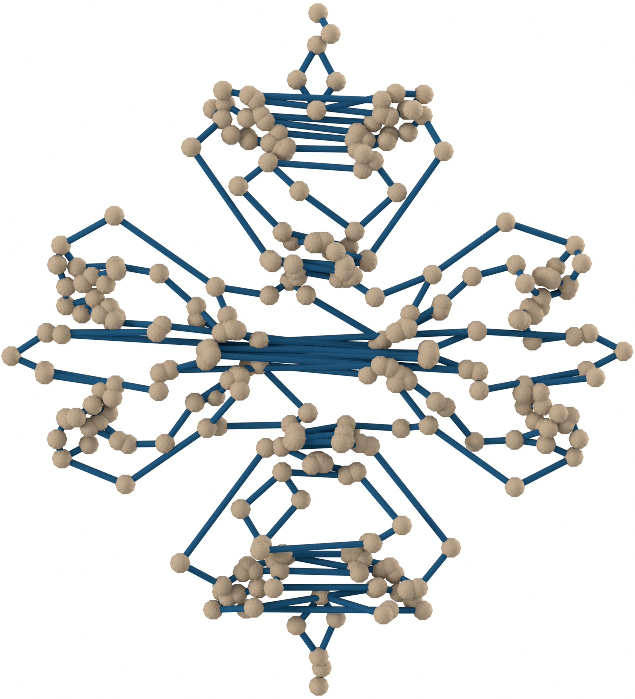}
    \end{minipage}}
    {\begin{minipage}[m]{0.095\linewidth}\centering
    \includegraphics[height=1.1cm]{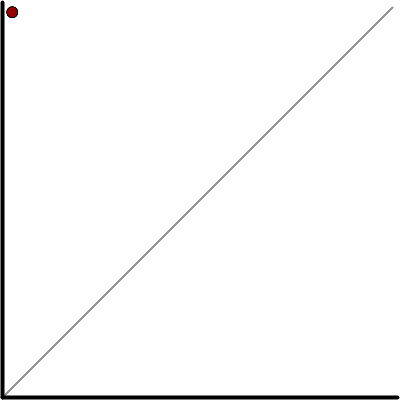}\hspace{10pt}
    \includegraphics[height=1.1cm]{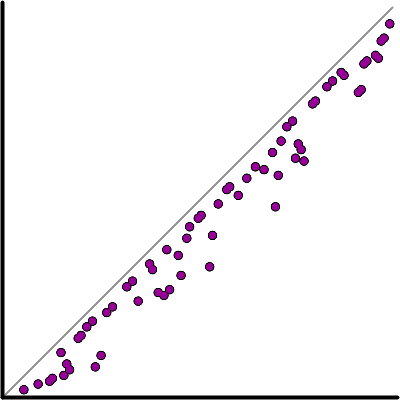}
    \end{minipage}}
    }
    \hfill

    \caption{The meshes colored by the scalar value (left), Reeb graphs (middle), $\Dg_0$ with up-forks in blue and down-forks in red (top), and $\eDg_1$ cycles in purple (bottom) are shown for evaluation.}
    \label{fig.result.mesh}
\end{figure}

\vspace{5pt}
\noindent We investigate the performance of the algorithms using the following:
\begin{itemize}
    \item Synthetic split trees, join trees, and Reeb graphs generated by a Python script. Given a positive integer $n$, where $n=\{100,500,1000,3000,5000\}$, the script creates a fork $G_1$ consisting of a node with valency~$3$ and 3 nodes with valency~$1$ linked to the $3$-valence node. At each iteration $i<n$, another fork is generated, and 1 or 2 of its $1$~valency nodes are glued to the nodes in $G_{i-1}$ with valency 1. Constraining the gluing to a single node at each iteration results in a split tree.
    \item Reeb graphs calculated on publicly available meshes in Figures~\ref{fig.result.mesh}
    and meshes provided by AIM@SHAPE Shape Repository. Reeb graphs were extracted using our own Reeb graph implementation in C++.
    \item Time-series of 120 Mapper graphs taken from the 2016 SciVis Contest\footnote{\url{https://www.uni-kl.de/sciviscontest/}}, a large time-varying multi-run particle simulation, in \figref{fig.result.scivis}. Our evaluation took one realization, smoothing length 0.44, run 50, and calculated the Mapper graphs for all 120 time-steps using the variable \textit{concentration}. Our video, available at \url{https://youtu.be/AcJX4GdzBZY}, shows the entire sequence. The Mapper graphs were generated using a Python script that follows the standard Mapper algorithm~\cite{singh2007topological}.
\end{itemize}

\paragraph{Overview of Results} The performance for the algorithms can be seen in Table~\ref{table.results}. These values were obtained by running the test 1000 times and storing the average compute time. The persistence diagrams of both the single-pass and multipass algorithms were compared in order to verify correctness. For most cases, the single-pass approach outperformed the multipass approach. The exceptions being the random split tree, random graph, and SciVis contest data, each of which we will discuss.

\begin{figure}[!t]
	\centering
	\subfigure[Random split/join tree\label{fig.result.perfplots.tree}]{\includegraphics[width=0.285\linewidth,height=2.5cm]{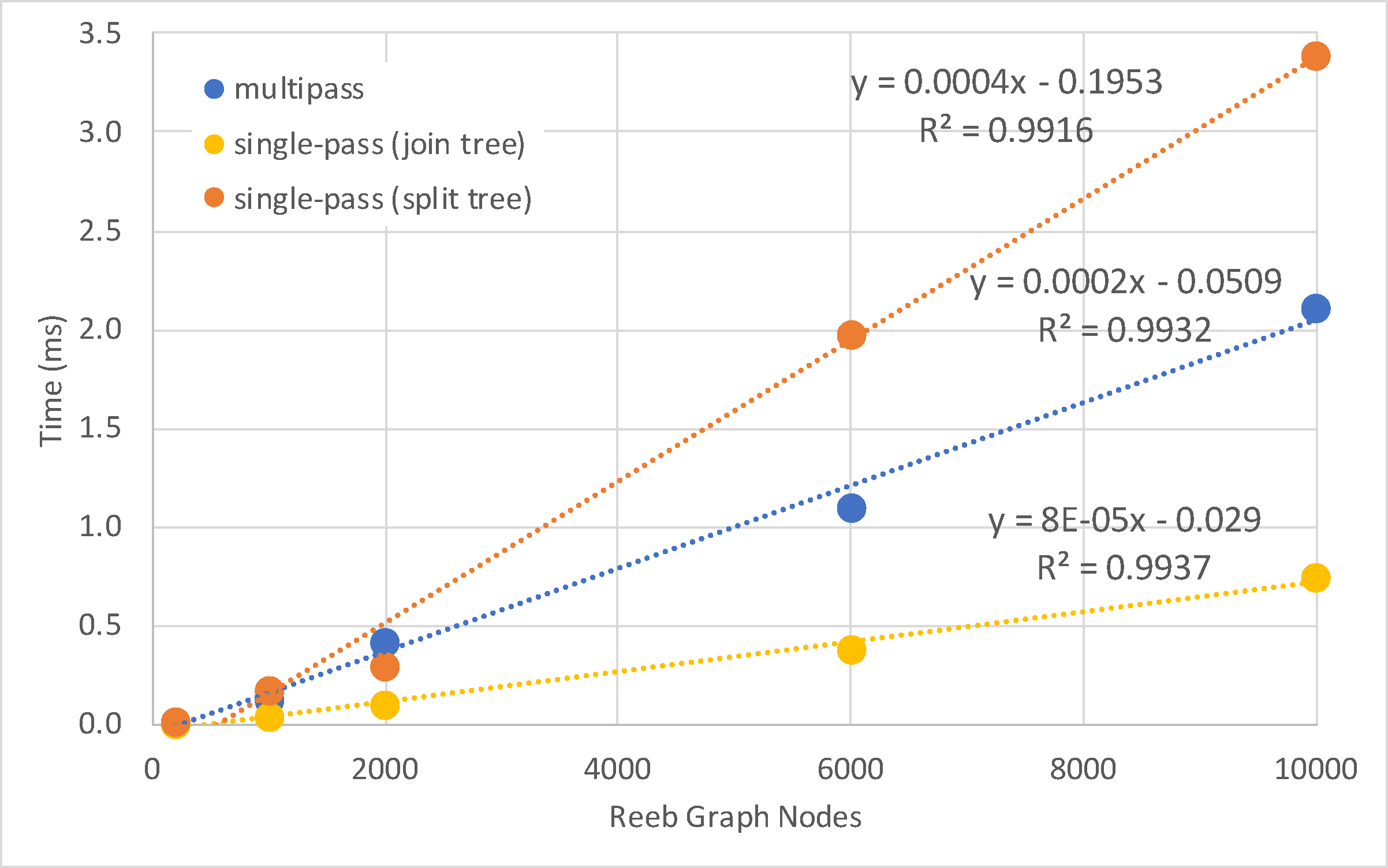}}
	\hfill
	\subfigure[Random graph\label{fig.result.perfplots.graph}]{\includegraphics[width=0.285\linewidth,height=2.5cm]{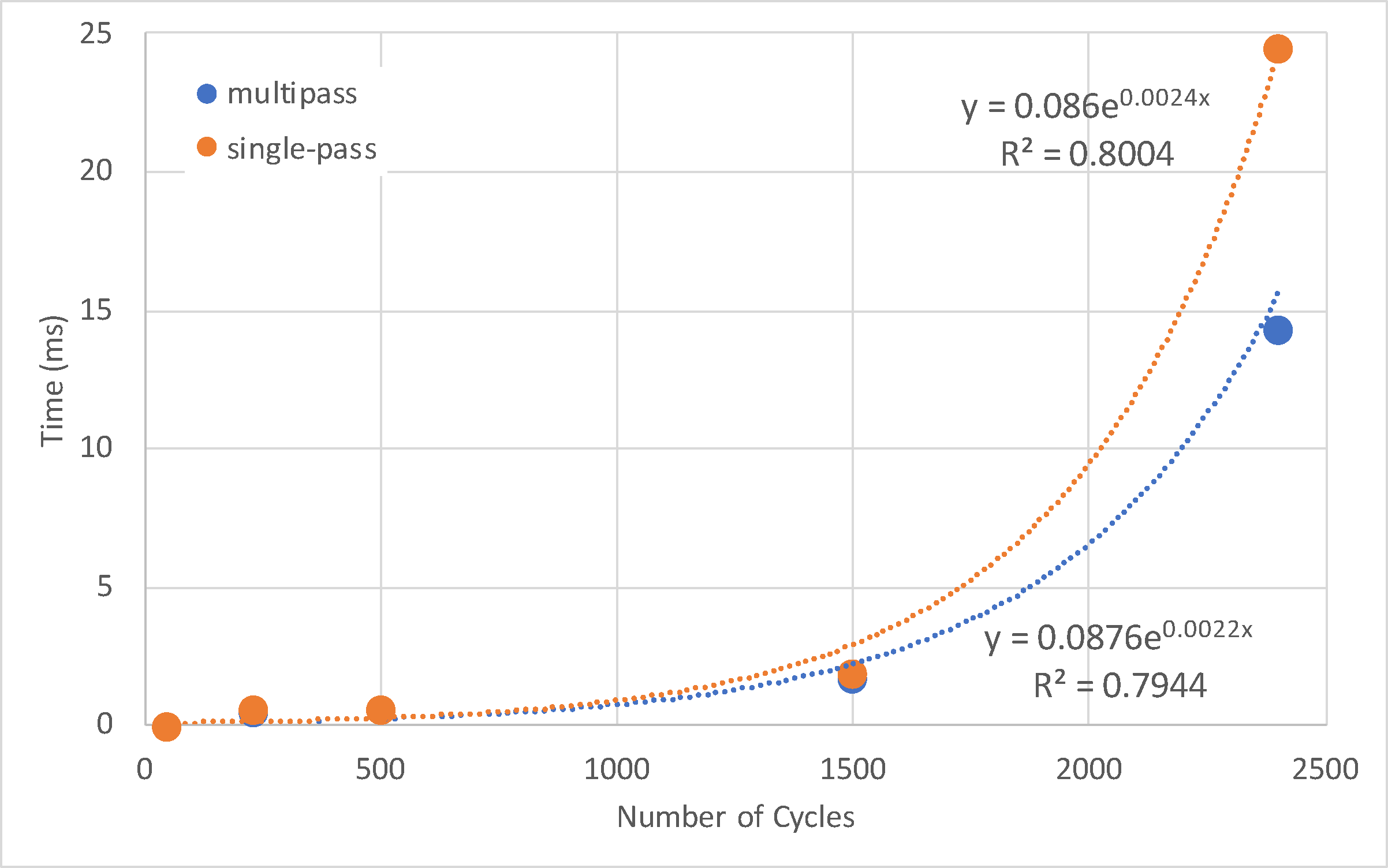}}
	\hfill
	\subfigure[Cutting cycles in random\_graph\_5000\label{fig.result.varyloops}]{\includegraphics[width=0.415\linewidth,height=2.5cm]{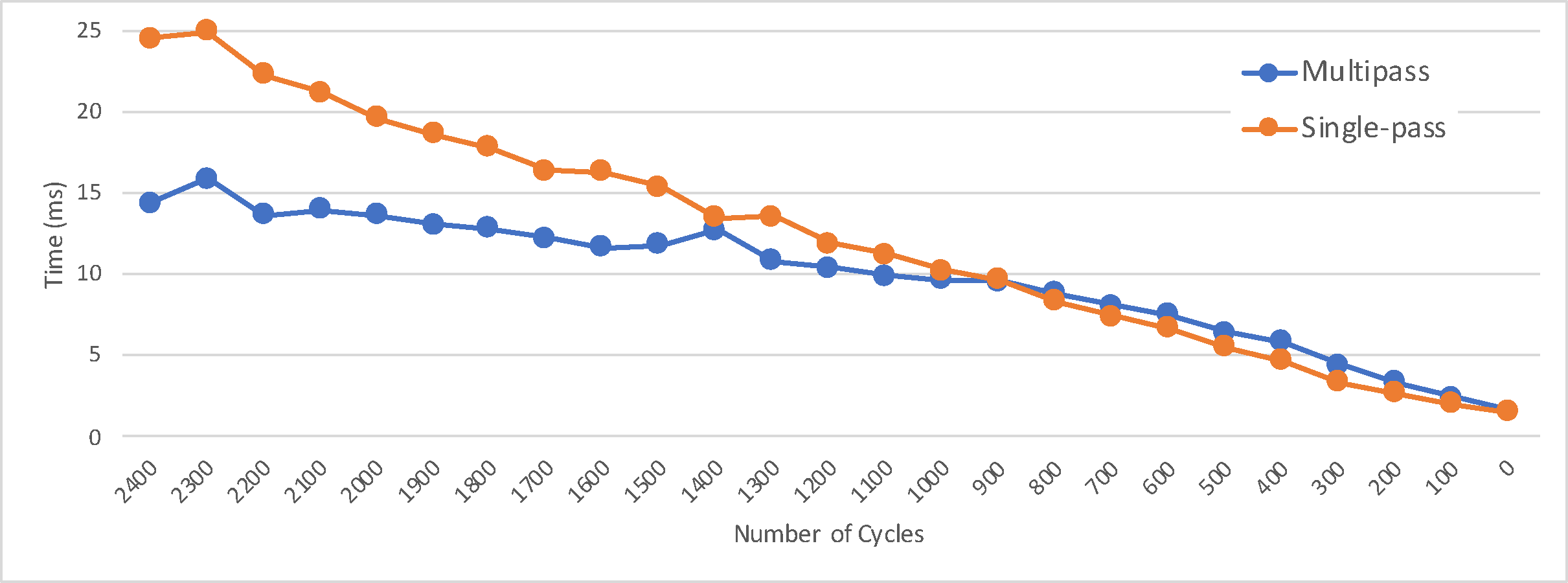}}	
	
    \caption{Plots of the compute time for various input sizes to (a) the random split/join tree and (b) the random graph for Table~\ref{table.results}. (c) Performance results when cutting cycles in the random\_graph\_5000. As more cycles that are cut, the single-pass algorithm begins to outperform the multipass variant.}
    \label{fig.result.perfplots}
\end{figure}

\paragraph{Random Split Tree vs.\ Join Tree} We compared the exact same tree structures as split trees and join trees by negating the function value of the input tree. The performance observed in Table~\ref{table.results} and \figref{fig.result.perfplots.tree} shows that the join tree performs significantly better than the split tree. The explanation for this is quite simple. The join tree consists of exclusively down-forks, while the split tree consists of exclusively up-forks. Since only up-forks generate virtual edges, the split tree created and processed many virtual edges, while the join tree has none. In fact, split trees represent one worst case by generating many unneeded virtual edges. From a practical standpoint, the algorithm can avoid situations like this by switching sweep directions (i.e.\ top-to-bottom), when the number of up-forks is significantly larger than the number of down-forks.

\paragraph{Random Graph} We next investigate the performance of randomly generated Reeb graphs, shown in Table~\ref{table.results} and \figref{fig.result.perfplots.graph}. These Reeb graphs consist predominantly of cycles. This represents another type of worst case, since many up-forks generate virtual edges, which are then merged into even more virtual edges at the down-forks. To verify this, we ran an experiment, as seen in \figref{fig.result.varyloops}, that randomly cuts $n$ cycles in the starting Reeb graph random\_graph\_5000 containing 2400 cycles. The break even was about 900 cycles (about $25\%$ essential and $75\%$ non-essential forks).

\paragraph{SciVis Contest Data} The SciVis contest data was ``cycle heavy'' as can be seen in the persistence diagram of \figref{fig.result.scivis}. Given the random graph analysis, it is unsurprising that the performance of the single-pass approach was lower than the multipass approach.

\section{Discussion \& Conclusion}

Pairing critical points is a key part of the TDA pipeline---the Reeb graphs capture complex structure, but direct representation is impractical. Critical point pairing enables a compact visual representation in the form of a persistence diagram. The value of representing a dataset with the persistence diagram is the simplicity and efficiency. Persistence diagrams avoid the occlusions problems of normal 3D datasets (e.g., the internal structure of \figref{fig.result.scivis}), and they avoid the potential confusion of direct representation of the Reeb graph (e.g., the Reeb graph of \figref{fig.result.mesh.flower}). In addition, they provide sharp visual cue for time-varying data (see our video).

Our results showed that although the single-pass algorithm tended to outperform the multipass algorithm, there was no clear winner. We point out some advantages and disadvantages for each.
The multipass algorithm has an advantage in simplicity of implementation. Once the merge tree and branch decomposition are implemented, the only necessity is repeated calls to those algorithms. This approach also has a potential advantage for (limited) parallelism. First, processing join and split trees in parallel, then all essential up-forks. 
The single-pass algorithm showed a slight edge in performance, particularly for data with a balance between essential and non-essential forks. The other significant advantage of the approach is that it is in fact a single-pass approach, only visiting critical points once. This is useful for streaming or time-varying data, where the critical points arrive in order, but analysis cannot wait for the entire data to arrive.

\paragraph{Acknowledgments} This project was supported in part by National Science Foundation (IIS-1513616 and IIS-1845204). Mesh data are provided by AIM@SHAPE Repository.

\bibliographystyle{splncs04}

\bibliography{p-n-p}

\end{document}